\documentclass[a4paper,11pt]{article}
\usepackage[margin=0.5in]{geometry}
\usepackage{enumerate}
\usepackage{jcappub}
\usepackage{mathtools}

\usepackage{amsmath}
\usepackage{amsthm}
\usepackage{amssymb}
\usepackage{natbib}
\usepackage{aas_macros}
\usepackage{xspace}
\usepackage[utf8]{inputenc}
\usepackage[T1]{fontenc}
\usepackage{hyperref}
\usepackage{caption}
\usepackage{verbatim}
\usepackage{lipsum}

\hypersetup{
	unicode,
	pdfauthor={Author One, Author Two, Author Three},
	pdftitle={Constraints on primordial non-Gaussianity from Quaia},
	pdfsubject={Constraints on primordial non-Gaussianity from Quaia},
	pdfproducer={LaTeX},
	pdfcreator={pdflatex}
}
\renewcommand{\arraystretch}{1.7}
\usepackage{graphicx, color}
\newcommand{\planck}{{\sl Planck}\xspace}
\newcommand{\quaia}{\textit{Quaia}\xspace}
\newcommand{\gaia}{{\sl Gaia}\xspace}
\newcommand{\unwise}{{\sl unWISE}\xspace}
\newcommand{\nv}{\hat{\bf n}}

\newcommand{\lmax}{\ell_{\rm max}}
\newcommand{\lmin}{\ell_{\rm min}}

\newcommand{\fNL}{f_{\rm NL}}
\newcommand{\lcdm}{$\Lambda$CDM\xspace}
\newcommand{\hpx}{{\tt HEALPix}\xspace}
\newcommand{\nmt}{{\tt NaMaster}\xspace}

\usepackage{xcolor}

\title{\boldmath Constraints on primordial non-Gaussianity from \quaia}

\author[a, b,c]{Giulio Fabbian,}
\author[d]{David Alonso,}
\author[e]{Kate Storey-Fisher,}
\author[d]{Thomas Cornish}
\affiliation[a]{Institut d’Astrophysique Spatiale, Universit\'e Paris-Saclay, CNRS, 91405, Orsay, France}
\affiliation[b]{School of Physics and Astronomy, Cardiff University, The Parade, Cardiff, Wales CF24 3AA, U.K.}
\affiliation[c]{Kavli Institute for Cosmology Cambridge, Madingley Road, Cambridge CB3 0HA, U.K.}
\affiliation[d]{Department of Physics, University of Oxford, Denys Wilkinson Building, Keble Road, Oxford OX1 3RH, U.K.}
\affiliation[e]{Kavli Institute for Particle Astrophysics and Cosmology, Stanford University, 452 Lomita Mall, Stanford, CA 94305, U.S.A.}
\emailAdd{giulio.fabbian@universite-paris-saclay.fr}
\emailAdd{David.Alonso@physics.ox.ac.uk}
\emailAdd{kstoreyf@stanford.edu}

\abstract{We analyse the large-scale angular clustering of quasars in the \gaia-\unwise quasar catalog, \quaia, and their cross-correlation with maps of the lensing convergence of the Cosmic Microwave Background (CMB), to constrain the level of primordial non-Gaussianity (PNG). Specifically, we target the scale-dependent bias that would be induced by PNG on biased tracers of the matter inhomogeneities on large scales. The \quaia sample is particularly well suited for this analysis, given the large effective volume covered, and our ability to map out the main potential sources of systematic contamination and mitigate their impact. Using the universality relation to characterise the response of the quasar overdensity to PNG ($p_\phi=1$), we report constraints on the local-type PNG parameter $\fNL$ of $\fNL=-20.5^{+19.0}_{-18.1}$ (68\% C.L.) by combining the quasar auto-correlation and its cross-correlation with CMB lensing in two tomographic redshift bins (or $\fNL=-28.7^{+26.1}_{-24.6}$ if assuming a lower response for quasars, $p_\phi=1.6$). The error on $\fNL$ can be further improved if the cross-correlation between the tomographic redshift bins is included. Using the CMB lensing cross-correlations alone, we find $\fNL=-13.8^{+26.7}_{-25.0}$ and $\fNL = -15.6^{+42.3}_{-34.8}$ for $p_\phi=1$ and $p_\phi=1.6$ respectively. These are the tightest constraints on $\fNL$ to date from angular clustering statistics and cross-correlations with CMB lensing.}
\begin{document}
\maketitle
\flushbottom

  \section{Introduction} \label{sec:intro}
    One of the most general predictions of inflation, our most successful theory describing the early stages of our Universe, is that the primordial metric fluctuations should be very close to Gaussian-distributed \cite{astro-ph/0406398,0907.5424}. This prediction is borne out by the statistics of the CMB temperature and polarization fluctuations, for which no significant departures from Gaussianity have been detected \cite{1905.05697,2504.00884}. In detail, however, different inflationary models predict such departures at different levels, and improving current constraints on the level of primordial non-Gaussianity (PNG) is vital to advance our understanding of the early Universe.
  
    Of particular relevance is the case of ``local'' PNG. In this case, the level of non-Gaussianity in the primordial gravitational potential $\Phi({\bf x})$ is parametrised as
    \begin{equation}\label{eq:fnl_def}
      \Phi({\bf x})=\phi_G({\bf x})+\fNL\left(\phi_G^2({\bf x})-\langle\phi_G^2\rangle\right),
    \end{equation}
    where $\phi_G({\bf x})$ is a Gaussian random field, $\langle\cdots\rangle$ denotes an ensemble average, and $\fNL$ quantifies the amplitude of the non-Gaussian contributions, caused by the non-linear $\phi_G^2$ term. While single-field models predict a negligibly small amplitude for local PNG \cite{astro-ph/0407059}, multi-field scenarios generally can produce PNG amplitudes of order $\fNL\sim 1$ \cite{byrnes2010}. A constraint on $\fNL$ at this level would therefore represent a powerful probe for a wider class of inflationary models including, e.g., curvaton, modulated reheating, models with an inhomogenous end to inflation, hybrid inflation or N-flation (see e.g. \cite{deputter2017} and references therein). Although current constraints on $\fNL$ are driven by the study of the CMB three-point function (with an uncertainty $\sigma(\fNL)=5$ \cite{2504.00884}), significant further progress in this direction is difficult due to cosmic variance. Instead, three-dimensional tracers of the large-scale structure, such as galaxies, offer a promising alternative to reach the desired $\sigma(\fNL)\lesssim 1$ threshold. The response of the small-scale power to long-wavelength fluctuations caused by the initial non-linearity in $\Phi$ modifies the relation between the gravitational potential (and its derivatives) and the overdensity of any biased LSS tracer. In particular, local PNG produces an additional, scale-dependent, contribution to the linear bias relation, which is most relevant on large scales \cite{0710.4560,png-lss-review}. PNG thus manifests itself in the clustering statistics of biased tracers already at the level of two-point functions, with additional information encoded in higher-order correlators. This makes PNG one of the most relevant science cases for LSS studies.

    The main challenges for local PNG searches with LSS are related to the $\sim1/k^2$ behaviour of the scale-dependent bias contribution (where $k$ is the Fourier-space wavenumber) \cite{0710.4560}. Firstly, since the $\fNL$ signature peaks on large scales, maximising the effective volume covered by the LSS sample under study is of critical importance. This makes high-redshift quasars strong candidates to constrain $\fNL$ \cite{1904.08859,2305.07650,2309.15814}, although other tracers have been used and proposed, including the Cosmic Infrared Background \cite{2210.01049}, radio continuum surveys \cite{1402.2290}, and Luminous Red Galaxies (LRGs) \cite{2412.10279,2411.17623}. Secondly, the depth and completeness of any galaxy sample is never perfectly homogeneous, and multiple systematic effects may lead to spurious fluctuations in the observed distribution of galaxies. Since many of these sources of contamination are associated with Galactic foregrounds or variations in observing conditions, their impact is often dominant on large angular scales and can lead to a significant bias on the inferred value of $\fNL$. Typical examples of systematics manifesting as a spurious modulation of power at the large scales, and that can mimic effects of primordial non-Gaussianities, are dust extinction, given the large apparent angular size of the Milky Way, or the expected variation of calibration or observing conditions between subsequent exposures of a telescopes carrying out a survey on large sky fractions \cite{weaverdyck2021}. Therefore, a robust measurement of $\fNL$ requires a tight control over the potential sources of systematic contamination in the data, or the use of data combinations that are resilient against them.

    In this paper we present constraints on $\fNL$ obtained by combining the clustering of quasars in the \gaia-\unwise quasar sample, \quaia \cite{2306.17749}, with the gravitational lensing of the CMB as reconstructed from the \planck temperature and polarization fluctuations \cite{2206.07773}. Specifically, we will employ a tomographic approach, measuring and analysing the angular power spectrum of the \quaia sources in bins of redshift, as well as their large-scale cross-correlation with CMB lensing. Although this approach discards information from large-scale radial modes in the quasar distribution, it allows for a completely self-consistent analysis while avoiding potential systematics in the spectro-photometric redshifts of the sample. Crucially, the data combination used here addresses the challenges described in the previous paragraph: the \quaia sample covers one of the largest effective volumes mapped by current LSS datasets. Since it is mostly based on data from space-borne experiments, the selection function of the sample and its associated spatial inhomogeneity can be accurately characterised, and the impact of residual contamination may be further mitigated through the use of the CMB lensing cross-correlations. It is also similar to other recent $\fNL$ analyses that exploited the projected clustering of photometric sources (quasars and LRGs) and CMB lensing cross-correlations, allowing us to compare our results with those using other state-of-the-art datasets. 

    This paper is structured as follows: Section \ref{sec:data} presents the datasets used (\quaia and \planck CMB lensing). Section \ref{sec:meth} describes the methods used to estimate clustering statistics and their uncertainties, and to quantify the presence of sky contamination. It also presents the model and likelihood used to interpret our data. Our results are then described in Section \ref{sec:res}, including the battery of tests carried out to quantify their robustness. We summarise these results and conclude in Section \ref{sec:conc}.
  
  \section{Data} \label{sec:data}
  \subsection{Quaia}\label{ssec:data.quaia}
    \begin{figure}
        \centering
        \includegraphics[width=0.8\textwidth]{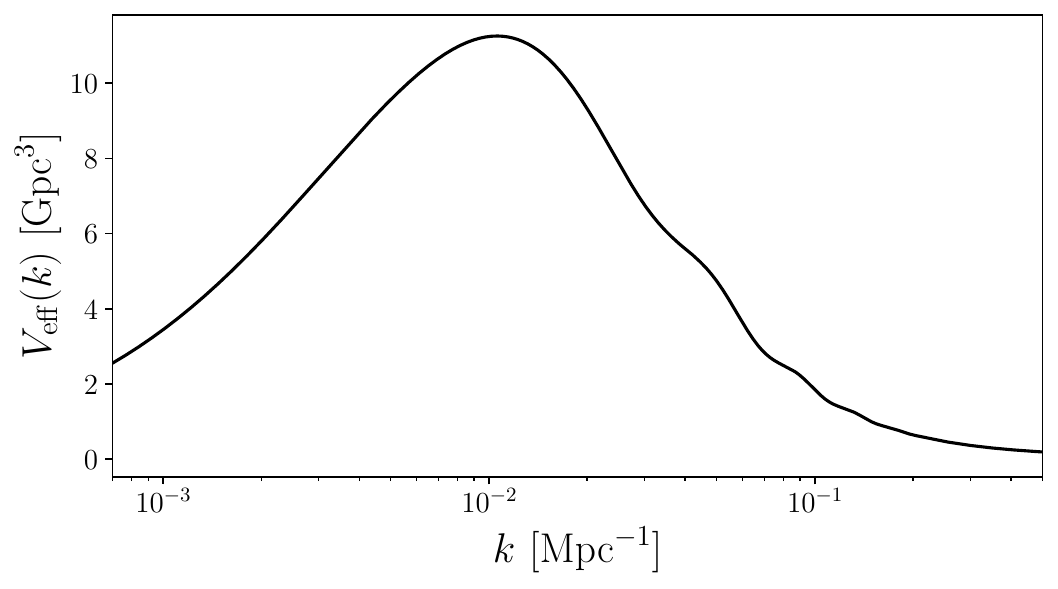}
        \caption{Left: effective volume $V_{\rm eff}$ covered by \quaia as a function of Fourier wavenumber $k$ (see \cite{astro-ph/9706198} for the definition of $V_{\rm eff}(k)$). }
        \label{fig:veff}
    \end{figure}
        \begin{figure}
        \centering
        \includegraphics[width=0.5\textwidth]{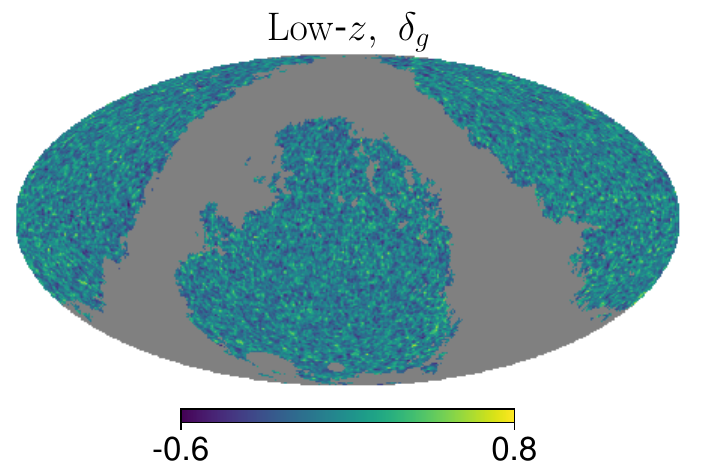}\includegraphics[width=0.5\textwidth]{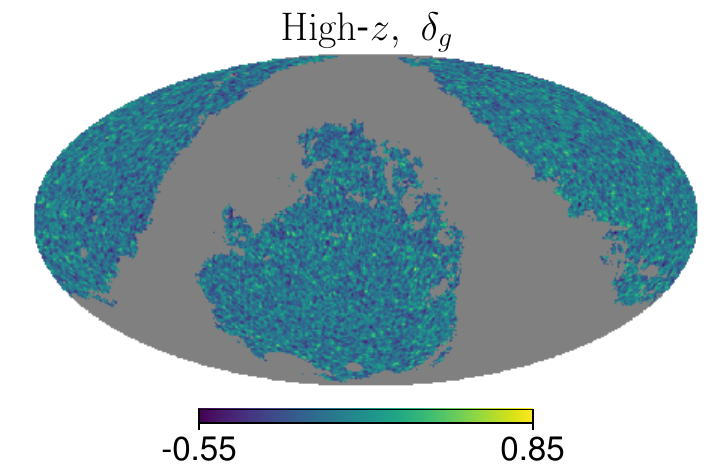}
        \includegraphics[width=0.5\textwidth]{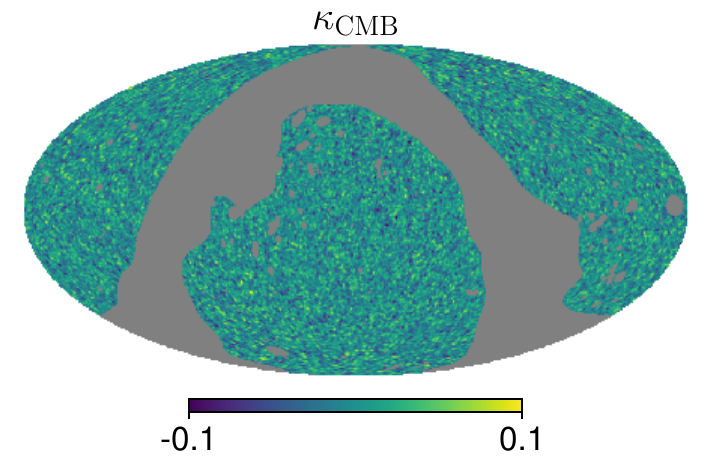}\includegraphics[width=0.458\textwidth]{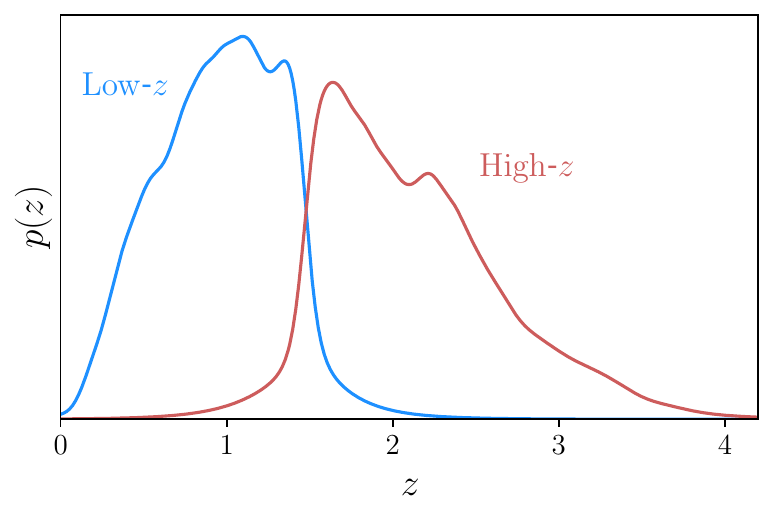}
        \caption{Top: maps of the QSO overdensity of \quaia used in this work. Bottom: the reference \planck PR4 CMB lensing map used in our analysis is shown on the left while the redshift distribution of the two tomographic redshift bins $p(z)$ is shown on the right. For visualization purpose, both the QSO and CMB lensing maps have been smoothed with a Gaussian kernel of 1 deg FWHM.}
        \label{fig:maps}
    \end{figure}
    
    \begin{figure}
        \centering
        \includegraphics[width=0.82\linewidth]{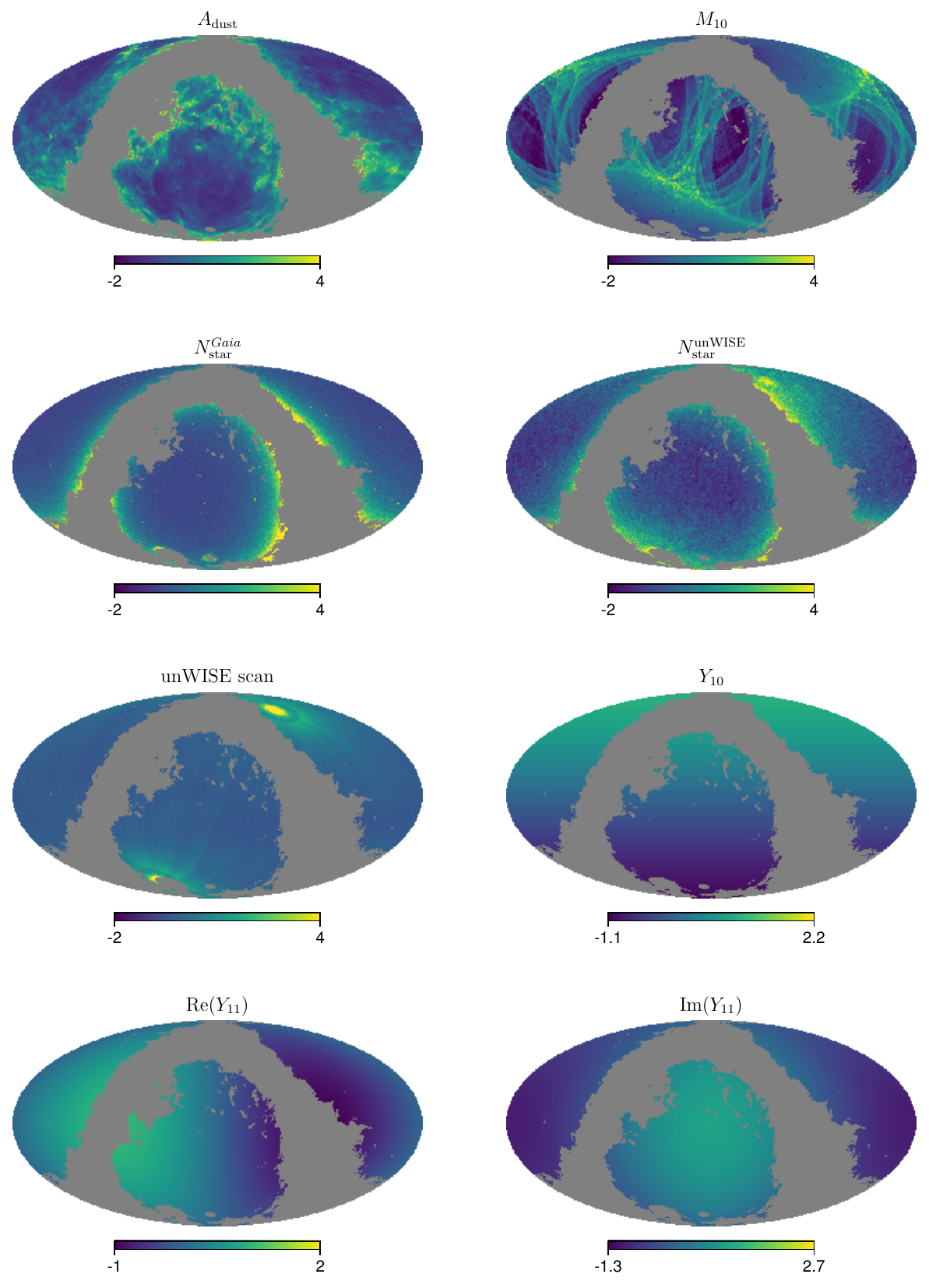}
        \caption{Systematic templates used to construct the \quaia selection function, and to quantify the impact of large-scale contamination. In all cases, we show the local deviation with respect to the mean as a fraction of the root-mean squared (rms) in each quantity, with mean and rms defined within the shown footprint. I.e. for systematic template $S(\nv)$, we show $(S(\nv)-\bar{S})/\sigma_S$, with $\bar{S}$ and $\sigma_S$ the mean and rms deviation within the footprint. In addition to these 8 templates, we use two additional maps of the source density in \gaia and \unwise in the region around the Magellanic clouds, not shown here. }
        \label{fig:sysmaps}
    \end{figure}
    We use data from the \gaia-\unwise quasar sample, \quaia \cite{2306.17749}. The catalog covers the entire sky and contains $\sim1.3$ million sources with magnitudes $G<20.5$. The catalog was constructed by cross-matching the quasar candidates in the third \gaia data release \cite{2208.00211} against photometry from the \unwise reprocessing of the Wide-Field Infrared Survey Explorer (WISE) \cite{1909.05444}, improving the level of star contamination and the quality of the estimated spectro-photometric redshifts.

    Crucial to this analysis, the \quaia sample traces one of the largest volumes covered by existing samples, several times larger than previous datasets, such as the eBOSS quasar sample \cite{2007.09000} (see Table 1 in \cite{2306.17749} and Fig.~\ref{fig:veff}). According to the definition of spanning and effective volume adopted in \cite{2306.17749}, which covered a specific redshift range for the sake of comparison with other data sets, contemporary data sets such as DESI have a larger effective volume than \quaia (by about $\sim30\%$), thanks to their 2.5 times higher number density. However, the DESI spanning volume is still $\sim 2.5$ times smaller than the one of \quaia, as it covers a reduced redshift range and a smaller sky fraction with effectively a comparable number of objects. Other samples derived from photometric full-sky surveys such as WISE (AGNs or unWISE galaxies \cite{wise, unwise}) span smaller volumes and have lower redshift accuracy. This allows us to access the very large scales over which the PNG signal peaks. As highly-biased tracers of the matter density, quasars are also particularly sensitive to PNG. Furthermore, since \quaia is only based on data from space-borne experiments, the list of potential contaminants affecting the observed number of quasars is significantly shorter than that of ground-based surveys. This allows for a tighter control over the survey selection function, as well as the potential sources of large-scale systematics.

    For our fiducial analysis we split the \quaia sample into two redshift bins corresponding to sources with spectro-photometric redshifts above and below the median redshift of the sample $z_p=1.47$. These are the same bins used in the analyses of \cite{2306.17748,2410.24134}, and we will refer to them as the ``Low-$z$'' and ``High-$z$'' redshift bins respectively. Splitting the catalog in these two redshift bin represents a good compromise to introduce tomographic capabilities while keeping shot noise in each redshift bin under control. In \cite{2402.05761} we found in fact that splitting the catalog in more than two redshift bins led us to a minor loss in the precision of derived cosmological constraints. The redshift distributions for both bins were estimated by stacking the Gaussian probability density functions of each source, which was found to be a good approximation in \cite{2306.17748}. The final redshift  distribution $p(z)$ for both bins is shown in Fig.~\ref{fig:maps}. The survey selection function for each bin, which characterises the fraction of sources missed by the catalog across the sky, was estimated as described in \cite{2306.17749}. The catalog and survey selection function were then used to construct quasar overdensity maps as described in \cite{2306.17748}. Since we are only interested in the clustering signal on relatively large scales, these maps were generated using the \hpx \footnote{\url{https://healpix.sourceforge.io}} pixelisation scheme with resolution parameter $N_{\rm side}=256$, corresponding to a pixel size of $\delta\theta_{\rm pix}\sim0.23^\circ$, and they are shown in Fig.~\ref{fig:maps}. As part of our robustness tests, we will also consider a single redshift bin containing all \quaia sources with $0<z_p\lesssim4.5$. 

    To quantify the presence of residual contamination in these maps, we consider template maps for 10 different potential systematics. These include: i) a dust extinction map constructed by \cite{2306.03926}, ii) two stellar contamination maps built by selecting random sources from \gaia and \unwise, iii) templates tracking the scanning patterns of \gaia and \unwise, iv) stellar density maps in the regions around the Large and Small Magellanic Clouds, and v) three dipole components, corresponding to the spherical harmonic functions with $\ell=1$\footnote{The dipole in the \quaia sample is being investigated in a separate work \cite{WilliamsInprep}.} ($Y_{1,0}, Y_{1,\pm 1}$). These systematic templates are shown in Fig.~\ref{fig:sysmaps}. Note that the \quaia selection function was constructed as a non-linear combination of the systematic templates i) -- iv) listed above. As we describe in Section \ref{ssec:meth.cls}, these templates (as well as the dipole components) are linearly deprojected from the galaxy overdensity maps as part of the power spectrum estimation pipeline, to further remove residual contamination.

  \subsection{Planck}\label{ssec:data.planck}
    We use maps of the CMB lensing convergence from the \planck experiment. Specifically, we use the lensing maps based on data from the \planck ``PR4'' release\footnote{\url{https://github.com/carronj/planck_PR4_lensing}}, generated with the {\tt NPIPE} pipeline \cite{2206.07773} (see Fig.~\ref{fig:maps}). In our fiducial analysis we use the Generalised Minimum Variance (GMV) map, constructed from a joint inverse-variance Wiener filtering of the CMB temperature and polarization maps accounting for inhomogeneous noise. In addition to this map, we use the so-called ``No-TT'' map, first introduced in \cite{2402.05761}, and constructed from a quadratic estimator that uses both temperature and polarization but downweights $TT$ correlations. This map is thus largely immune to the presence of unpolarized extragalactic foregrounds, such as the Cosmic Infrared Background (CIB) and the thermal Sunyaev-Zel'dovich effect. Contamination from the CIB is of particular relevance for the high-redshifts probed by \quaia, although no conclusive evidence of such contamination was found in \cite{2402.05761}.

    These maps were preprocessed as described in \cite{2402.05761}: the convergence harmonic coefficients and the analysis mask are rotated to Equatorial coordinates, and downgraded to our base resolution (including only multipoles up to $\lmax=3N_{\rm side}-1$ to avoid aliasing effects).

\section{Methods}\label{sec:meth}
  \subsection{Power spectrum measurements}\label{ssec:meth.cls}
    \begin{figure}
        \centering
        \includegraphics[width=0.99\textwidth]{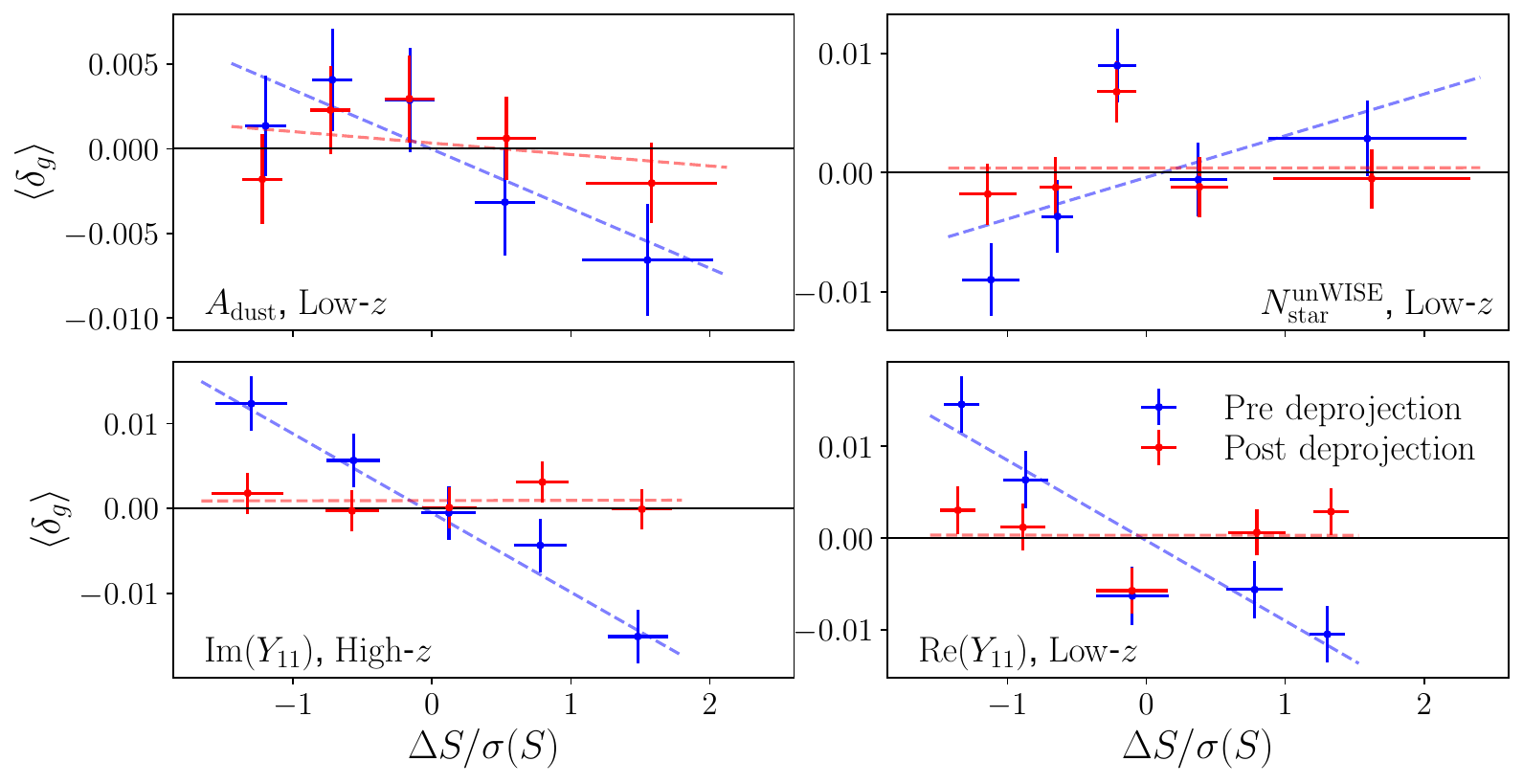}
        \caption{Map-level empirical correlation between a selection of contaminant templates and the galaxy overdensity $\langle \delta_g \rangle$ in the  \quaia data. Results are shown for overdensity maps without any contaminant deprojection (blue) and after linear deprojection (red). Deprojection eliminates any residual linear correlation between the data and these particular contaminant templates. To illustrate this point, the best-fit linear trend, found by fitting a linear model to the data in each individual plot, is shown as dashed lines (although note that linear deprojection applies this to the full linear subspace spanned by all the systematic templates). The lower panels show that we find a strong dipole component, which is also removed via deprojection.}
        \label{fig:syscorr}
    \end{figure}
    The effects of PNG on 2D projected statistics are most prominent on the largest angular scales. The estimate of 2-point statistics on those scales is complicated not only by the presence of observational systematics but also by partial sky coverage and survey geometry effects. As such we investigated the use of two different pipelines that make different assumptions and simplifications when accounting for instrumental systematics and masking effects on the measured power spectrum. 

    \subsubsection{Pseudo-$C_\ell$ pipeline}\label{sssec:meth.pcl}
      The first pipeline estimates the angular power spectrum with a state-of-the art pipeline based on a pseudo-$C_\ell$ (PCL) power spectrum estimation method \cite{astro-ph/0105302} as implemented in the publicly available code \nmt \cite{namaster}. The PCL estimator is in principle suboptimal, particularly on large angular scales, for data with steep power spectra or complex sky footprints \cite{1306.0005}. However, in \cite{2410.24134} we showed that this approximation is in fact quite accurate for \quaia, since the sample is shot-noise dominated across most scales.

      To suppress the presence of contamination from systematics, and to determine the angular scales over which our measurements may be safely used, we linearly deproject the angular fluctuations of the 10 templates described in Section \ref{ssec:data.quaia} from the QSO overdensity maps using the approach described in \cite{1609.03577,namaster}. Such mode-deprojection removes modes in the map that are proportional to any of the systematic templates deprojected or their linear combination without distinguishing between overlapping signal modes that behave in a similar way. As such, the resulting power spectrum is biased due this mode loss. While for typical cases when the number of systematics templates is low and the resulting mode loss is negligible on small angular scales, for large scale analyses this bias needs to be accounted for; otherwise, this mode loss could bias the inferred value of $\fNL$.

\begin{figure}
    \centering
    \includegraphics[width=0.8\linewidth]{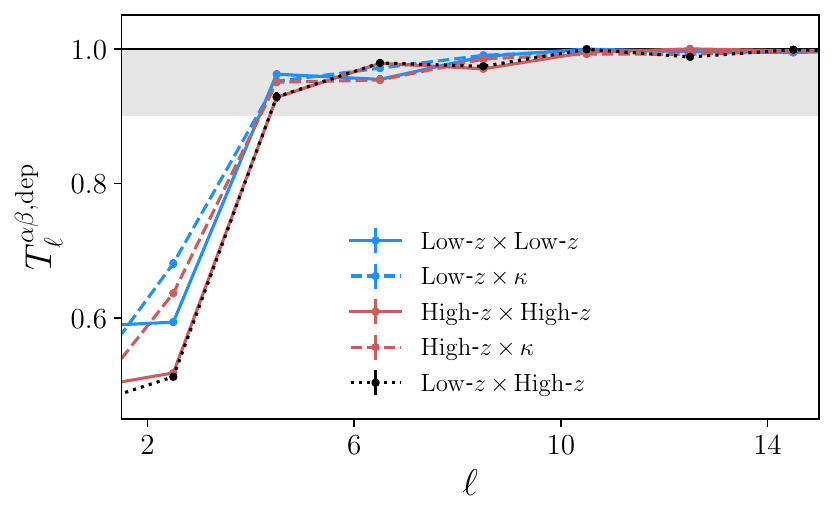}
    \caption{Systematics deprojection transfer function for the different auto and cross-correlations relevant for this work. }
    \label{fig:tf_deproj}     
\end{figure}

      While the deprojection bias term can be corrected analytically if the true underlying power spectrum is known \cite{1609.03577,namaster}, we followed the approach of \cite{2410.24134} and estimated the deprojection bias as an effective multiplicative transfer function from simulated mock data. This is similar to the procedure used in CMB analyses, when the final maps have been subjected to non-trivial filtering operations (either in the time, pixel, or Fourier domains) to mitigate the impact of specific systematics (e.g. correlated noise, scan-synchronous signals, etc.) \cite{1910.02608,2502.00946,2503.14452}. This approach has the advantage of being more robust against the assumed ``true'' power spectrum used when estimating the deprojection bias analytically, which in the case of studies involving large, cosmic-variance-dominated scales (as is our case) may not be known in detail. This method is discussed in detail in \cite{CornishInprep}. As discussed in \cite{2410.24134}, we found that the mode loss is important only on scales $\ell\lesssim 6$ and is generally smaller for the cross-correlations with CMB lensing than for the quasar auto-correlations. The effect is, nevertheless, small, in particular on scales used in our cosmological analysis. In fact, the mode loss induced by deprojection affects mainly multipoles below $\ell\simeq 4$ at $\sim 40\%$ level, and at  better than 5\% level  for $\ell\gtrsim 6$, with the large loss on largest scales induced by the deprojection of the $Y_{1,0}$ and $Y_{1,1}$ templates needed to account for the \quaia dipole. In Figure~\ref{fig:tf_deproj} we show the transfer functions for all the relevant spectra used in the following.
      
In Section \ref{sec:res} we will quantify the impact of the choices made regarding contaminant deprojection on our final constraints. While in Sec.~\ref{sssec:meth.qml} we describe an alternative analysis pipeline that is immune to transfer function effects and that was used to further validate the robustness of our analysis. We also verified that deprojecting mean-subtracted quadratic versions of the systematics templates did not change our conclusions.

      To further illustrate the impact of deprojection in this analysis, Fig.~\ref{fig:syscorr} shows the correlation between the galaxy overdensity field and the systematic templates before and after deprojection (blue and red points, respectively). We show particular combinations of redshift bin and systematic templates that best show the effects of deprojection. In each panel, the points with error bars show the mean value of $\delta_g$ (and its standard deviation) across the map after binning the map pixels by the local value of a given systematic template $S$. The results are shown as a function of the fluctuation in the systematic map from its mean across the footprint as a fraction of its standard deviation. The dashed lines in the figure show the best-fit linear trend describing the data. By construction, deprojection eliminates any linear correlation between $\delta_g$ and any map described by a linear combination of the systematic templates. We can also see that the residual linear correlation with the templates that were used to construct the \quaia selection function are relatively small (top panels). In turn, the data shows a clear linear trend against the dipole components. This correlation may be due to the local kinematic dipole in the observed density of distant quasars \cite{2311.14938,WilliamsInprep}, or a consequence of residual large-scale systematic contamination not captured by any of the other templates. Regardless of its origin, we choose to remove this dipole component altogether in order to suppress its contribution to the power spectrum uncertainties at higher multipoles via mode-coupling (see Section \ref{sssec:meth.qml}). 
      
 We mask out pixels in the \quaia maps for which the corresponding \quaia selection function is $<0.5$ and use the selection function as pixel weights, and used the \planck lensing mask as mask for $\kappa$. For reference, our pixel cut based on the selection function reduces the observable sky area from $\sim 87\%$ to $61\%$ and $52\%$ for the Low-$z$ and High-$z$ redshift bins respectively. We binned all power spectra into bin of width $\Delta \ell = 2 $ in the range $\ell \leq 40$, $\Delta\ell = 5$ in the range $40 \leq \ell < 60$, and $\Delta\ell = 30$ in the range $\ell \geq 60$.

      As a last notable ingredient of this pipeline, we accounted for the Monte Carlo correction of the normalization of the $\kappa$ field that arises from the presence of inhomogeneous noise of the \planck data in our reference footprint. For this purpose we made use of the lensing reconstruction simulations made available with the PR4 lensing maps and followed the procedure discussed in \cite{2306.17748,2305.07650}. Such normalization becomes very large for the first two bandpowers but is smaller than $10\%$ for smaller angular scales.

  \subsubsection{Quadratic maximum likelihood pipeline}\label{sssec:meth.qml}
    As an alternative, we developed a second pipeline based on a quadratic maximum likelihood (QML) power spectrum estimation \cite{astro-ph/9611174}. This pipeline allows us to test many of the assumptions inherent to the first pipeline. Firstly, QML power spectrum estimation methods effectively act as quadratic functions of the pixels and operates in pixel space. As such the estimate of the mode-coupling coefficients caused by the presence of a mask is not subject to the potential impact of finite band-limits in harmonic space.

    Secondly, if the covariance matrix of the pixels is known, and if the field under study is Gaussianly-distributed, the resulting estimator is manifestly optimal: all information available in the data is encoded in the estimated spectrum. By construction, the PCL estimator assumes a diagonal pixel covariance, and may be suboptimal when recovering spectra with a red tilt, exhibiting significant correlations across distant pixels. This is often particularly important on the largest angular scales. As in our analysis we are dealing with auto and cross-power spectra estimates we used the  QML power spectrum estimators of \cite{1807.02484} as implemented in the publicly available code \texttt{xQML}\footnote{\url{https://gitlab.in2p3.fr/xQML/xQML}}. 
  
    This implementation writes down the quadratic estimate of the power spectrum $\hat{C}^{AB,{\rm QML}}_\ell$ of two pixelized maps of scalar fields $\mathbf{d}_A,\,\mathbf{d}_B$ assuming that the cross-covariance between both fields is negligible when compared with their auto-covariances. In this case, the estimators for all auto- and cross-correlations are decoupled and may be calculated independently as
    \begin{equation}\label{eq:qml}
      \hat{C}^{AB,{\rm QML}}_{\ell} \equiv \sum \left[W^{-1}\right]_{\ell\ell^{\prime}}\hat{y}^{AB}_{\ell^{\prime}}.
    \end{equation}
   The elements of Eq. \ref{eq:qml} are:
    \begin{eqnarray}
      \hat{y}^{AB}_{\ell^{\prime}} &=& \mathbf{d}_A^T \mathbf{E}_\ell \mathbf{d}_B - b_\ell^{AB},\\
      \mathbf{E}_\ell &=& \frac{1}{2}\left(\mathbf{C}^{AA}\right)^{-1}\mathbf{P}_\ell\left(\mathbf{C}^{BB}\right)^{-1},\label{eq:qml1}\\
      W_{\ell\ell^{\prime}} &=& \frac{1}{2}{\rm Tr}\left[\left(\mathbf{C}^{AA}\right)^{-1}\mathbf{P}_\ell\left(\mathbf{C}^{BB}\right)^{-1}\mathbf{P}_{\ell^\prime}\right], \label{eq:qml2}
    \end{eqnarray}
    where $\mathbf{P}_\ell$ is a matrix involving the $\ell$-th Legendre polynomial:
    \begin{equation}
      \mathbf{P}_{\ell,{ij}} = \frac{2\ell+1}{4\pi}P_\ell(\cos{\theta_{ij}}),
    \end{equation}
    with $\theta_{ij}$ being the angular separation between pixels $i$ and $j$.
    In the previous Equations, ${\bf C}^{XY}$ is the pixel covariance between fields $X$ and $Y$, which can be decomposed into signal and noise components
    \begin{equation}
      \mathbf{C}^{AB} \equiv \langle \mathbf{d}_A\,\mathbf{d}_B^T\rangle = \mathbf{S}^{AB} +\mathbf{N}^{AB}.
    \end{equation}

    \noindent $b_\ell^{AB}={\rm Tr}\left[\mathbf{E}_\ell\mathbf{N}^{AB}\right]$ is the noise bias estimate. In the case of cross-spectra $A=\delta_g, B=\kappa$ the noise is uncorrelated between maps and as such the cross noise covariance matrix $\mathbf{N}^{AB}=0$ and $b_\ell^{AB}=0$. In turn, $C_\ell^{gg}$ is affected by shot-noise. We estimated the shot noise bias computing the average auto spectrum of 100 jackknife null maps of QSO overdensities derived from random equal splits of the \quaia catalog\footnote{We also verified that the noise-debiased  $C_\ell^{gg}$ agrees with the average of the cross-spectra of the aforementioned maps derived of the data splits which is inherently noise-bias free.}. This allows us to minimize the residual noise biases arising from mismatches between the assumed data covariance and the true data one. 

    The computational time of the QML estimator roughly scales as the number of the observed pixel to the third power, driven by the complexity of the brute force matrix inversion involved in the algorithm. As such, it would be very hard to carry out the analysis at the same resolution of the PCL pipeline. We therefore reprocessed the \quaia catalog to produce maps of the QSO overdensity on a $N_{\rm side}=16$ \hpx grid. The \planck CMB $\kappa$ estimates are given in terms of harmonic coefficients. To have a low resolution version of the CMB $\kappa$ map, we resampled on the same grid only the Fourier modes $\ell\leq 3N_{\rm side}=48$ (the maximum multipole supported by the grid) convolving them with the corresponding pixel window function. We reprocessed the simulated $\kappa$ map of the \planck data release and use those to compute $\mathbf{N}^{\kappa\kappa}$.  We then used the publicly available random realization of the \quaia catalog (which includes unclustered objects drawn from a uniform Poisson sampling of sky locations according to the selection function) to compute the expected number of objects in each pixel $p$, $\bar{n}_p$, and assumed a diagonal QSO noise covariance matrix $\mathbf{N}^{gg}_{pp^\prime} = \delta_{pp^\prime}/\bar{n}_p$ so that pixels where the selection function averaged in the pixel is lower are downweighted. We generated the signal covariance $\mathbf{S}$ from the fiducial $C_\ell$s of \quaia sources and $\kappa$ adopting $\fNL = 0$, the fiducial cosmology of the \planck FFP10 simulations' cosmology, and the best-fit bias model and redshift distributions of \cite{2306.17748}.
    
    In the QML pipeline we treated the presence of systematics using mode deprojection via template marginalization \cite{1992ApJ...398..169R,1306.0005}. In this context the map-level systematics are removed by assigning infinite variance to contaminating templates, making therefore the power spectrum estimate insensitive to the templates and is equivalent to marginalizing over the contamination amplitude of each of them. Mathematically, this is equivalent to the linear deprojection approach implemented in the PCL estimator \cite{1609.03577}. For a list of $k$ systematics templates ${\bf t}_k$ this involves modifying the covariance matrices entering equations \eqref{eq:qml1}, \eqref{eq:qml2} as:
    \begin{equation}
      \mathbf{C} \rightarrow \mathbf{C} + \sum_k \beta_k {\bf t}_k {\bf t}_k^T,
    \end{equation}
    where the sum is done over the number of templates and $\beta_k$ are arbitrarily large numbers. We included the same templates employed in the PCL case downgraded to the same resolution of the data. This includes the three dipole templates, which were incorporated by adding a term proportional to the $\ell=1$ Legendre polynomial $\Delta C_{ij}\propto P_1(\cos \theta_{ij})$ with a large prefactor. We also considered an additional template made by the \planck Zodiacal light map but we found this template to have no effect for the systematic marginalization. 
    
    Contrary to the PCL implementation of linear deprojection, the marginalisation approach does not bias the estimator, since the mode loss inherent to deprojection is consistently incorporated in the mode-coupling matrix (Eq. \ref{eq:qml2}). We verified this empirically by applying the estimator, including mode-deprojection, to simulated data that do not contain systematic effects. In the QML pipeline we adopted the same binning for overlapping multipoles with the PCL pipeline to minimize the computational cost which becomes quickly important, while keeping enough resolution in the power spectrum estimate. However, the final power spectra combined the QML $C_\ell$  estimates up to $\ell \leq 16$ and use the estimates derived on a pseudo-$C_\ell$ power spectrum estimation on smaller scales, similarly to what is done for the PCL pipeline. The $\ell=16$ transition was chosen as we observed that the error of the QML and PCL estimates became effectively the same for the mock data on small angular scales when the latter did not deproject the dipole moments (see discussion Sec.~\ref{ssec:meth.cov}). We also did not include any deprojection transfer function for the PCL data points as it is negligible on those scales. For the computation of the power spectra across all scales we used a single galactic mask given by the product of the two selection function and lensing mask employed in the PCL pipeline. 

\begin{figure}
    \centering
    \includegraphics[width=0.5\linewidth]{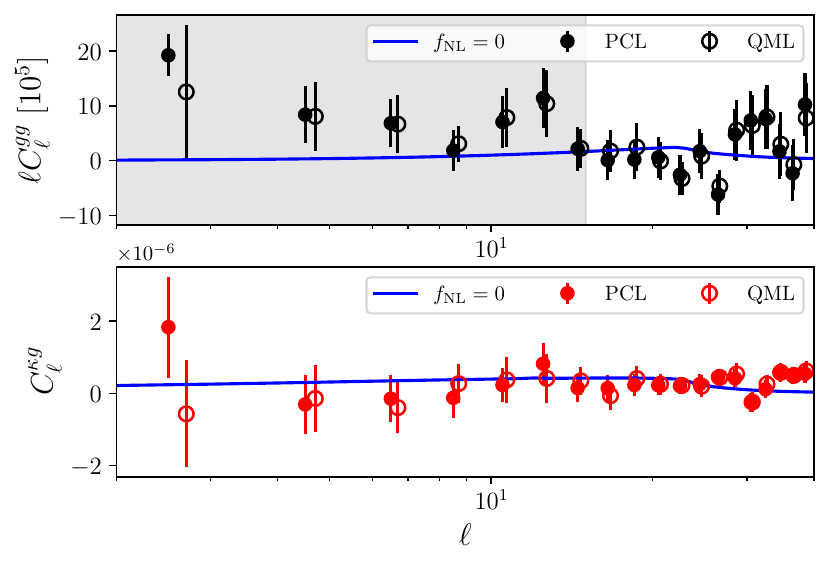}\includegraphics[width=0.45\linewidth]{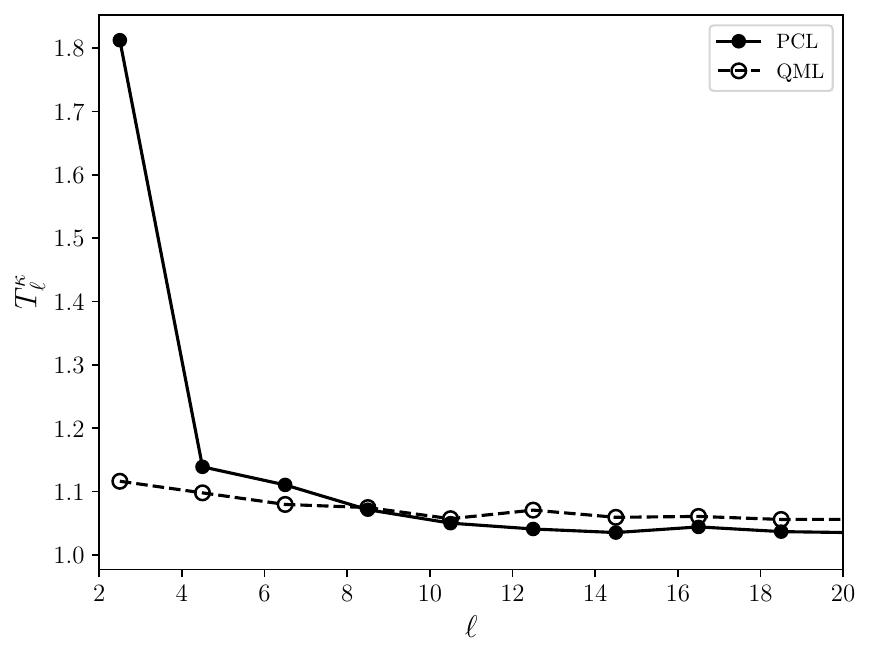}
    \caption{Left: comparison between PCL and QML pipeline used for this work for the largest angular scales where the treatment of systematics, estimation and covariance differs the most. We show the auto- (top) and cross-correlation with CMB lensing (bottom) for a single tomographic bin for simplicity. The shaded area identifies scales not included in the cosmological analysis. Right: comparison of the CMB lensing Monte Carlo correction for the QML and PCL pipelines. The QML estimate here includes a marginalization over the large $\ell=1$ mode in the CMB lensing field that better controls its leakage to higher multipoles.}
    \label{fig:qml-pcl-comparison}
\end{figure}

    Similarly to the PCL pipeline, we accounted for the Monte Carlo correction of the normalization of the CMB $\kappa$ field (denoted $T_\ell^{\kappa}$ hereafter for consistency with \cite{2410.24134}) using the same procedure but processing the \planck PR4 lensing simulations in the same way we did with the data. As the evaluation of the transfer function requires the computation of the auto-spectrum of the input signal of the $\kappa$ maps as well as their cross correlation with the corresponding reconstructed field $\kappa_{\rm rec}$ for the same realization, we set the $\mathbf{N}^{\kappa\kappa} = 0$ and computed  $\mathbf{S}^{\kappa\kappa}$  from the theoretical power spectrum $C_\ell^{\kappa\kappa}$ while we computed the $\mathbf{C}^{\kappa_{\rm rec}\kappa_{\rm rec}}$ from the PR4 simulations directly. As the PR4 simulation include a large dipolar component due to the kinematic motion of our local frame with respect to the CMB, we included a marginalisation of the $P_1$ template for computing both of these cross-spectra. Contrary to the PCL case, the CMB lensing transfer function is always well below $10\%$ at all angular scales because the QML estimator minimizes the impact of coupling of the dipolar modulations of the lensing field connected to the local kinematic dipole. We show for comparison the different transfer functions in Fig.~\ref{fig:qml-pcl-comparison}.\\
    
    Because the QML pipeline is computationally expensive and less flexible, in the following we report the main results obtained for the PCL pipeline. However, we carried out extensive consistency tests in particular for the case involving one single tomographic bin that we discuss in Appendix \ref{app:qml}. In Figure~\ref{fig:qml-pcl-comparison} we show a comparison of the estimated large scale angular power spectrum for both pipelines. Despite the radical differences in terms of systematics treatment and assumptions implemented in the estimators and in the covariance modelling, the agreement of the measurements at all scales is very good, as it can be seen in the figure, and as are the final constraints.
    
  \subsection{Covariances}\label{ssec:meth.cov}
    For the two fiducial power spectrum estimation pipelines, we adopt different fiducial models for the data covariances. For the PCL pipeline we take a similar approach to \cite{2410.24134} and estimate the covariance matrix of the data from the power spectra of 1000 Gaussian simulations of the quasar overdensity and CMB convergence maps, based on the measured power spectra of the data as input to such simulations. This automatically incorporates the additional variance caused by the presence of large scale systematics in the \quaia maps. This approach neglects in particular the noise inhomogeneity in both $\kappa$ and \quaia maps, as well as the fact that the input power spectra are noisy, non-smooth realizations of the underlying true power spectra. As in \cite{2410.24134}, we verified that both these effects have a negligible impact on the power spectrum uncertainties and on the final constraints. As a consistency test we also used an analytical Gaussian covariance based on the Narrow Kernel Approximation \cite{1906.11765}, as implemented in \nmt, which is less noisy but also in principle less accurate for the largest angular scales.
    
    For the QML pipeline, we also adopted a simulation-based approach but revised some of the simplifications of the PCL implementation. We used the official \planck PR4 simulated reconstructed $\kappa$ maps, which contain the effect of correlated inhomogeneous noise of the \planck satellite measurements, and generated correlated Gaussian realizations of the \quaia maps starting from the signal realizations of the PR4 simulations and a fiducial cross-correlation model assuming the same \planck cosmology of the PR4 simulations, the eBOSS QSO bias model of \cite{1705.04718} and the measured redshift distribution of the \quaia sources. We then added to this signal map a shot-noise realization based on the random \quaia catalogues that are publicly available\footnote{The public random catalogues contain 10 times more objects than the \quaia sample, but we only draw a number equal to the size of the data sample in each realisation.}, and a dipolar modulation with the same amplitude and direction observed in the \quaia maps. As the QML pipeline adopts a mix of pseudo-$C_\ell$ and QML estimates for the largest scales, we resampled each realizations on the native \hpx grid resolution of the PCL and QML pipelines and processed these maps as we did with the data. The resulting simulated data vectors were then used to compute the data power spectrum covariance. The resulting covariance matrix incorporates more accurately the correlation between badpowers and any other non-Gaussian contribution to the covariance. However they do not include the extra variance due to the excess power induced by the \quaia systematics. To overcome this issue  in particular for $\ell \leq 30$ we assumed a Gaussian approximation for these multipoles but calibrated the diagonal of the covariance measuring the effective number of degrees of freedom $\nu_b$ for each multipole bin $b$ using the mock data as \cite{astro-ph/0105302}
    \begin{equation}
      Var\left(C^{AB,{\rm QML}}_{b}\right)_s = \frac{1}{\nu_b}\left(\langle C_b^{AA,{\rm QML}}\rangle_s\langle C_b^{BB,{\rm QML}}\rangle_s+\langle C_b^{AB,{\rm QML}}\rangle_s\right)
    \end{equation}
    where the $s$ subscript denotes quantities averaged over the number of realizations of our simulated data set and $A,B\in\{g,\kappa\}$. We then used the estimated $\nu_b$ to rescale the covariance by the power spectrum measured on data as
    \begin{equation}
      Var\left(C^{AB,{\rm QML}}_{b}\right) = \frac{1}{\nu_b}\left(C_b^{AA,{\rm QML}} C_b^{BB,{\rm QML}}+C_b^{AB,{\rm QML}}\right),
    \end{equation}
    so that potential non-Gaussian bandpower correlation is preserved and the extra variance due to observational systematics is accounted for. We note that for scales where the systematics in the auto-correlation do not clearly dominate, we adopted the largest between the analytical and the Monte Carlo error bar. For both the PCL and QML pipeline we applied a correction factor to the inverse covariance matrix to account for the finite number of simulations used to calculate the covariance, as described in \cite{astro-ph/0608064}. As we start from the PR4 reconstructed $\kappa$ realizations we are limited to 400 realizations for the QML pipeline (compared to the 1000 realisations used for the PCL estimator).

  \subsection{Low-$\ell$ scale cuts}\label{ssec:meth.scalecuts}
    The set of simulations produced for the QML pipeline covariance estimation was also used to define the large-scale cuts adopted in our analysis. For each of the simulated maps we analysed the realizations with both the PCL and QML pipelines and applied systematics subtraction (via deprojection or marginalization) as we implemented for the real data. We then calculated the distribution of differences in each power spectrum multipole bin before and after systematics deprojection, for both $C_\ell^{gg}$ and $C_\ell^{\kappa g}$, and for both analysis pipelines. We then compared these simulated distributions to the differences observed in the real data after systematics deprojection. Since our simulations do not include systematic effects, any observed change in power due to deprojection is random—arising from chance correlations and mode loss. We therefore decided to exclude power spectrum bins where the observed shift in the data had a probability to exceed (PTE) below 5\% compared to the simulation-based distribution. We computed the PTE using a two-sided distribution to account for the sign of the shift, given that systematics deprojection preferentially reduces power.   
    
\begin{figure}[!htbp]
      \includegraphics[width=\textwidth]{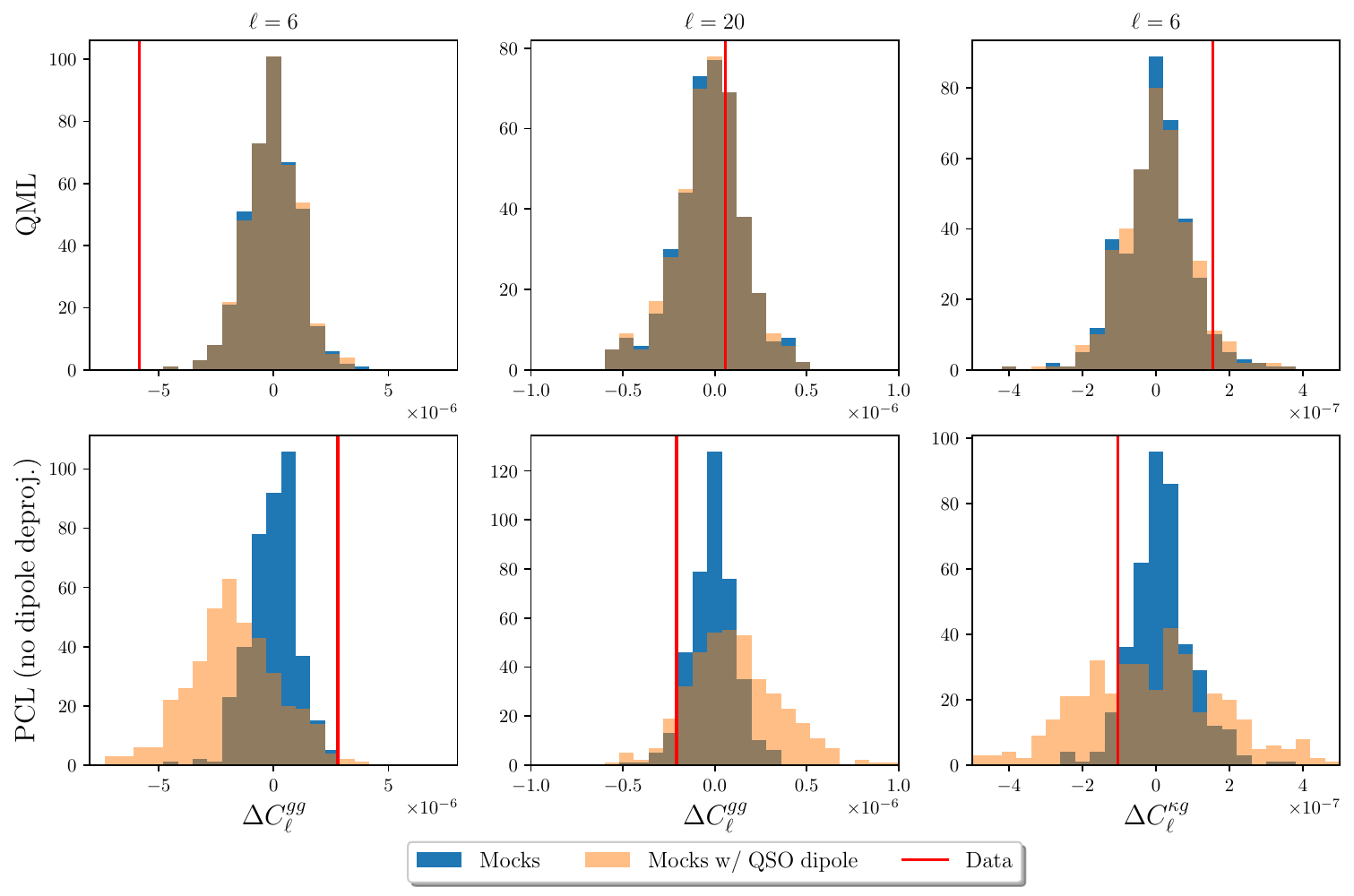}
      \caption{Examples of the difference in the measured power spectra induced by systematics deprojection for different multipole bins and spectra. In solid red we show the measurements on data, while the histograms show the expected distribution of such shifts estimated from simulations with no systematics. Orange histograms show distributions evaluated adding a dipole in the simulated QSO maps while the blue histograms are derived from simulations with no QSO dipole. The two rows show the results for the QML (top) and PCL (bottom) pipeline. For $\ell=6$ the probability of data and simulations being consistent is $<1\%$, hinting for the presence of systematics in the data (left). For $\ell=20$  (middle) shifts are consistent with random fluctuations. $\ell=15$ is the largest scales for which the observed shifts are consistent with random fluctuations at the 95\% C.L. and we therefore select this as our scale cut. For $C_\ell^{\kappa g}$ (right) no sign of systematics can be observed even for scales contaminated in the QSO auto-correlation.  For illustrative purposes, we did not remove the QSO dipole for the PCL panels. As can be seen, accounting for the QSO dipole is crucial to avoid biases and extra variance induced by the leakage of the dipole through the observational masks. We did so in our fiducial PCL analysis deprojecting the $Y_{10},Y_{\pm1}$ templates. }\label{fig:scale_cuts}
    \end{figure}

    For both the QML and PCL pipeline we found that scales $\ell \leq 15$ shift in an anomalous way for $C_\ell^{gg}$ and as such are affected by systematics. We therefore adopted this conservative scale cut in the analysis of the QSO auto correlation as baseline. On the other hand, we found that the shift in all the $C_\ell^{\kappa g}$ bandpowers were consistent with random fluctuations. As such we decided to include all the angular scales down to $\lmin=2$ in the analysis of the CMB lensing cross-correlation. We show a summary of these results in Fig.~\ref{fig:scale_cuts}.
    
    This scale cut definition checks for anomalous shifts while removing part of the sample variance, as opposed to an assessment simply based on the amplitude of the power shifts induced by systematics subtraction in relation to the statistical uncertainties. Nevertheless, as we show in Sec.~\ref{sec:res}, systematic deprojection shifts the bandpowers by an amount that is subdominant with respect of the statistical error on the scales used in this analysis. We note that we defined this set of scale cuts looking at the overall QSO clustering based on one single redshift bin. However, in the final cosmological constraints we assessed the impact of the scale cuts modifying the $\lmin$ included in the analysis and show that we recover consistent results and thus the scale cuts applied are robust. 
    
    In Fig.~\ref{fig:scale_cuts} we also show the importance of properly treating the large dipole of the QSO distribution in our pipelines. For this purpose we show results obtained using mock data including a dipole in the QSO distribution as well as results based on dipole-free simulations. As it can be seen in the bottom panel for the case of the PCL pipeline, an untreated dipole generates additional variance and biases in the observed distributions of power spectra shifts induced by systematics deprojection compared to the case where mock data are dipole free. 
     Including the $\ell=1$ spherical harmonics templates in the systematic deprojection removes biases and extra variance associated to the QSO dipole with the price of having to model this through a transfer function. In the case of the QML pipeline shown in the top panel conversely, the dipole marginalization effectively makes the presence of the QSO dipole irrelevant.      
     We will test the robustness to all these aspects of the analysis in the following.
  
  \subsection{Primordial non-Gaussianity and projected fields}\label{ssec:meth.fnl}
    We focus our analysis on the impact of local-type PNG on the clustering of biased tracers. In the presence of local PNG, the primordial gravitational potential can be characterised as in Eq. \ref{eq:fnl_def}. As initially shown by \cite{0710.4560}, local PNG modifies the clustering of galaxies on large scales. Focusing only on large scales, we can model the three-dimensional quasar overdensity in Fourier space at the tree level as
    \begin{equation}
      \Delta_g({\bf k},z)=\left[b(z)+\frac{2\delta_c}{\alpha(k,z)}\left(b(z)-p_\phi\right)\,\fNL\right]\,\Delta_m({\bf k},z),
    \end{equation}
    where $\delta_c\equiv1.686$ is the collapse threshold \cite{1974ApJ...187..425P}, and $b(z)$ is the galaxy bias, which we explicitly show is redshift-dependent. The scale-dependent factor $\alpha(k)$ allows us to connect the 3D matter overdensity $\Delta_m$ with the primordial gravitational potential as
    \begin{equation}
      \alpha(k,z)=\frac{2c^2k^2T(k)\,D(z)}{3\Omega_mH_0^2},
    \end{equation}
    where $T(k)$ is the matter transfer function, normalised to $T(k\rightarrow0)\rightarrow1$, and $D(z)$ is the linear growth factor for matter perturbations, normalised to $(1+z)\,D(z)\rightarrow 1$ at high redshifts, during the matter-dominated era. $\Omega_m$ is the fractional amount of non-relativistic matter at $z=0$, and $H_0$ is the expansion rate today. Finally, $p_\phi$ characterises the modification to the response function of the galaxy overdensity to a large-wavelength fluctuation caused by $\fNL$. $p_\phi$ is assumed to range from $p_\phi=1$, corresponding to the spherical collapse prediction for a Universal mass function \cite{0805.3580}, to $p_\phi\sim 1.6$ for recently-merged haloes (as found in $N$-body simulations) \cite{0805.3580,2006.09368,2107.06887}. The latter value was argued to be potentially more appropriate for quasars \cite{0805.3580}.

    In this work we analyse the \emph{projected} clustering of quasars. Within the linear bias model presented above, the observed 2D overdensity $\delta_g(\nv)$ in the sky position $\nv$ is given by \cite{2305.07650}
    \begin{equation}
      \delta_g(\nv)=\delta_g^{\rm int}(\nv)+\delta_g^{\rm PNG}(\nv)+\delta_g^{\rm RSD}(\nv)+\delta_g^{\rm mag}(\nv).
    \end{equation}
    Here, $\delta_g^{\rm int}(\nv)$ is the contribution from the intrinsic clustering of galaxies in the absence of PNG:
    \begin{equation}
      \delta_g^{\rm int}(\nv)\equiv\int dz\,p(z)\,b(z)\,\Delta_m(\chi\nv,z),
    \end{equation}
    where $p(z)$ is the redshift distribution of the sample and  $\chi$ is the comoving radial distance. $\delta^{\rm PNG}_g(\nv)$ is the scale-dependent contribution due to PNG
    \begin{equation}
      \delta_g^{\rm PNG}(\nv)\equiv\int dz\,p(z)\,\fNL\left(b(z)-p_\phi\right)\,\Delta^{\rm PNG}_g(\chi\nv,z),
    \end{equation}
    with
    \begin{equation}
      \Delta_g^{\rm PNG}({\bf k},z)\equiv\frac{2\delta_c}{\alpha(k,z)}\Delta_m({\bf k},z).
    \end{equation}
    Finally, $\delta_g^{\rm RSD}(\nv)$ and $\delta_g^{\rm mag}(\nv)$ are the contributions due to redshift-space distortions (RSD) and magnification bias:
    \begin{align}
      \delta_g^{\rm RSD}(\nv)=-\int \frac{dz}{H(z)}\,p(z)\,\partial_\parallel v_\parallel,
      \hspace{12pt}
      \delta_g^{\rm mag}=\int d\chi\,W_{\rm mag}(\chi)\,\left[\frac{1}{2}\nabla^2_\perp(\phi+\psi)\right](\chi\nv,z(\chi)),
    \end{align}
    where $v_\parallel$ is the peculiar velocity field along the line of sight, $\partial_\parallel$ is the line-of-sight gradient, $\psi$ and $\phi$ are the metric potential in the Newtonian gauge, $\nabla^2_\perp$ is the Laplacian operator on the plane perpendicular to the line of sight, and
    \begin{equation}
      W_{\rm mag}(\chi)\equiv\chi\int_{z(\chi)}^\infty dz'\,p(z')(5s(z')-2)\,\left(1-\frac{\chi}{\chi(z')}\right).
    \end{equation}
    Here $s(z)$ is the logarithmic slope of the cumulative number of sources as a function of magnitude for the sample under study. RSD and magnification are the leading contributions to the fluctuations in the observed number counts of sources caused by perturbations to the comoving volume elements and source fluxes in the presence of inhomogeneities \cite{1105.5280,1105.5292,1505.07596}. The remaining so-called ``relativistic effects'' are subdominant and can lead to a small increment or decrement of power on large scales, commensurate with a PNG signal of $|\fNL|\sim 1$. While this may become important with future, more sensitive searches, the effect of these terms can be safely neglected in this analysis \cite{1505.07596,1507.03550,2412.06553}.

    We also study the cross-correlation of the \quaia sample with maps of the CMB lensing convergence $\kappa(\nv)$, given by
    \begin{equation}
      \kappa(\nv)=\int_0^{\chi_{\rm LSS}} d\chi\,\chi\,\left(1-\frac{\chi}{\chi_{\rm LSS}}\right)\left[\frac{1}{2}\nabla_\perp^2(\phi+\psi)\right](\chi\nv,z(\chi)),
    \end{equation}
    where $\chi_{\rm LSS}$ is the distance to the last-scattering surface, at $z=1089$.

    The theoretical prediction for the cross-spectrum between any of these two projected fields can be written as
    \begin{equation}\label{eq:cl_theory}
      C_\ell^{\alpha\beta}=\frac{2}{\pi}\iiint d\chi_1\,d\chi_2\,dk\,k^2\,P(k,z(\chi_1),z(\chi_2))\,\Delta^\alpha_\ell(k,\chi_1)\,\Delta^\beta_\ell(k,\chi_2),
    \end{equation}
    where $P(k,z_1,z_2)$ is the unequal-time matter power spectrum, and $\Delta^\alpha_\ell(k,\chi)$ is the line-of-sight source function for field $\alpha$. We approximate the unequal-time power spectrum assuming perfect correlation between different times
    \begin{equation}
      P(k,z_1,z_2)=\sqrt{P(k,z_1)\,P(k,z_2)},
    \end{equation}
    which is sufficiently accurate \cite{1612.00770,1905.02078}. Within the linear bias model used here, and assuming a flat \lcdm model to relate $v_r$ and $\phi+\psi$ with $\Delta_m$, the source function for the observed quasar overdensity is \cite{1105.5292,png-lss-review}
    \begin{align}
      &\Delta^g_\ell(k,\chi)=\Delta^{\rm int}_\ell(k,\chi)+\Delta^{\rm PNG}_\ell(k,\chi)+\Delta^{\rm RSD}_\ell(k,\chi)+\Delta^{\rm mag}_\ell(k,\chi),\\\label{eq:Delta_int}
      &\Delta^{\rm int}_\ell(k,\chi)\equiv b(z)\,H(z)\,p(z)\,j_\ell(k\chi),\\\label{eq:Delta_PNG}
      &\Delta^{\rm PNG}_\ell(k,\chi)\equiv (b(z)-p)\fNL\,\frac{2\delta_c}{\alpha(k,z)}\,H(z)\,p(z)\,j_\ell(k\chi),\\
      &\Delta^{\rm RSD}_\ell(k,\chi)\equiv -H(z)\,p(z)\,f(z)\,j_\ell''(k\chi),\\
      &\Delta^{\rm mag}_\ell(k,\chi)\equiv\frac{\ell(\ell+1)}{(k\chi)^2}\frac{3}{2}H_0^2\Omega_m\,(1+z)\,W_{\rm mag}(\chi)\,j_\ell(k\chi),
    \end{align}
    and the CMB lensing source function is
    \begin{equation}
      \Delta^\kappa_\ell(k,\chi)=\frac{\ell(\ell+1)}{(k\chi)^2}\frac{3}{2}H_0^2\Omega_m\,(1+z)\,\chi\left(1-\frac{\chi}{\chi_{\rm LSS}}\right)\,\Theta(\chi<\chi_{\rm LSS}).
    \end{equation}
    Here, $j_\ell(x)$ is the spherical Bessel function of order $\ell$, $\Theta(<x)$ is the Heavyside function, $f(z)$ is the growth rate, and $H(z)$ is the expansion rate at redshift $z$. In all cases above, $z$ implicitly denotes the redshift at the comoving distance $\chi$ in the background.

    As Eq. \ref{eq:cl_theory} shows, calculating the angular power spectrum involves, in principle, a triple integral over $\chi_1$, $\chi_2$, and $k$. This can be significantly sped up using the Limber approximation \cite{1953ApJ...117..134L,loverde2008}. This approximation becomes less accurate on large scales, where most of the sensitivity to $\fNL\neq0$ lies, and therefore we only adopt it at $\ell>200$, where it is very accurate \cite{loverde2008,simon2007}. For smaller values of $\ell$ we use the non-Limber calculation as implemented in the Core Cosmology Library (CCL) \cite{ccl}, specifically using the {\tt FKEM} technique \cite{2212.04291}, first presented in \cite{1911.11947}. Nevertheless, the impact of the Limber approximation was found to be highly subdominant for the \quaia sample on the scales explored here in \cite[See Figure 3 in][]{2410.24134}.

\begin{figure*}[htbp]
    \includegraphics[width=\textwidth]{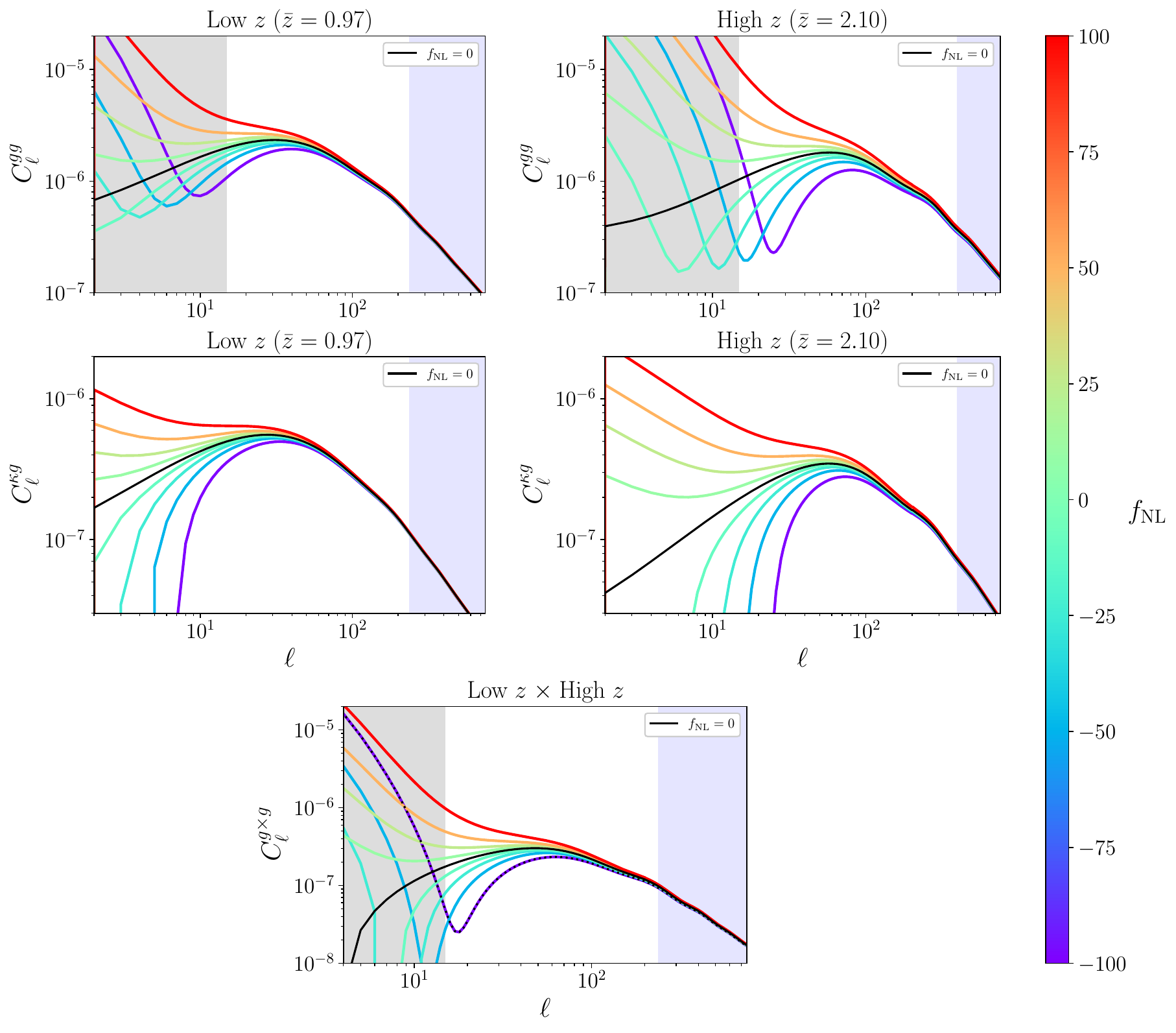}
    \caption{Impact of local primordial non-Gaussianity on the \quaia auto- and cross-correlations used in this analysis. All predictions shown here were calculated assuming the fiducial \planck cosmology, the \quaia redshift distribution and a PNG response parameter $p_{\phi}=1$. Results are shown for the for the quasar auto-correlations, the cross-correlations with CMB lensing (middle), and for the cross-bin power spectrum (bottom). The shaded bands identify the low-$\ell$ (grey) and high-$\ell$  (blue) scale cuts adopted in our fiducial analysis setup.}
    \label{fig:fnl_examples}
\end{figure*}

   Figure~\ref{fig:fnl_examples} shows the impact of local PNG on the angular power spectra used in this analysis for different values of $\fNL$. It is interesting to note the distinct effect of large negative values of $\fNL$ on the auto-correlations and the CMB lensing cross-correlations. Since the scale-dependent bias appears squared in the auto-correlation, negative values of $\fNL$ lead to a power \emph{enhancement} on the largest scales, and a suppression on intermediate scales. In turn, negative $\fNL$ always suppresses the cross-correlation. The latter therefore allows us to better distinguish between positive and negative values of $\fNL$, thanks to its monotonic trend with $\fNL$ and the possibility to access the largest scales with minimal systematic contamination. In order to better disentangle the sign of $\fNL$ from the auto-correlation, it is important to observe the characteristic signature of a potential negative $\fNL$ value at the smallest possible angular scales, which have a lower sample variance and less systematic contamination compared to its largest scales. This can only be ensured by tomographic analyses like the one carried out in this work using high redshift samples, as it can be seen comparing the top left and right panel of Fig.~\ref{fig:fnl_examples}.

  \subsection{Likelihood}\label{ssec:meth.like}
    To derive constraints on $\fNL$ we use a Gaussian power spectrum-level likelihood of the form
    \begin{equation}
      -2\log p({\bf d}|\vec{\theta}) = ({\bf d}-{\bf t}(\vec{\theta}))^T{\sf C}^{-1}({\bf d}-{\bf t}(\vec{\theta}))+K.
    \end{equation}
    Here, ${\bf d}$ is the data vector that contains our power spectrum measurements. ${\bf t}$ is the theoretical prediction for ${\bf d}$, dependent on a set of model parameters $\vec{\theta}$. ${\sf C}$ is the covariance matrix of ${\bf d}$, and $K$ is an irrelevant normalization constant.

    The data vector ${\bf d}$ contains our measurements of the quasar auto-correlation, and the quasar-lensing cross-correlation for each redshift bin. Each power spectrum is used over a different range of scales $(\lmin,\lmax)$, defined as follows: the large-scale cut is chosen, as described in Section \ref{ssec:meth.cls}, to ensure that our measurements are free from systematic contamination in the quasar density maps. Specifically, in our fiducial analysis we use $\lmin$ for the quasar-lensing cross-correlations, and $\lmin=15$ in the quasar auto-correlations. The small-scale cut $\lmax$ is chosen to ensure the validity of the linear bias model adopted in this analysis. Specifically, we define a different scale cut for each redshift bin as $\lmax=k_{\rm max}\,\bar{\chi}$, where $k_{\rm max}=0.07\,{\rm Mpc}^{-1}$, and $\bar{\chi}$ is the comoving distance to the mean redshift of each bin. This corresponds to $\lmax = 236$ and 390 in the low- and high-redshift bins respectively.

    Modelling the power spectrum likelihood as a multivariate Gaussian is only an approximation: even in the simplest case of Gaussian maps, the associated power spectra would follow a Wishart distribution \cite{0801.0554}. As we demonstrated in \cite{2410.24134}, the distribution of the quasar-lensing cross-correlation measurements is well described by a Gaussian at all scales. This is also the case for the quasar auto-correlation on the smaller scales ($\ell\geq15$) over which we use it to avoid clustering systematics, due to the central limit theorem.

    To construct ${\bf t}(\vec{\theta})$ we follow the model described in Section \ref{ssec:meth.fnl}. The redshift distribution of each bin was constructed as described in Section \ref{ssec:data.quaia}. The redshift-dependent bias in the $i$-th redshift bin is modelled as
    \begin{equation}
      b(z)=A_i\,b_*(z),
    \end{equation}
    where $b_*(z)$ is the fitting function for the bias of quasars in the sample studied by \cite{1705.04718}:
    \begin{equation}
      b_*(z)=0.278((1+z)^2-6.565)+2.393,
    \end{equation}
    and $A_i$ is a free parameter of the model (one for each redshift bin). To quantify the impact of this choice of redshift evolution, we will also repeat our analysis for the alternative model presented in \cite{2402.05761} (P24 hereafter), which provides a better fit for the \quaia sample:
    \begin{equation}\label{eq:bz_P24}
      b_*(z)=1.26\frac{D(0)}{D(z)},
    \end{equation}
    where $D(z)$ is the linear growth factor for the best-fit \lcdm cosmology found by \planck \cite{1807.06209}. Note that the cosmological analysis of the \quaia sample presented in \cite{2306.17748} found cosmological parameters that are slightly different from (but consistent with) the best-fit \planck values.

    Our fiducial analysis will use the auto- and cross-correlations from the two redshift bins described in Section \ref{ssec:data.quaia}. Thus, our model is described by three free parameters: $\vec{\theta}\equiv\{A_1,A_2,\fNL\}$, where $A_1$ and $A_2$ are the bias amplitudes in the low- and high-redshift bins, respectively. We sample this likelihood using the affine-invariant Markov chain Monte Carlo (MCMC) algorithm implemented in {\tt emcee} \cite{emcee}.

    In order to significantly speed up the evaluation of the theoretical prediction described above, we approximate the dependence of the angular power spectra on these parameters as follows: first, we write the source function for the quasar intrinsic clustering (Eq. \ref{eq:Delta_int}) in the $i$-th redshift bin as:
    \begin{equation}
      \Delta^{{\rm int},i}_\ell(k,\chi)=A_i\,\bar{b}_{*,i}\frac{b_*(z)}{\bar{b}_{*,i}}\,H(z)\,p_i(z)\,j_\ell(k\chi(z))\equiv b_i\,\,\bar{\Delta}_\ell^{{\rm int},i}(k,\chi),
    \end{equation}
    which defines the parameter-independent source function $\bar{\Delta}^{{\rm int},i}_\ell(k,\chi)$, and where $\bar{b}_{*,i}$ is the mean value of the fiducial bias function $b_*(z)$ over the redshift distribution of the $i$-th bin
    \begin{equation}
      \bar{b}_{*,i}\equiv \int dz\,p_i(z)\,b_*(z).
    \end{equation}
    We have also defined $b_i\equiv A_i\bar{b}_{*,i}$. Then, we approximate the PNG source function (Eq. \ref{eq:Delta_PNG}) as
    \begin{align}\nonumber
      \Delta^{{\rm PNG},i}_\ell(k,\chi)
      &=(b(z)-p)\fNL\frac{2\delta_c}{\alpha(k,z)}H(z)\,p_i(z)\,j_\ell(k\chi)\\\nonumber
      &\simeq (b_i-p)\fNL\,\frac{b_*(z)-p}{\bar{b}_{*,i}-p}\,\frac{2\delta_c}{\alpha(k,z)}H(z)\,p_i(z)\,j_\ell(k\chi)\\
      &\equiv (b_i-p)\fNL\,\bar{\Delta}^{{\rm PNG},i}_\ell(k,\chi),
    \end{align}
    which defines the parameter-independent PNG source function $\bar{\Delta}^{{\rm PNG},i}_\ell(k,\chi)$. Evidently, this approximation is only appropriate if the bias amplitude parameters $A_i$ are close to 1. Since in all the cases explored here the best-fit values of $A_i$ never deviate from this value by more than $\sim10\%$, we find that this approximation is indeed sufficiently accurate. Specifically, for a 10\% deviation in $A_i$, this approximation incurs a very small error that leads to a difference in the $\chi^2$ of the associated model of $\Delta\chi^2\simeq0.01$, which is negligible.
    \begin{figure}
        \centering
        \includegraphics[width=0.8\textwidth]{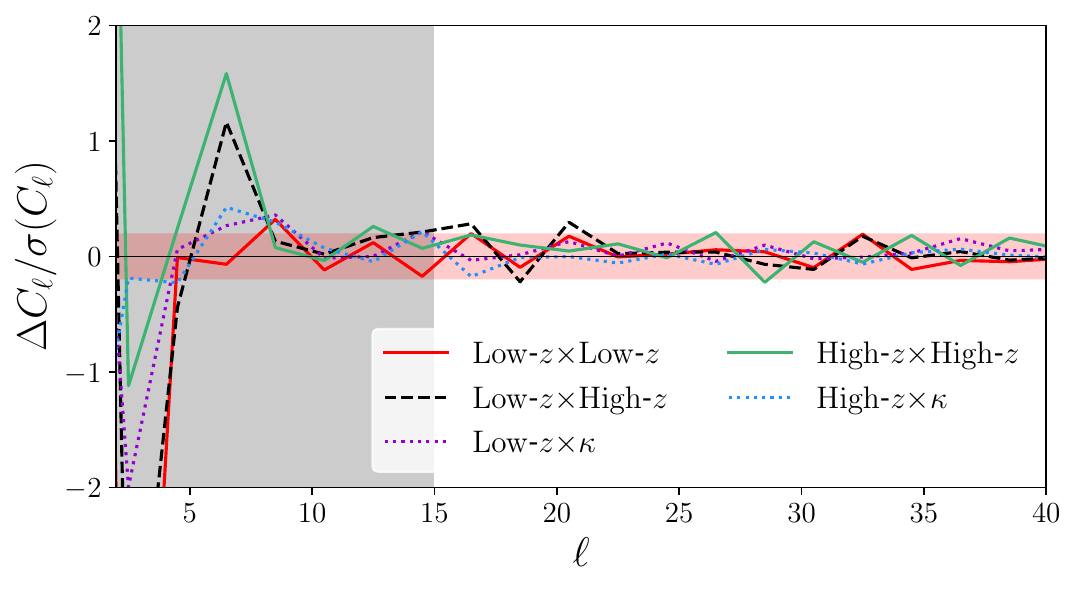}
        \caption{Difference between power spectra measured with and without linear contaminant deprojection as a fraction of the statistical uncertainties for $\ell<40$. Results are shown for the quasar auto-correlations (solid lines), including the cross-bin correlation (black dashed lines -- not used in the fiducial analysis), as well as the quasar--CMB lensing cross-correlations (dotted lines). On scales $\ell\geq15$ (gray shaded area), corresponding to the fiducial scale cut applied to the quasar auto-correlations, the differences remain below 20\% of the measurement errors (shown as the red-shaded horizontal band).}
        \label{fig:dcl_deproj}
    \end{figure}

    The main advantage of this approximation is that it allows us to write the quasar auto-correlation in bin $i$ as:
    \begin{align}\nonumber
      C_\ell^{gg}
      =&\, b_i^2\,\bar{C}^{{\rm int},{\rm int}}_\ell+2b_i(b_i-p)\fNL\,\bar{C}^{{\rm int},{\rm PNG}}_\ell+2b_i\bar{C}^{{\rm int},({\rm RSD}+{\rm mag})}_\ell\\
      &+(b_i-p)^2\fNL^2\,\bar{C}^{{\rm PNG},{\rm PNG}}_\ell+2(b_i-p)\fNL\bar{C}^{{\rm PNG},({\rm RSD}+{\rm mag})}_\ell\\
      &+\bar{C}^{({\rm RSD}+{\rm mag}),({\rm RSD}+{\rm mag})}_\ell
    \end{align}
    where all barred $C_\ell$s (e.g. $\bar{C}^{{\rm int},{\rm int}}_\ell$) denote power spectra involving the parameter-independent source functions. Likewise, the quasar-lensing cross-correlation is
    \begin{align}\nonumber
      C_\ell^{g\kappa}
      =&\, b_i\,\bar{C}^{{\rm int},\kappa}_\ell+(b_i-p)\fNL\,\bar{C}^{{\rm PNG},\kappa}_\ell+\bar{C}^{({\rm RSD}+{\rm mag}),\kappa}_\ell.
    \end{align}
    Although the barred power spectra are slow to calculate, particularly without the Limber approximation, they need only be computed once before sampling the likelihood (since they do not depend on the model parameters). Once this is done, the likelihood is a simple polynomial function of the parameters, which can be evaluated in $\sim0.3$ milliseconds on a standard laptop. This allows us to run Monte Carlo chains with tens of thousands of samples at a negligible computational cost.

\section{Results}\label{sec:res}
  \subsection{Power spectrum measurements}\label{ssec:res.cls}
    \begin{figure}
        \centering
        \includegraphics[width=0.85\textwidth]{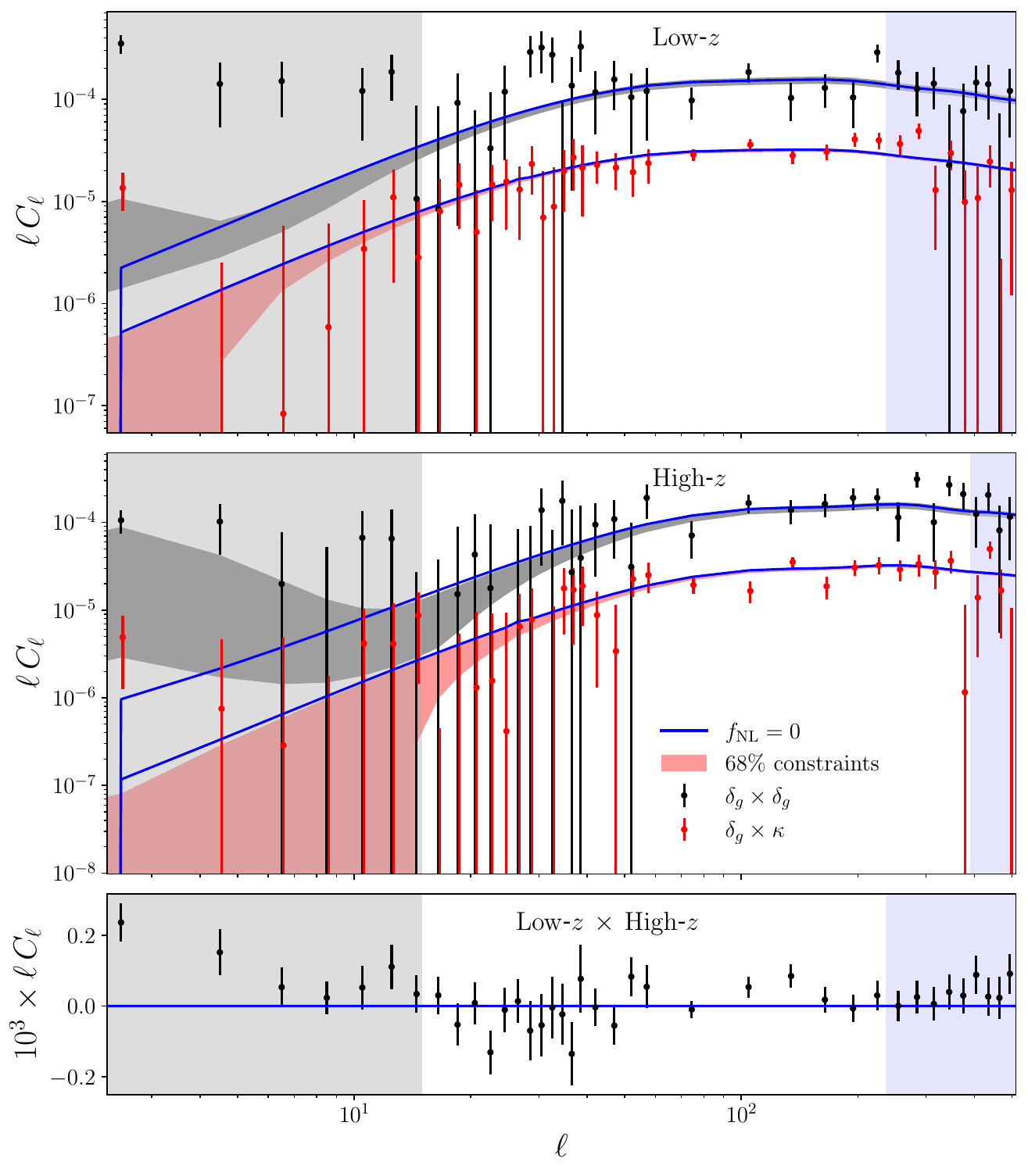}
        \caption{Measurements, after contaminant deprojection, of the quasar auto-correlation (black points) and their CMB lensing cross-correlation (red points) for the Low-$z$ and High-$z$ redshift bins (top and middle panels, respectively), and the cross-correlation between the redshift bins (bottom panel). The shaded red and black bands show the 68\% confidence level (C.L.) bounds on the power spectra within our 3-parameter model, including $\fNL$ and the linear bias in both redshift bins. On large scales, this is dominated by the uncertainty on $\fNL$, corresponding to $\fNL=-20.2\pm18.4$ (constraints found when including data from both redshift bins). The best-fit prediction is shown as solid lines. The blue lines show the theoretical prediction for the best-fit bias parameters but fixing $\fNL=0$. The blue shaded vertical band marks the high-$\ell$ scale cut applied for both $C_\ell^{gg}$ and $C_\ell^{g\kappa}$, while the gray vertical band shows the large-scale cut imposed on the quasar auto-correlations.}
        \label{fig:cls}
    \end{figure}
    We measure the auto-correlation of the quasar overdensity in the two redshift bins, as well as the cross-bin correlation, and their cross-correlation with the CMB lensing map using the pseudo-$C_\ell$ approach described in Section \ref{ssec:meth.cls}. Crucially for this analysis, we mitigate the impact of observational systematics on large scales by linearly deprojecting the ten contaminant templates presented in Section \ref{ssec:data.quaia}. These trace the scanning patterns of both \gaia and \unwise, stellar abundance, and dust extinction, as well as three dipole moments. As discussed in \ref{ssec:meth.cls}, we find that the \quaia overdensity maps are most strongly correlated with the dipole, likely due to a combination of the kinematic dipole and residual large-scale contamination. Deprojecting the dipole components has a beneficial effect in the power spectrum measurements, reducing the variance caused by mode coupling on the smallest multipoles. 
    
    We correct for the loss of power caused by deprojection through the transfer function approach discussed in Section \ref{ssec:meth.cls}. To quantify the potential presence of residual contamination in the data after this procedure, we compared the power spectrum measurements with and without contaminant deprojection. The difference between both measurements on large scales is shown in Fig.~\ref{fig:dcl_deproj} as a fraction of the statistical uncertainties. We find that, after $\ell=15$, the differences are subdominant (less than 20\% of the statistical uncertainties), further justifying our choice of large-scale cut for the quasar auto-correlations. It is interesting to note that the cross-correlation between both bins (not used in the fiducial analysis) exhibits deviations between the cases with and without deprojection above $0.2\sigma$ extending to smaller scales. 

    Our measurements of the quasar auto-correlations and cross-correlations with CMB lensing are shown in Fig.~\ref{fig:cls}. The shaded bands show the $1\sigma$ bounds for the theoretical prediction within our 3-parameter model ($\{A_1,A_2,\fNL\}$, see Section \ref{ssec:meth.like}). The vertical blue band marks the small scales excluded from our analysis of both $C_\ell^{\rm gg}$ and $C_\ell^{g\kappa}$, while the gray vertical band shows the large-scale cut used for $C_\ell^{gg}$. Within this range of scales, the model is able to provide a good fit to the data, with a best-fit $\chi^2=117.9$ for a total of 109 degrees of freedom, corresponding to a probability-to-exceed (PTE) of $0.26$.

  \subsection{Constraints on $\fNL$}\label{ssec:res.fNL}
    \begin{figure}
        \centering
        \includegraphics[width=0.6\textwidth]{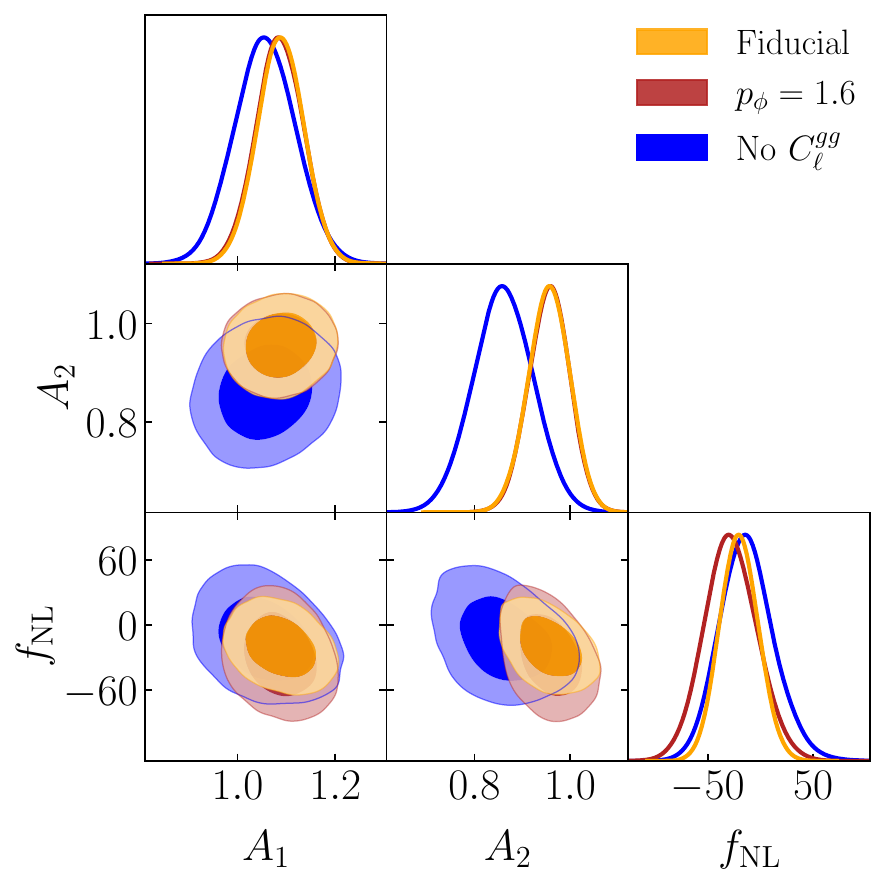}
        \caption{Constraints on the 3 parameters of our model: the bias amplitudes in each redshift bin $A_1$ and $A_2$, and the PNG amplitude $\fNL$. Constraints are shown for our fiducial analysis, using both auto- and cross-correlations and assuming a universality relation for the bias response to PNG, with  $p_\phi=1$ (yellow contours), for the case of a merger-driven population ($p_\phi=1.6$,  red contours), and for an analysis using only the CMB lensing cross-correlations (blue contours).}
        \label{fig:triangle}
    \end{figure}
    The constraints on $\fNL$ and the two bias amplitude parameters $A_1$ and $A_2$, using our fiducial analysis choices, are shown in Fig.~\ref{fig:triangle} (yellow contours). We find a constraint on the PNG amplitude parameter of
    \begin{equation}\label{eq:fnl_fid}
      \fNL=-20.5^{+19.0}_{-18.1}\,\,(p_\phi=1,\,\,68\%\,\,{\rm C.L.}),
    \end{equation}
    compatible with $\fNL=0$ at the $1.1\sigma$ level. The galaxy bias parameters are also in relatively good agreement (within $\sim2\sigma$ of unity) with the fiducial bias model. As mentioned in the previous section, the resulting best-fit model is a good fit to the data. 
    
    Fig.~\ref{fig:triangle} also shows the constraints found assuming a PNG response corresponding to a population dominated by recent mergers as may be the case for quasar populations, with $p_\phi=1.6$, as discussed in \ref{ssec:meth.fnl}. In this case, the constraints are degraded slightly, as expected, with
    \begin{equation}\label{eq:fnl_p1p6}
      \fNL=-28.7^{+26.1}_{-24.6},\,\,(p_\phi=1.6,\,\,68\%\,\,{\rm C.L.}),
    \end{equation}
    in similar agreement with $\fNL=0$. Finally, the figure shows the constraints found using only the quasar-CMB lensing cross-correlations, which are more robust to the presence of residual contamination in the galaxy overdensity maps. The resulting constraints, assuming $p_\phi=1$, are
    \begin{equation}\label{eq:fnl_nogg}
      \fNL=-13.8^{+26.7}_{-25.0},\,\,(p_\phi=1,\,\,{\rm no}\,\,C_\ell^{gg},\,\,68\%\,\,{\rm C.L.}),
    \end{equation}
    We observe a marginal shift towards $\fNL=0$ with respect to our fiducial result, combined with a $\sim40\%$ increase in the statistical uncertainties. Repeating  this analysis assuming $p_\phi=1.6$ we obtain
    \begin{equation}\label{fnl_nogg_p1p6}
      \fNL=-15.6^{+42.3}_{-34.8},\,\,(p_\phi=1.6,\,\,{\rm no}\,\,C_\ell^{gg},\,\,68\%\,\,{\rm C.L.}),
    \end{equation}
  with a degradation in the constraints comparable to the one obtained for the $p_{\phi}=1$ case.
    In all cases, we find the posterior distribution on all parameters to be mostly symmetric and close to Gaussian. Nevertheless, in all cases we report the median and two-sided 68\% confidence interval around it.

    \begin{table}
      \centering
      \renewcommand{\arraystretch}{1.5}
      \normalsize
      \begin{tabular}{|l|l|l|l|l|l|}
        \hline
        Analysis setting & $\fNL$ & $A_1$ & $A_2$ & $\chi^2/N_{\rm dof}$ & PTE \\[0.5ex]
        \hline
1. ${\rm Fiducial}$ & $-20.5^{+19.0}_{-18.1}$ & $1.09 \pm 0.05$ & $0.96 \pm 0.04$ & $117.9 / 109$ & $0.3$ \\
2. $p_\phi=1.6$ & $-28.7^{+26.1}_{-24.6}$ & $1.09 \pm 0.05$ & $0.96 \pm 0.04$ & $118.0 / 109$ & $0.3$ \\
3. ${\rm No}\,\,C_\ell^{gg}$ & $-13.8^{+26.7}_{-25.0}$ & $1.06 \pm 0.06$ & $0.86 \pm 0.06$ & $56.0 / 60$ & $0.6$ \\
4. ${\rm No}\,\,C_\ell^{g\kappa}$ & $-61.8^{+37.0}_{-35.6}$ & $1.11 \pm 0.06$ & $1.05 \pm 0.06$ & $55.5 / 46$ & $0.2$ \\
5. ${\rm Low}$-$z$ & $-17.8^{+67.1}_{-59.3}$ & $1.07 \pm 0.06$ & N.A. & $52.2 / 49$ & $0.4$ \\
6. ${\rm High}$-$z$ & $-23.2^{+19.5}_{-19.3}$ & N.A. & $0.96 \pm 0.04$ & $60.9 / 59$ & $0.4$ \\
7. $+{\rm cross}$-${\rm bin}$ & $-28.3^{+17.7}_{-16.8}$ & $1.08 \pm 0.05$ & $0.97 \pm 0.04$ & $146.9 / 131$ & $0.2$ \\
8. $1\,\,{\rm bin}$ & $-46.9^{+24.0}_{-22.3}$ & $1.03 \pm 0.04$ & N.A. & $50.0 / 55$ & $0.7$ \\
9. $G<20$ & $-12.1^{+27.2}_{-25.0}$ & $1.18 \pm 0.07$ & $0.94 \pm 0.07$ & $110.9 / 109$ & $0.4$ \\
10. $\ell^{gg}_{\rm min}=10$ & $-17.7^{+19.0}_{-18.1}$ & $1.09 \pm 0.05$ & $0.95 \pm 0.04$ & $126.6 / 115$ & $0.2$ \\
11. $\ell^{g\kappa}_{\rm min}=4$ & $-29.0^{+19.8}_{-19.6}$ & $1.09 \pm 0.05$ & $0.96 \pm 0.04$ & $110.7 / 107$ & $0.4$ \\
12. $\ell^{g\kappa}_{\rm min}=10$ & $-38.2^{+24.4}_{-24.0}$ & $1.10 \pm 0.05$ & $0.97 \pm 0.05$ & $109.2 / 101$ & $0.3$ \\
13. ${\rm No\,\,deproj.}$ & $-13.5^{+19.0}_{-18.7}$ & $1.10 \pm 0.05$ & $0.95 \pm 0.04$ & $116.0 / 109$ & $0.3$ \\
14. $k_{\rm max}=0.1$ & $-28.5^{+18.4}_{-17.9}$ & $1.12 \pm 0.04$ & $0.96 \pm 0.04$ & $142.3 / 127$ & $0.2$ \\
15. ${\rm Analytic\,\,cov.}$ & $-21.0^{+18.8}_{-18.8}$ & $1.09 \pm 0.05$ & $0.94 \pm 0.04$ & $124.0 / 109$ & $0.2$ \\
16. ${\rm With\,\,RSD}$ & $-20.1^{+18.6}_{-18.6}$ & $1.09 \pm 0.05$ & $0.96 \pm 0.04$ & $117.9 / 109$ & $0.3$ \\
17. ${\rm With\,\,mag.}$ & $-21.2^{+18.5}_{-18.4}$ & $1.09 \pm 0.05$ & $0.95 \pm 0.04$ & $118.3 / 109$ & $0.3$ \\
18. ${\rm P24}\,\,b(z)$ & $-23.6^{+19.3}_{-18.8}$ & $0.88 \pm 0.04$ & $0.99 \pm 0.05$ & $118.7 / 109$ & $0.2$ \\
19. ${\rm No}\,\,\kappa\,\,{\rm TF}$ & $-29.4^{+17.7}_{-17.6}$ & $1.07 \pm 0.05$ & $0.96 \pm 0.05$ & $121.3 / 109$ & $0.2$ \\
20. ${\rm No}$-${\rm TT}$ & $-47.4^{+21.8}_{-21.5}$ & $1.06 \pm 0.05$ & $1.07 \pm 0.05$ & $89.8 / 109$ & $0.9$ \\
21. ${\rm No}$-${\rm TT},\,{\rm No}\,\,C_\ell^{gg}$ & $-24.0^{+28.7}_{-26.4}$ & $1.02 \pm 0.09$ & $1.07 \pm 0.10$ & $46.4 / 60$ & $0.9$ \\
22. $p_\phi=1.6$,\,${\rm No}\,\,C_\ell^{gg}$ & $-15.6^{+42.3}_{-34.8}$ & $1.06 \pm 0.06$ & $0.86 \pm 0.06$ & $56.1 / 63$ & $0.6$ \\
        \hline
      \end{tabular}
      \caption{Constraints on the model parameters: the PNG amplitude $\fNL$ and the bias scaling amplitudes for the different redshift bins ($A_1$ and $A_2$). Results are shown for our fiducial analysis (first row), as well as all other variations used to explore the stability of our results to different analysis choices and potential systematics; detailed descriptions are in the text. The best-fit $\chi^2$, number of degrees of freedom, and associated PTE are listed in the last two columns.}
      \label{tab:results}
    \end{table}

    \begin{figure}
        \centering
        \includegraphics[width=0.95\textwidth]{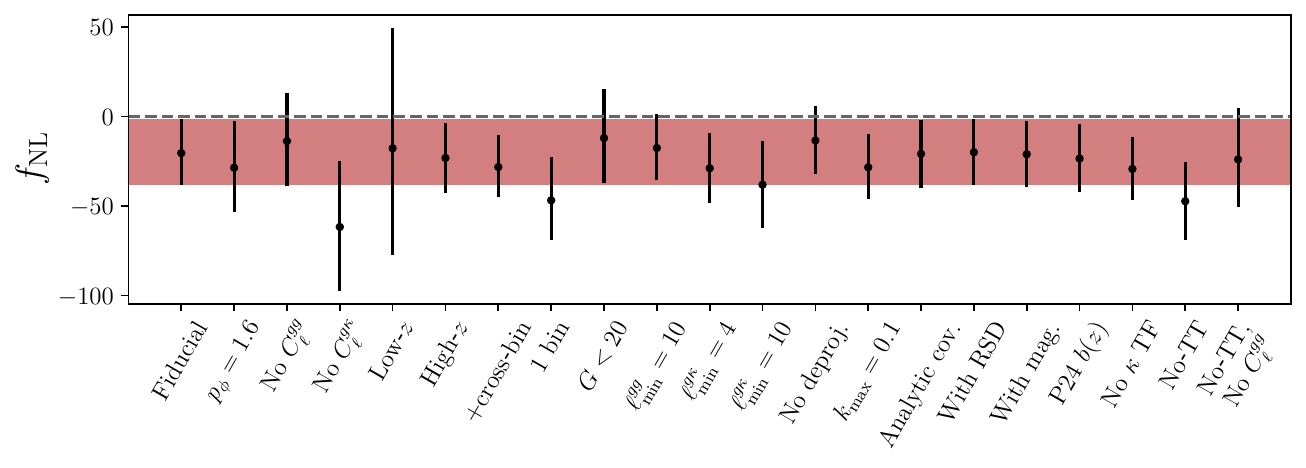}
        \caption{Constraints on $\fNL$ obtained in our fiducial analysis (shaded horizontal band), and adopting alternative analysis strategies that test the robustness of our results. The numerical values shown here are listed in Table \ref{tab:results}, and the different tests are described in the main text.}
        \label{fig:fnl_tests}
    \end{figure}
    These constraints are robust to our specific analysis choices and against several potential sources of systematic errors. We carry out a number of tests to quantify this, repeating our analysis for variations in the data vector and theory model. The resulting constraints obtained in these cases are listed in Table \ref{tab:results} and the impact on the $\fNL$ measurements is shown in Fig.~\ref{fig:fnl_tests}. 

    First, we repeat our analysis focusing on different subsets of the data to quantify the level of internal consistency of our constraints. The result of considering only the galaxy-lensing cross-correlation was reported above. Removing these cross-correlations and using only the quasar auto-correlations (row 4 in Table \ref{tab:results}) results in a lower central value for $\fNL$, accompanied by a 90\% increase in the statistical uncertainties. This is one of the largest shifts we observe across all our tests, and is mostly driven by the marginally higher bias value favoured by the quasar auto-correlation in the High-$z$ bin. 
    
\begin{figure}
        \centering
        \includegraphics[width=0.7\linewidth]{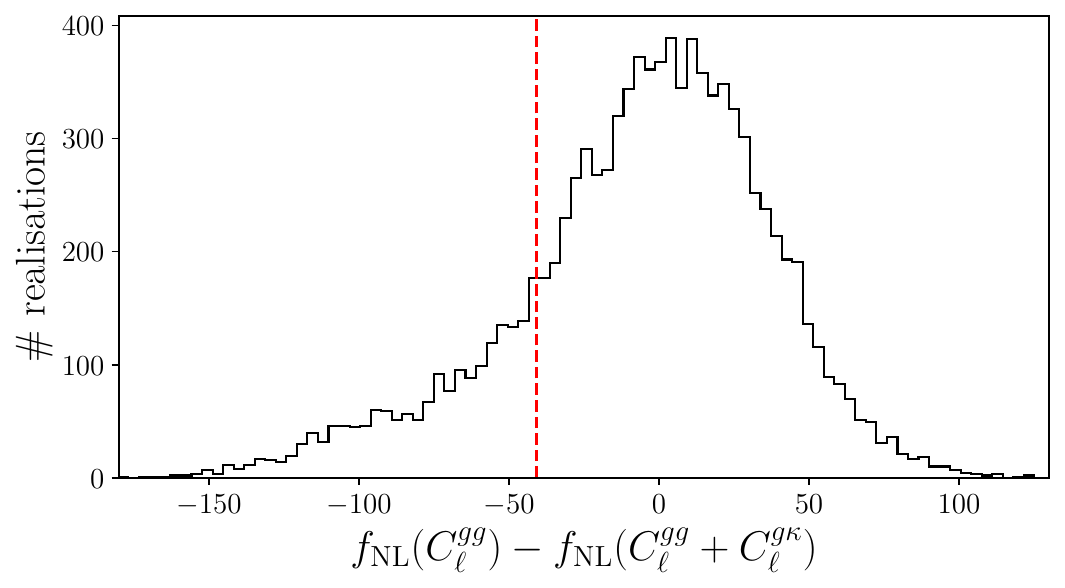}
        \includegraphics[width=0.7\linewidth]{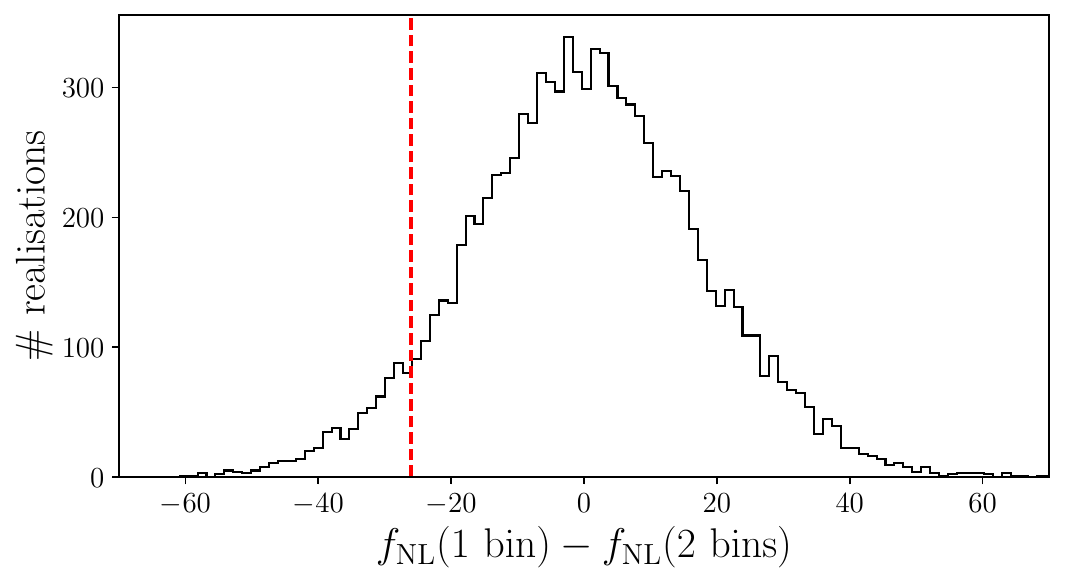}
        \caption{{\sl Top panel:} distribution of the difference between the best-fit values of $\fNL$ found using only the galaxy auto-correlation and that found using both auto- and cross-correlations, obtained from a suite of $10^4$ Gaussian simulations. The red dashd line shows the value obtained from the real data. {\sl Bottom panel:} as the top panel for the difference between the best-fit values found using one or two redshift bins.}
        \label{fig:diff_fNL}
    \end{figure}

    To quantify the significance of this shift, and the potential presence of internal tension in the data, we carry out the following frequentist exercise. We generate simulated realisations of our full $C_\ell^{gg}+C_\ell^{g\kappa}$ data vector by drawing from a multivariate normal distribution with a mean given by the theoretical prediction corresponding to our fiducial best-fit parameters, and the covariance matrix of the data. For each realisation, we then find the best-fit parameters of our model, consider either the full set of auto- and cross-correlations, or only the quasar auto-correlations. We then study the distribution of the difference between the best-fit values of $\fNL$ found in either case. The distribution reconstructed from $10^4$ such realisations is shown in the top panel of Fig.~\ref{fig:diff_fNL}, with the $\fNL$ difference found in the real data shown as a vertical line. We find that 18.6\% of all realisations recover downward shifts in $\fNL$ that are larger than that found in the data. The shift we recover is therefore not statistically significant. The large increase in statistical uncertainties when using only the quasar auto-correlations also shows that our constraints are dominated by the $C_\ell^{g\kappa}$ cross-correlation.
    
    We then consider the impact of restricting our analysis to only one of the two redshift bins (rows 5 and 6 for the Low-$z$ and High-$z$ bins, respectively). Discarding the high-redshift bin results in a dramatic $\sim3.5$-fold increase in the statistical uncertainties, clearly showing that our constraints are driven by the high-redshift data. This is reasonable, as this sample covers a larger volume, and the low-$\ell$ angular power spectra are able to access smaller physical wavenumbers $k$, more sensitive to PNG. In either case, the resulting $\fNL$ constraints are compatible with our fiducial measurement. Finally, we repeat our analysis including the cross-correlation between quasars in both redshift bins. The resulting constraints on $\fNL$ (row 7 in Table \ref{tab:results}) shift downwards by $\sim0.4\sigma$, while the errors shrink by approximately $10\%$. The impact of including this cross-correlation is interesting for several reasons. First, in principle this cross-correlation should allow us to recover some information contained in long-wavelength radial modes that is lost by considering only projected statistics \cite{1207.6487}. At the same time, cross-bin correlations are less robust to various forms of systematics. Besides genuine extra power on very large scales, the only other possible contributions to the signal in this spectrum come from the overlap between the redshift distributions of both bins, or from magnification bias. On the other hand, a non-zero cross-correlation could also be sourced by correlated residual sky contamination in both bins. Therefore, the fact that our fiducial results do not change significantly when including the cross-bin correlation, is further reassurance that these sources of systematic uncertainty (photometric redshift errors, uncertainties in the magnification bias slope, and residual sky contamination) are under control in our sample. The goodness of fit for this case also supports this conclusions \cite[see also the discussion in][]{2306.17748}). However, in order to be conservative as we do not have a fully overlapping spectroscopic sample to accurately test or recalibrate our redshift distribution, we did not quote this as our baseline.

    We also explored the impact our analysis varying our selection of the \quaia sources. First, we repeated our analysis using a single redshift bin covering the range $z_p<5$ (row 8). The resulting constraints shift towards negative $\fNL$ by approximately $1\sigma$, with uncertainties that grow by $\sim30\%$. To quantify the significance of this shift, we repeat the analysis described above to study the shift observed when excluding the lensing cross-correlations. We generate Gaussian realisations of our power spectrum measurements for the single-bin and two-bin versions of the analysis, fully accounting for the covariance between both datasets at the level of both signal and shot noise (we estimate the cross-covariance between the single-bin and two-bin power spectra following the same approach used for our fiducial measurements). We then find the best-fit values of $\fNL$ in both datasets, and recover the distribution of the difference between them. The result is shown in the bottom panel of Fig.~\ref{fig:diff_fNL}, with the difference found in the real data marked as a vertical red dashed line. We find that $13\%$ of all realisations recover larger shifts than those found in the data. This shift is therefore only mildly significant ($1.5\sigma$), although there are several potential reasons for it. For instance, the single-bin case is potentially more strongly affected by the inclusion of small-scale structures at low redshifts within our scale cuts, as well as more sensitive to the assumed evolution of the quasar bias within the bin. A misestimation of the effective galaxy bias due to either of these effects would be compensated by a shift in $\fNL$. As a second test of our quasar selection, we obtained constraints after imposing a brighter magnitude cut on the \quaia sample, using only sources with $G>20$; this is a cleaner sample with better redshift uncertainties and lower systematics. The resulting constraints (row 9) are compatible with our fiducial measurement, with a measurement error that grows by $\sim40\%$ due to the impact of shot noise.

    We find that our constraints are broadly stable against variations in the large-scale cut imposed on the data vector, both for $C_\ell^{gg}$ and $C_\ell^{g\kappa}$ (rows 10-12). In particular, using a less conservative scale cut for $C_\ell^{gg}$ (row 10) leads to a shift in $\fNL$ towards zero. Since potential contamination from observational systematics would most likely manifest itself on the largest scales probed, this is indirect confirmation that our results are robust against this contamination. As further evidence of this robustness, repeating our analysis on galaxy overdensity maps on which no deprojection of contaminant templates is performed (row 13) does not change our results significantly. Changing the small-scale cut used from $k_{\rm max}=0.07\,{\rm Mpc}^{-1}$ to $k_{\rm max}=0.1\,{\rm Mpc}^{-1}$ also does not strongly affect our constraints (row 14).

    To test the stability of our constraints against the procedure used to estimate the covariance matrix of our measurements, we repeat our analysis using the analytical Gaussian covariance matrix estimated by \nmt. The narrow-kernel approximation used in this estimate is known to lose accuracy on large scales, where the PNG signal is concentrated. The methodology used here also ignores the impact of inhomogeneity in the associated maps (e.g. caused by the spatially-varying \quaia selection function). In spite of this, the constraints found are in very good agreement with our fiducial measurement (row 15).

    Our constraints are also robust against potential errors in the theoretical predictions. Our fiducial model ignores both magnification bias and redshift-space distortions (RSDs), and assumes a specific functional form for the redshift evolution of the linear galaxy bias. Accounting for RSDs leads to virtually no change in the constraints (row 16). The inclusion of magnification bias, using the values of $s(z)$ measured for the \quaia sample in \cite{2306.17748}  (see Fig.~3), also has a negligible effect on our results (row 17). Using the bias evolution model of P24 instead (see Eq. \ref{eq:bz_P24}) also does not change our conclusions significantly (row 18).

    Finally, we study the impact of potential systematics in the CMB lensing maps used in our analysis. First, we quantify the effect of correcting for the mode loss in lensing reconstruction by repeating our analysis without correcting for the associated transfer function described in Section \ref{ssec:data.planck}. Failing to include this transfer function results in a downward shift in $\fNL$ by $\sim0.5\sigma$ (row 19). Thus, we do not expect any potential misestimation of the lensing reconstruction transfer function to affect our constraints significantly. Secondly, we test for the potential presence of residual contamination from extragalactic foregrounds in the CMB lensing map. Most importantly, the presence of Cosmic Infrared Background residuals could correlate strongly with the high-redshift quasars, leading to a biased cross-correlation measurement. This could then bias our estimate of $\fNL$, either by directly modifying the shape of the cross-correlation on large scales, or by affecting our estimate of the quasar bias. As was done in \cite{2306.17748,2402.05761}, to test for this systematic, we make use of an alternative CMB lensing map constructed avoiding $TT$ contributions in the quadratic estimator, described in Section \ref{ssec:data.planck}. Since only combinations involving polarization data contribute to this map, it should be largely free from extragalactic foreground contamination, at the cost of a degraded sensitivity. Repeating our analysis using this map, using both $C_\ell^{gg}$ and $C_\ell^{g\kappa}$, leads to the constraints found in row 20 of Table \ref{tab:results}. The constraints are shifted by $\sim-1\sigma$ with respect to our fiducial measurements. However, this shift is not caused by the change in CMB lensing map, but is instead due to the preference of $C_\ell^{gg}$ for a more negative value of $\fNL$, discussed above (see row 4): the lower sensitivity of the No-TT map downweights the contribution of $C_\ell^{g\kappa}$, which then leads to this shift. We can verify this by removing the quasar auto-correlations from the analysis. The resulting constraints on $\fNL$ (row 21) are in good agreement with our fiducial results (row 1), and with the $C_\ell^{g\kappa}$-only analysis carried out with the GMV $\kappa$ map (row 3).

    In summary, none of these tests reveal a significant systematic in our fiducial $\fNL$ constraints, or a lack of internal consistency in our measurements. The best-fit $\chi^2$ and associated PTE for these alternative analyses are listed in the two last columns of Table \ref{tab:results}. In all cases we find that our best-fit model provides a reasonable description of the data, with PTEs ranging from 0.15 to 0.9.

\section{Discussion and Conclusions}\label{sec:conc}
  \begin{figure}
    \centering
        \includegraphics[width=\textwidth]{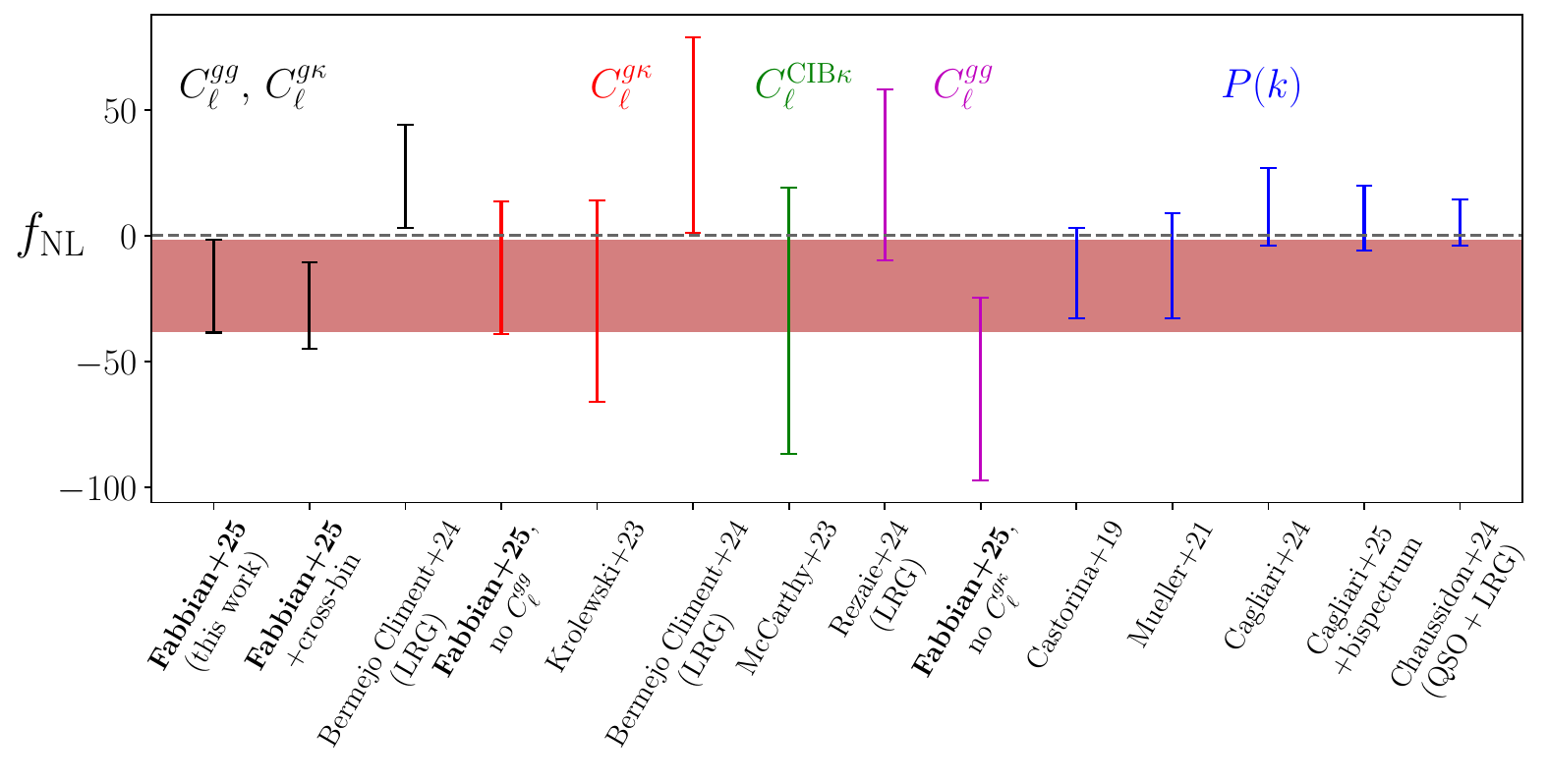}
    \caption{Constraints on $\fNL$ at 68\% C.L. found in this work in comparison with recent bounds from other groups in the literature \cite{1904.08859,2106.13725,2210.01049,2305.07650,2309.15814,2411.17623,2412.10279,2502.14758,2307.01753}. Black, red, and blue error bars show constraints found from the combination of $C_\ell^{gg}$ and $C_\ell^{g\kappa}$, from $C_\ell^{g\kappa}$ alone, and from the three-dimensional clustering of galaxies, respectively. The green error bars show the constraints found from the cross-correlation of CIB maps and CMB lensing \cite{2210.01049} and the magenta one from $C_\ell^{gg}$ alone. Unless stated the constraints used samples of QSOs.}
    \label{fig:fnl_other}
  \end{figure}

  We have presented constraints on local-type primordial non-Gaussianity from the clustering of quasars in the \quaia sample, using both their projected auto-correlation and their cross-correlation with maps of the CMB lensing convergence by \planck. We validated our power spectrum estimation pipeline against an independent implementation of the quadratic minimum-variance estimator, and quantified the impact of systematic contamination in the quasar overdensity maps via linear deprojection techniques. We find the PNG amplitude $\fNL$ to be compatible with zero, at the $\sim 1.1\sigma$ level in our fiducial analysis, and at $1.0\sigma$ with the \planck estimate of $\fNL^{local}=-0.9\pm 5.1$ \cite{1905.05697}. The measurement uncertainties, assuming an $\fNL$ response with $p_\phi=1$ or $p_\phi=1.6$, are $\sigma(\fNL)\simeq18$ and $\sigma(\fNL)\simeq25$, respectively (see Eqs. \ref{eq:fnl_fid} and \ref{eq:fnl_p1p6}). These results are consistent with those found using only the quasar-CMB lensing cross-correlation, which is potentially more robust against systematic contamination (see Eq. \ref{eq:fnl_nogg}). Exploiting the quasar cross-correlation between different redshift bins allows us to further reduce the statistical uncertainties by $\sim10\%$.

  Our results are robust against the most likely sources of systematic uncertainty, both observational and theoretical. This includes our choice of scale cuts, the presence of residual contamination in the galaxy overdensity and CMB lensing maps, the impact of magnification bias and redshift-space distortions, and assumptions about bias evolution. The results also display a good level of internal consistency. Although we observe a shift in $\fNL$ towards more negative values when considering only the quasar auto-correlations, we showed that the shift is not statistically significant. Our constraints are driven by the large-scale cross-correlation with CMB lensing, and by the High-$z$ quasar sample.

  It is interesting to compare our results with the most recent PNG constraints from the literature. Figure~\ref{fig:fnl_other} summarises the current state of the art. The results presented here are consistent with those found from the projected clustering of quasars and luminous red galaxies (LRGs) in the DESI Legacy Survey \cite{2305.07650,2412.10279}, both in auto- and cross-correlation, as well as the cross-correlation between CMB lensing and the Cosmic Infrared Background \cite{2210.01049}. Our analysis gives the tightest constraints to date on $\fNL$ from CMB lensing cross-correlations, and from projected datasets overall. Using the three-dimensional clustering of quasars allows for tighter constraints on $\fNL$, however. In particular, although the precision obtained in this analysis is similar to that achieved by early analyses of quasar clustering in the extended Baryon Oscillation Spectroscopic Survey (eBOSS) \cite{1904.08859,2106.13725}, the uncertainties on $\fNL$ obtained here are between 15\% and 50\% larger than those found using the latest eBOSS quasar data \cite{2309.15814,2502.14758} with optimized redshift-dependent weights, or the combination of LRG and quasar clustering from the first data release of the Dark Energy Spectroscopic Instrument \cite{2411.17623}. Nevertheless, the CMB lensing cross-correlation analysis offers the key advantage of being less sensitive to the impact of residual large-scale sky contamination, and \quaia is an independent sample that complements the measurements from these data sets. 
  
  Although the large angular scales of \quaia are easier to model compared to ground-based data sets, modes $\ell\lesssim 15$ in the projected QSO distribution are still affected by systematics. We carried out a forecast with our likelihood pipeline to assess the impact of systematics in our final statistical uncertainties. For this purpose, we replaced our measured data with theoretical predictions based on a cosmological model with $\fNL=0$, and considered two cases: first, we used the same covariance of the data, and secondly, a modified covariance that includes only contributions of signal and shot noise at all scales. We found that given the assumed scale cuts, the derived marginalized error bar on $\fNL$ is consistent in both cases. This shows that our constraints (which are driven by the sources at $z\gtrsim 1.5$) are at present limited by the shot noise in \quaia, the \planck lensing noise, and by the choice of the scale cuts rather than the additional error induced by systematics. Conversely, for the 1-bin case (and for the analysis of the low-$z$ bin) systematics limit more significantly the results. We also verified that improvements expected from the addition from the cross-correlation between the two tomographic redshift bins should be around 7\%, in the ballpark of what we saw in the data. If \quaia can be improved in future work or using future \gaia data releases, and therefore a lower $\lmin^{gg}$ can be included in the analysis, the variance of $C_\ell^{g\kappa}$ would be consequently reduced and the constraining power if $C_\ell^{gg}$ would increase. In this scenario, $\fNL$ could be constrained with a precision $\sigma(\fNL)\approx 9.5$ for a tomographic analysis with 2 redshift bins and $\lmin^{gg}=2$ combining $C_\ell^{g\kappa}$ and $C_\ell^{gg}$. Using $C_\ell^{gg}$ alone would give $\sigma(\fNL)\approx 11.5$ and would become the most constraining observable, with the \planck CMB lensing noise as limiting factor. Using CMB lensing from upcoming deep CMB polarization surveys covering large sky fractions such as Simons Observatory \cite{so,aso} will allow to overcome this limitation, as shown recently for the case of ACT \cite{embil-villagra2025}.
  
  We also note that the analysis presented here could be improved in a number of ways, all of which could lead to a significant increase in the constraining power of the \quaia dataset. Firstly, we have not attempted to derive and apply redshift-dependent weights for the \quaia sources that would maximise the sensitivity to $\fNL$. The use of optimal weights has been shown to lead to very significant improvements in 3D analyses \cite{1904.08859,2502.14758}, and similar approaches have been advanced for projected data \cite{1707.08950,1710.09465}. Secondly, although the precision of the spectro-photometric redshifts in \quaia should prevent a detailed measurement of the small-scale clustering along the line of sight, it should be sufficient to recover long-wavelength radial modes carrying additional information about PNG. More quantitatively, the mean normalised error in \quaia is approximately $\sigma_z/(1+z)\simeq0.06$. Translating this into a radial wavenumber as $k_{\parallel,{\rm max}}\equiv\pi\,H(z)/\sigma_z$, $k_{\parallel,{\rm max}}$ ranges between $\sim0.01\,{\rm Mpc}^{-1}$ and $0.016\,{\rm Mpc}^{-1}$ across the redshift range covered by \quaia. Although this is not enough to resolve the radial BAO, it should allow \quaia to constrain the clustering of quasars on horizon-sized radial scales over which PNG plays a significant role. Finally, we could further constrain PNG by including information from higher-order statistics, namely the galaxy bispectrum \cite{0705.0343,2010.14523,2201.11518,2204.01781,2502.14758}. Deploying efficient estimators of the projected bispectrum \cite{2409.07980}, targeting both the quasar auto-correlation and cross-correlations with CMB lensing, could therefore allow us to sharpen the constraints presented here using \quaia.

\begin{acknowledgments}
  We would like to thank Nestor Arsenov, Jos\'e Bermejo-Climent, Carlos Hern\'andez-Monteagudo, David W. Hogg, Andras Kovacs, An\v ze Slosar, Abby Williams and the members of the \quaia team for useful comments and discussions. GF thanks Gabriella Contardo for fostering this collaboration. GF acknowledges the support of the STFC Ernest Rutherford fellowship, of the European Research Council under the Marie Sk\l{}odowska Curie actions through the Individual Global Fellowship No.~892401 PiCOGAMBAS,  of the European Union’s Horizon 2020 research and innovation program (Grant agreement No. 851274), and of the Simons Foundation for the initial phase of this work. DA acknowledges support from the Beecroft Trust. KSF acknowledges the support of the Kavli Foundation.
  
\end{acknowledgments}

\section*{Software and Data Availability}
This work relied on open-source software packages, including \nmt~\citep{namaster}, \texttt{CCL}~\citep{ccl}, \texttt{SciPy}~\citep{scipy}, \texttt{NumPy}~\citep{numpy}, \texttt{Matplotlib}~\citep{matplotlib}, \texttt{emcee}~\citep{emcee}, \texttt{healpy}~\citep{healpy}, \hpx~\citep{healpix} and \texttt{xQML}~\citep{xqml}.
We make our key data products and code available at \url{https://github.com/gfabbian/quaia-fnl}; the rest of the analysis code will be made available upon request to the authors.

\appendix
\section{QML results and consistency with PCL pipeline}\label{app:qml}
\begin{figure}[!htbp]
    \centering
    \includegraphics[width=0.95\linewidth]{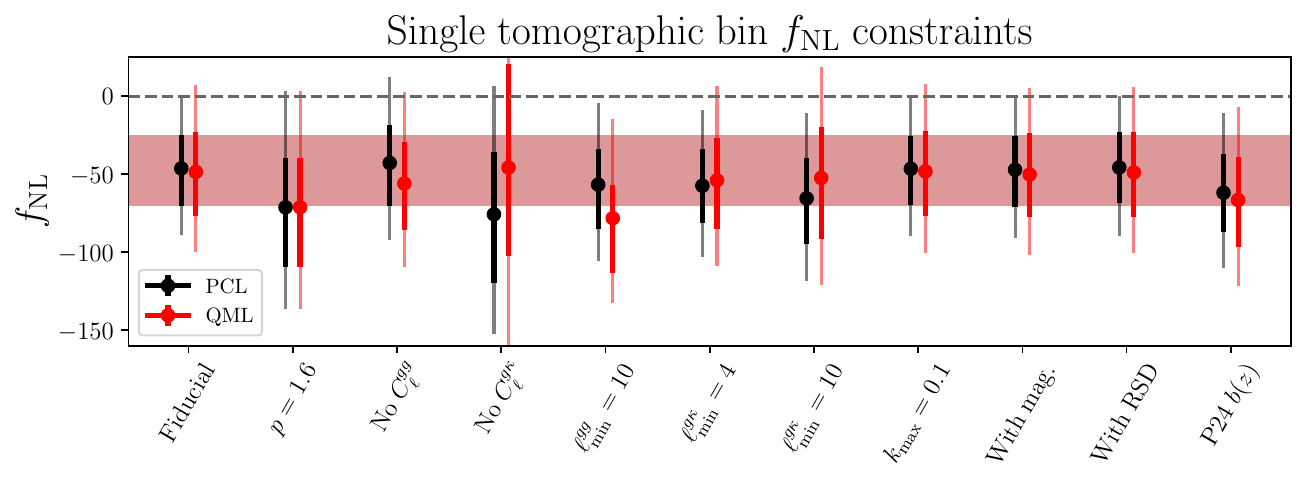}\\
    \includegraphics[width=0.95\linewidth]{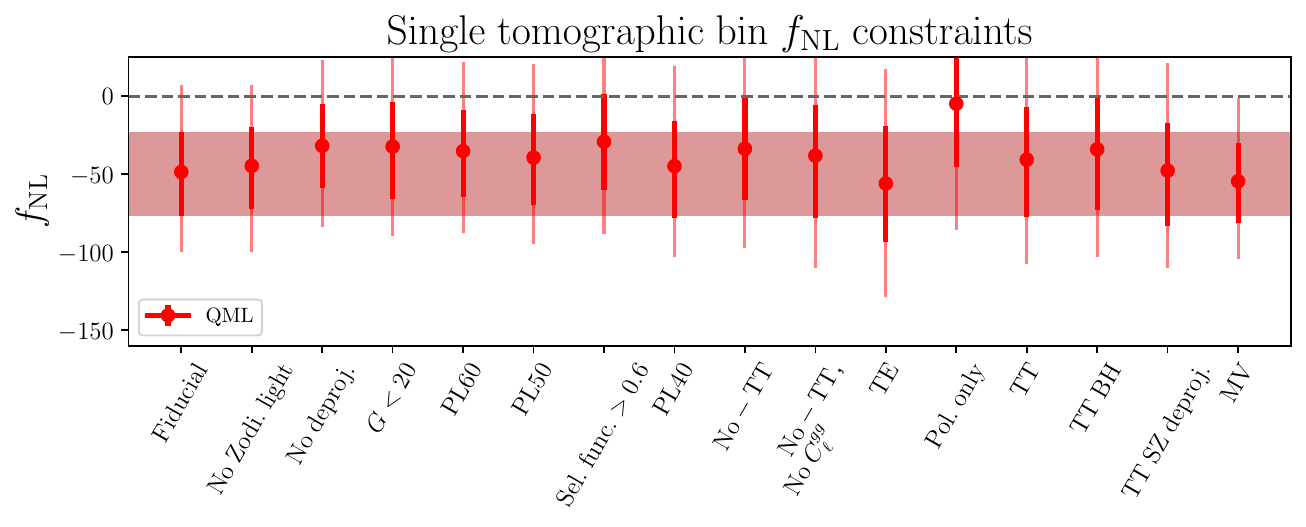}
    \caption{Top: $\fNL$ constraints from a single tomographic bin analysis of \quaia derived with the QML pipeline compared to the results obtained with the PCL pipeline for common analysis setups of the $G<20.5$ sample of Fig.~\ref{fig:fnl_tests}. The results show excellent agreement both in terms of central value and error bars. Bottom: additional consistency tests on $\fNL$ constraints ran for the QML pipeline mostly focus on stability with respect to changes in analysis mask, CMB lensing map or catalog definition. For both top and bottom panel, the shared error bars indicate the $95\%$ confidence level errors after marginalization over the QSO bias amplitude.}
    \label{fig:qml_fnl_tests}
\end{figure}

In Fig.~\ref{fig:qml-pcl-comparison} we show that the measurements of the angular power spectrum derived from the QML and PCL pipeline show a very good level of agreement despite implementing very different approaches to remove systematics and to estimate the data points and their covariance. As the PCL pipeline is numerically less costly and more easily adaptable, we use it to carry out the tomographic analysis reported in the main text. Before doing so, we performed several tests to assess the level of consistency on the estimated $\fNL$ in the case of a single tomographic redshift bin across both pipeline. \\*
In the upper panel of Fig.~\ref{fig:qml_fnl_tests} we show a detailed comparison of the same analysis setups considered in Fig.~\ref{fig:fnl_tests} and Table \ref{tab:results}. As it can be seen both the QML and PCL pipeline give very consistent results. The differences in the estimated value of $\fNL$ are negligible. The QML pipeline shows a slightly larger error bar of the order of 10\% or less, consistent with the difference in the the number of realizations employed to estimate the measurements' covariances in the two pipelines. As the QML pipeline also models the noise properties of the measurements using more realistic simulations, these results also prove that the data-based approach used to model the covariance in the PCL pipeline is sufficiently accurate to capture the error of the measurements. The only minor differences across pipeline can be seen when we relax the scale cuts used for $C_\ell^{gg}$ and include larger angular scales where the systematics subtraction is very different. However, even in this case,  the shifts in the estimated $\fNL$ amounts to $\lesssim 0.5\sigma$. 

The lower panel of Fig.~\ref{fig:qml-pcl-comparison} here shows the stability of the $\fNL$ estimates of the QML with respect to the change of catalog (using only the brightest $G<20$ QSOs), systematics templates as well as CMB lensing reconstruction method. Specifically we tested that removing the Zodiacal light template from the list of systematics templates that are marginalized over, does not lead to a noticeable difference in the results. For the mask variations we considered the official \planck galactic masks that retain the 60\% (PL60) and 40\% (PL40) of the sky available on the \planck legacy archive. To this we considered an extra mask obtained thresholding the brightest pixels of the \planck 353 GHz map to retain 50\% of the sky (PL50) as well a mask including sky locations where the \quaia selection function exceeds 60\%. This roughly retains a 40\% sky fraction. Finally we also considered additional variations of the \planck PR4 CMB lensing estimation compared those in Sec.~\ref{sec:res}. In particular, we included CMB lensing maps derived using the minimum variance combination (MV) that treats suboptimally the CMB noise inhomogeneity \cite{1807.06210}, using CMB polarization data only, or correlating CMB temperature and E-mode polarization only (TE). We also included CMB lensing maps reconstructed using CMB temperature data only (TT) with and without specific extragalactic foreground mitigation techniques such as bias hardening (BH) against point source emission, or deprojection of the SZ effect in the component separation step prior to lensing reconstruction. All these variation of lensing reconstruction and related simulation (that we used for the covariance estimation) are publicly available\footnote{\url{https://data.cmb-s4.org/planck_pr4.html}}. For all these cases we found shifts in the parameter within fractions of the error. The polarization only reconstruction experiences larger shifts, however the CMB lensing map is considerably noisier and the final error on $\fNL$ is 50\% larger. As such it is not statistically significant. Because we applied the different galactic masks on top of the mask used for the fiducial analysis (and therefore the data sets are effectively nested), we tested the consistency of the parameters using the criteria outlined in \cite{1911.07754}, and found no indication of anomalies. We also found that using data for which no systematics marginalization is done, led to minor changes in the results, showing the robustness of our scale cuts and lack of systematics in $C_\ell^{g\kappa}$ that drives the constraints. For the cases  considered here, the estimated value of $\fNL$ is also consistent with 0 at $\sim 2\sigma$ level (95\% C.L.). All the results of the tests we considered are summarized in Table \ref{tab:results-qml}.

    \begin{table}[!htbp]
      \centering
      \renewcommand{\arraystretch}{1.5}
      \normalsize
      \begin{tabular}{|l|l|l|l|l|l|}
        \hline
        Analysis setting & $\fNL$ & $A_1$ & $\chi^2/N_{\rm dof}$ & PTE \\[0.5ex]
        \hline
        
${\rm Fiducial}$ & $-48.6^{+25.5}_{-28.5}$ & $ 1.02 \pm 0.04$& $54.3 / 55$ & $0.50$ \\
$p=1.6$ & $-71.3^{+31.7}_{-38.0}$ & $ 1.03 \pm 0.04$& $50.1 / 55$ & $0.66$ \\
${\rm No}\ C_{\ell}^{gg}$ & $-56.0^{+26.5}_{-29.7}$ & $ 1.00 \pm 0.05$& $24.6 / 30$ & $0.74$ \\
${\rm No}\ C_{\ell}^{g\kappa}$ & $-45.8^{+66.6}_{-56.5}$ & $ 1.03 \pm 0.06$& $23.2 / 23$ & $0.45$ \\
$G<20$ & $-32.3^{+28.4}_{-33.5}$ & $ 1.05 \pm 0.06$& $43.2 / 55$ & $0.87$ \\
$\ell_{\rm min}^{gg}=10$ & $-78.2^{+21.6}_{-35.4}$ & $ 1.04 \pm 0.04$& $61.6 / 58$ & $0.35$ \\
$\ell_{\rm min}^{g\kappa}=4$ & $-53.9^{+27.3}_{-31.3}$ & $ 1.02 \pm 0.04$& $54.3 / 54$ & $0.46$ \\
$\ell_{\rm min}^{g\kappa}=10$ & $-52.4^{+32.9}_{-39.1}$ & $ 1.02 \pm 0.05$& $53.8 / 51$ & $0.37$ \\
${\rm No\  deproj.}$ & $-31.9^{+27.1}_{-27.1}$ & $ 1.01 \pm 0.04$& $58.2 / 55$ & $0.36$ \\
$k_{\rm max}=0.1$ & $-48.2^{+26.1}_{-28.6}$ & $ 1.01 \pm 0.04$& $69.0 / 65$ & $0.34$ \\
${\rm With\  RSD}$ & $-48.9^{+26.1}_{-28.3}$ & $ 1.02 \pm 0.04$& $54.3 / 55$ & $0.50$ \\
${\rm With\  mag.}$ & $-50.2^{+26.8}_{-27.0}$ & $ 1.01 \pm 0.04$& $54.4 / 55$ & $0.50$ \\
${\rm P24}$ $b(z)$ & $-66.6^{+27.6}_{-30.4}$ & $ 0.93 \pm 0.04$& $55.5 / 55$ & $0.46$ \\
${\rm No\ TT}$ & $-33.7^{+32.6}_{-32.9}$ & $ 1.04 \pm 0.05$& $47.3 / 55$ & $0.76$ \\
${\rm No\ TT}$ ${\rm No}$ $C_{\ell}^{gg}$ & $-38.1^{+32.6}_{-39.6}$ & $ 1.01 \pm 0.08$& $19.5 / 30$ & $0.93$ \\
${\rm TT}$ & $-40.7^{+33.5}_{-37.0}$ & $ 1.00 \pm 0.05$& $49.5 / 55$ & $0.68$ \\
${\rm TT  BH}$ & $-34.1^{+34.4}_{-38.8}$ & $ 0.99 \pm 0.05$& $49.3 / 55$ & $0.69$ \\
${\rm TT\  SZ \ deproj.}$ & $-47.7^{+30.5}_{-35.2}$ & $ 1.00 \pm 0.05$& $50.6 / 55$ & $0.64$ \\
${\rm TE}$ & $-56.0^{+37.2}_{-37.7}$ & $ 1.03 \pm 0.05$& $46.0 / 55$ & $0.80$ \\
${\rm Pol.\  only}$ & $-4.9^{+40.6}_{-40.3}$ & $ 1.02 \pm 0.05$& $43.3 / 55$ & $0.87$ \\
${\rm MV}$ & $-54.5^{+24.2}_{-27.0}$ & $ 1.02 \pm 0.04$& $49.2 / 55$ & $0.69$ \\
${\rm PL60}$ & $-35.3^{+26.4}_{-29.2}$ & $ 1.00 \pm 0.04$& $52.7 / 55$ & $0.56$ \\
${\rm PL50}$ & $-39.4^{+27.6}_{-30.6}$ & $ 0.98 \pm 0.04$& $56.5 / 55$ & $0.42$ \\
${\rm PL40}$ & $-44.9^{+29.1}_{-32.9}$ & $ 0.99 \pm 0.05$& $46.8 / 55$ & $0.78$ \\
${\rm Sel.\  func.}>0.6$ & $-29.2^{+30.6}_{-30.7}$ & $ 0.99 \pm 0.04$& $60.1 / 55$ & $0.30$ \\
${\rm No\  Zodi.\  light}$ & $-44.7^{+25.0}_{-27.4}$ & $ 1.02 \pm 0.04$& $54.4 / 55$ & $0.50$ \\        
         \hline
      \end{tabular}
      \caption{Constraints on the model parameters obtained with the QML pipeline for the benchmark case of a single tomographic redshift bin as in Table \ref{tab:results}. Results are shown for our fiducial analysis (first row), as well as all other variations used to explore the stability of our results to different analysis choices and potential systematics. Error bars are 68\% C.L. values.}
      \label{tab:results-qml}
    \end{table}

\bibliography{main}{}

@string{june = {June}}

@article{0705.0343,
 adsurl = {https://ui.adsabs.harvard.edu/abs/2007PhRvD..76h3004S},
 archiveprefix = {arXiv},
 author = {{Sefusatti}, Emiliano and {Komatsu}, Eiichiro},
 doi = {10.1103/PhysRevD.76.083004},
 eid = {083004},
 eprint = {0705.0343},
 journal = {\prd},
 month = {October},
 number = {8},
 pages = {083004},
 primaryclass = {astro-ph},
 title = {{Bispectrum of galaxies from high-redshift galaxy surveys: Primordial non-Gaussianity and nonlinear galaxy bias}},
 volume = {76},
 year = {2007}
}

@article{0710.4560,
 adsurl = {https://ui.adsabs.harvard.edu/abs/2008PhRvD..77l3514D},
 archiveprefix = {arXiv},
 author = {{Dalal}, Neal and {Dor{\'e}}, Olivier and {Huterer}, Dragan and {Shirokov}, Alexander},
 doi = {10.1103/PhysRevD.77.123514},
 eid = {123514},
 eprint = {0710.4560},
 journal = {\prd},
 month = {June},
 number = {12},
 pages = {123514},
 primaryclass = {astro-ph},
 title = {{Imprints of primordial non-Gaussianities on large-scale structure: Scale-dependent bias and abundance of virialized objects}},
 volume = {77},
 year = {2008}
}

@article{0801.0554,
 adsurl = {https://ui.adsabs.harvard.edu/abs/2008PhRvD..77j3013H},
 archiveprefix = {arXiv},
 author = {{Hamimeche}, Samira and {Lewis}, Antony},
 doi = {10.1103/PhysRevD.77.103013},
 eid = {103013},
 eprint = {0801.0554},
 journal = {\prd},
 month = {May},
 number = {10},
 pages = {103013},
 primaryclass = {astro-ph},
 title = {{Likelihood analysis of CMB temperature and polarization power spectra}},
 volume = {77},
 year = {2008}
}

@article{0805.3580,
 adsurl = {https://ui.adsabs.harvard.edu/abs/2008JCAP...08..031S},
 archiveprefix = {arXiv},
 author = {{Slosar}, An{\v{z}}e and {Hirata}, Christopher and {Seljak}, Uro{\v{s}} and {Ho}, Shirley and {Padmanabhan}, Nikhil},
 doi = {10.1088/1475-7516/2008/08/031},
 eid = {031},
 eprint = {0805.3580},
 journal = {\jcap},
 month = {August},
 number = {8},
 pages = {031},
 primaryclass = {astro-ph},
 title = {{Constraints on local primordial non-Gaussianity from large scale structure}},
 volume = {2008},
 year = {2008}
}

@article{0907.5424,
 adsurl = {https://ui.adsabs.harvard.edu/abs/2009arXiv0907.5424B},
 archiveprefix = {arXiv},
 author = {{Baumann}, Daniel},
 doi = {10.48550/arXiv.0907.5424},
 eid = {arXiv:0907.5424},
 eprint = {0907.5424},
 journal = {arXiv e-prints},
 month = {July},
 pages = {arXiv:0907.5424},
 primaryclass = {hep-th},
 title = {{TASI Lectures on Inflation}},
 year = {2009}
}

@article{1105.5280,
 adsurl = {https://ui.adsabs.harvard.edu/abs/2011PhRvD..84f3505B},
 archiveprefix = {arXiv},
 author = {{Bonvin}, Camille and {Durrer}, Ruth},
 doi = {10.1103/PhysRevD.84.063505},
 eid = {063505},
 eprint = {1105.5280},
 journal = {\prd},
 month = {September},
 number = {6},
 pages = {063505},
 primaryclass = {astro-ph.CO},
 title = {{What galaxy surveys really measure}},
 volume = {84},
 year = {2011}
}

@article{1105.5292,
 adsurl = {https://ui.adsabs.harvard.edu/abs/2011PhRvD..84d3516C},
 archiveprefix = {arXiv},
 author = {{Challinor}, Anthony and {Lewis}, Antony},
 doi = {10.1103/PhysRevD.84.043516},
 eid = {043516},
 eprint = {1105.5292},
 journal = {\prd},
 month = {August},
 number = {4},
 pages = {043516},
 primaryclass = {astro-ph.CO},
 title = {{Linear power spectrum of observed source number counts}},
 volume = {84},
 year = {2011}
}

@article{emcee,
 adsurl = {https://ui.adsabs.harvard.edu/abs/2013PASP..125..306F},
 archiveprefix = {arXiv},
 author = {{Foreman-Mackey}, Daniel and {Hogg}, David W. and {Lang}, Dustin and {Goodman}, Jonathan},
 doi = {10.1086/670067},
 eprint = {1202.3665},
 journal = {\pasp},
 month = {March},
 number = {925},
 pages = {306},
 primaryclass = {astro-ph.IM},
 title = {{emcee: The MCMC Hammer}},
 volume = {125},
 year = {2013},
 url = {https://emcee.readthedocs.io/}
}

@article{1207.6487,
 adsurl = {https://ui.adsabs.harvard.edu/abs/2012MNRAS.427.1891A},
 archiveprefix = {arXiv},
 author = {{Asorey}, Jacobo and {Crocce}, Martin and {Gazta{\~n}aga}, Enrique and {Lewis}, Antony},
 doi = {10.1111/j.1365-2966.2012.21972.x},
 eprint = {1207.6487},
 journal = {\mnras},
 month = {December},
 number = {3},
 pages = {1891-1902},
 primaryclass = {astro-ph.CO},
 title = {{Recovering 3D clustering information with angular correlations}},
 volume = {427},
 year = {2012}
}

@article{WilliamsInprep,
 author = {{Williams}, A. and others},
 journal = {in preparation.},
 year = {2025}
}

@article{CornishInprep,
    author = "Cornish, Thomas and Alonso, David and Leistedt, Boris and Wolz, Kevin",
    title = "{Systematics mitigation for catalogue-based angular power spectra}",
    archivePrefix = "arXiv",
    primaryClass = "astro-ph.CO",
      journal = {arXiv e-prints},
         year = 2025,
        month = oct,
          eid = {},
        pages = {},
          doi = {10.48550/arXiv.2510.19912},
archivePrefix = {arXiv},
       eprint = {2510.19912},
 primaryClass = {astro-ph.CO},

}

@article{1306.0005,
 adsurl = {https://ui.adsabs.harvard.edu/abs/2013MNRAS.435.1857L},
 archiveprefix = {arXiv},
 author = {{Leistedt}, Boris and {Peiris}, Hiranya V. and {Mortlock}, Daniel J. and {Benoit-L{\'e}vy}, Aur{\'e}lien and {Pontzen}, Andrew},
 doi = {10.1093/mnras/stt1359},
 eprint = {1306.0005},
 journal = {\mnras},
 month = {November},
 number = {3},
 pages = {1857-1873},
 primaryclass = {astro-ph.CO},
 title = {{Estimating the large-scale angular power spectrum in the presence of systematics: a case study of Sloan Digital Sky Survey quasars}},
 volume = {435},
 year = {2013}
}

@article{1402.2290,
 adsurl = {https://ui.adsabs.harvard.edu/abs/2014MNRAS.442.2511F},
 archiveprefix = {arXiv},
 author = {{Ferramacho}, L.~D. and {Santos}, M.~G. and {Jarvis}, M.~J. and {Camera}, S.},
 doi = {10.1093/mnras/stu1015},
 eprint = {1402.2290},
 journal = {\mnras},
 month = {August},
 number = {3},
 pages = {2511-2518},
 primaryclass = {astro-ph.CO},
 title = {{Radio galaxy populations and the multitracer technique: pushing the limits on primordial non-Gaussianity}},
 volume = {442},
 year = {2014}
}

@article{1505.07596,
 adsurl = {https://ui.adsabs.harvard.edu/abs/2015ApJ...814..145A},
 archiveprefix = {arXiv},
 author = {{Alonso}, David and {Bull}, Philip and {Ferreira}, Pedro G. and {Maartens}, Roy and {Santos}, M{\'a}rio G.},
 doi = {10.1088/0004-637X/814/2/145},
 eid = {145},
 eprint = {1505.07596},
 journal = {\apj},
 month = {December},
 number = {2},
 pages = {145},
 primaryclass = {astro-ph.CO},
 title = {{Ultra-large-scale Cosmology in Next-generation Experiments with Single Tracers}},
 volume = {814},
 year = {2015}
}

@article{1507.03550,
 adsurl = {https://ui.adsabs.harvard.edu/abs/2015PhRvD..92f3525A},
 archiveprefix = {arXiv},
 author = {{Alonso}, D. and {Ferreira}, P.~G.},
 doi = {10.1103/PhysRevD.92.063525},
 eid = {063525},
 eprint = {1507.03550},
 journal = {\prd},
 month = {September},
 number = {6},
 pages = {063525},
 primaryclass = {astro-ph.CO},
 title = {{Constraining ultralarge-scale cosmology with multiple tracers in optical and radio surveys}},
 volume = {92},
 year = {2015}
}

@article{1609.03577,
 adsurl = {https://ui.adsabs.harvard.edu/abs/2017MNRAS.465.1847E},
 archiveprefix = {arXiv},
 author = {{Elsner}, Franz and {Leistedt}, Boris and {Peiris}, Hiranya V.},
 doi = {10.1093/mnras/stw2752},
 eprint = {1609.03577},
 journal = {\mnras},
 month = {February},
 number = {2},
 pages = {1847-1855},
 primaryclass = {astro-ph.CO},
 title = {{Unbiased pseudo-C$_{{\ensuremath{\ell}}}$ power spectrum estimation with mode projection}},
 volume = {465},
 year = {2017}
}

@article{1612.00770,
 adsurl = {https://ui.adsabs.harvard.edu/abs/2017PhRvD..95f3522K},
 archiveprefix = {arXiv},
 author = {{Kitching}, T.~D. and {Heavens}, A.~F.},
 doi = {10.1103/PhysRevD.95.063522},
 eid = {063522},
 eprint = {1612.00770},
 journal = {\prd},
 month = {March},
 number = {6},
 pages = {063522},
 primaryclass = {astro-ph.CO},
 title = {{Unequal-time correlators for cosmology}},
 volume = {95},
 year = {2017}
}

@article{1705.04718,
 adsurl = {https://ui.adsabs.harvard.edu/abs/2017JCAP...07..017L},
 archiveprefix = {arXiv},
 author = {{Laurent}, Pierre and {Eftekharzadeh}, Sarah and {Le Goff}, Jean-Marc and {Myers}, Adam and {Burtin}, Etienne and {White}, Martin and {Ross}, Ashley J. and {Tinker}, Jeremy and {Tojeiro}, Rita and {Bautista}, Julian and {Brinkmann}, Jonathan and {Comparat}, Johan and {Dawson}, Kyle and {du Mas des Bourboux}, H{\'e}lion and {Kneib}, Jean-Paul and {McGreer}, Ian D. and {Palanque-Delabrouille}, Nathalie and {Percival}, Will J. and {Prada}, Francisco and {Rossi}, Graziano and {Schneider}, Donald P. and {Weinberg}, David and {Y{\`e}che}, Christophe and {Zarrouk}, Pauline and {Zhao}, Gong-Bo},
 doi = {10.1088/1475-7516/2017/07/017},
 eid = {017},
 eprint = {1705.04718},
 journal = {\jcap},
 month = {July},
 number = {7},
 pages = {017},
 primaryclass = {astro-ph.CO},
 title = {{Clustering of quasars in SDSS-IV eBOSS: study of potential systematics and bias determination}},
 volume = {2017},
 year = {2017}
}

@article{1707.08950,
 adsurl = {https://ui.adsabs.harvard.edu/abs/2018MNRAS.473.4306A},
 archiveprefix = {arXiv},
 author = {{Alonso}, David},
 doi = {10.1093/mnras/stx2644},
 eprint = {1707.08950},
 journal = {\mnras},
 month = {February},
 number = {4},
 pages = {4306-4317},
 primaryclass = {astro-ph.CO},
 title = {{Science-driven 3D data compression}},
 volume = {473},
 year = {2018}
}

@article{1710.09465,
 adsurl = {https://ui.adsabs.harvard.edu/abs/2018PhRvD..97l3540S},
 archiveprefix = {arXiv},
 author = {{Schmittfull}, Marcel and {Seljak}, Uro{\v{s}}},
 doi = {10.1103/PhysRevD.97.123540},
 eid = {123540},
 eprint = {1710.09465},
 journal = {\prd},
 month = {June},
 number = {12},
 pages = {123540},
 primaryclass = {astro-ph.CO},
 title = {{Parameter constraints from cross-correlation of CMB lensing with galaxy clustering}},
 volume = {97},
 year = {2018}
}

@article{1807.02484,
 adsurl = {https://ui.adsabs.harvard.edu/abs/2018PhRvD..98j3526V},
 archiveprefix = {arXiv},
 author = {{Vanneste}, S. and {Henrot-Versill{\'e}}, S. and {Louis}, T. and {Tristram}, M.},
 doi = {10.1103/PhysRevD.98.103526},
 eid = {103526},
 eprint = {1807.02484},
 journal = {\prd},
 month = {November},
 number = {10},
 pages = {103526},
 primaryclass = {astro-ph.CO},
 title = {{Quadratic estimator for CMB cross-correlation}},
 volume = {98},
 year = {2018}
}

@article{1807.06209,
 adsurl = {https://ui.adsabs.harvard.edu/abs/2020A&A...641A...6P},
 archiveprefix = {arXiv},
 author = {{Planck Collaboration} and {Aghanim}, N. and {Akrami}, Y. and {Ashdown}, M. and {Aumont}, J. and {Baccigalupi}, C. and {Ballardini}, M. and {Banday}, A.~J. and {Barreiro}, R.~B. and {Bartolo}, N. and {Basak}, S. and {Battye}, R. and {Benabed}, K. and {Bernard}, J. -P. and {Bersanelli}, M. and {Bielewicz}, P. and {Bock}, J.~J. and {Bond}, J.~R. and {Borrill}, J. and {Bouchet}, F.~R. and {Boulanger}, F. and {Bucher}, M. and {Burigana}, C. and {Butler}, R.~C. and {Calabrese}, E. and {Cardoso}, J. -F. and {Carron}, J. and {Challinor}, A. and {Chiang}, H.~C. and {Chluba}, J. and {Colombo}, L.~P.~L. and {Combet}, C. and {Contreras}, D. and {Crill}, B.~P. and {Cuttaia}, F. and {de Bernardis}, P. and {de Zotti}, G. and {Delabrouille}, J. and {Delouis}, J. -M. and {Di Valentino}, E. and {Diego}, J.~M. and {Dor{\'e}}, O. and {Douspis}, M. and {Ducout}, A. and {Dupac}, X. and {Dusini}, S. and {Efstathiou}, G. and {Elsner}, F. and {En{\ss}lin}, T.~A. and {Eriksen}, H.~K. and {Fantaye}, Y. and {Farhang}, M. and {Fergusson}, J. and {Fernandez-Cobos}, R. and {Finelli}, F. and {Forastieri}, F. and {Frailis}, M. and {Fraisse}, A.~A. and {Franceschi}, E. and {Frolov}, A. and {Galeotta}, S. and {Galli}, S. and {Ganga}, K. and {G{\'e}nova-Santos}, R.~T. and {Gerbino}, M. and {Ghosh}, T. and {Gonz{\'a}lez-Nuevo}, J. and {G{\'o}rski}, K.~M. and {Gratton}, S. and {Gruppuso}, A. and {Gudmundsson}, J.~E. and {Hamann}, J. and {Handley}, W. and {Hansen}, F.~K. and {Herranz}, D. and {Hildebrandt}, S.~R. and {Hivon}, E. and {Huang}, Z. and {Jaffe}, A.~H. and {Jones}, W.~C. and {Karakci}, A. and {Keih{\"a}nen}, E. and {Keskitalo}, R. and {Kiiveri}, K. and {Kim}, J. and {Kisner}, T.~S. and {Knox}, L. and {Krachmalnicoff}, N. and {Kunz}, M. and {Kurki-Suonio}, H. and {Lagache}, G. and {Lamarre}, J. -M. and {Lasenby}, A. and {Lattanzi}, M. and {Lawrence}, C.~R. and {Le Jeune}, M. and {Lemos}, P. and {Lesgourgues}, J. and {Levrier}, F. and {Lewis}, A. and {Liguori}, M. and {Lilje}, P.~B. and {Lilley}, M. and {Lindholm}, V. and {L{\'o}pez-Caniego}, M. and {Lubin}, P.~M. and {Ma}, Y. -Z. and {Mac{\'\i}as-P{\'e}rez}, J.~F. and {Maggio}, G. and {Maino}, D. and {Mandolesi}, N. and {Mangilli}, A. and {Marcos-Caballero}, A. and {Maris}, M. and {Martin}, P.~G. and {Martinelli}, M. and {Mart{\'\i}nez-Gonz{\'a}lez}, E. and {Matarrese}, S. and {Mauri}, N. and {McEwen}, J.~D. and {Meinhold}, P.~R. and {Melchiorri}, A. and {Mennella}, A. and {Migliaccio}, M. and {Millea}, M. and {Mitra}, S. and {Miville-Desch{\^e}nes}, M. -A. and {Molinari}, D. and {Montier}, L. and {Morgante}, G. and {Moss}, A. and {Natoli}, P. and {N{\o}rgaard-Nielsen}, H.~U. and {Pagano}, L. and {Paoletti}, D. and {Partridge}, B. and {Patanchon}, G. and {Peiris}, H.~V. and {Perrotta}, F. and {Pettorino}, V. and {Piacentini}, F. and {Polastri}, L. and {Polenta}, G. and {Puget}, J. -L. and {Rachen}, J.~P. and {Reinecke}, M. and {Remazeilles}, M. and {Renzi}, A. and {Rocha}, G. and {Rosset}, C. and {Roudier}, G. and {Rubi{\~n}o-Mart{\'\i}n}, J.~A. and {Ruiz-Granados}, B. and {Salvati}, L. and {Sandri}, M. and {Savelainen}, M. and {Scott}, D. and {Shellard}, E.~P.~S. and {Sirignano}, C. and {Sirri}, G. and {Spencer}, L.~D. and {Sunyaev}, R. and {Suur-Uski}, A. -S. and {Tauber}, J.~A. and {Tavagnacco}, D. and {Tenti}, M. and {Toffolatti}, L. and {Tomasi}, M. and {Trombetti}, T. and {Valenziano}, L. and {Valiviita}, J. and {Van Tent}, B. and {Vibert}, L. and {Vielva}, P. and {Villa}, F. and {Vittorio}, N. and {Wandelt}, B.~D. and {Wehus}, I.~K. and {White}, M. and {White}, S.~D.~M. and {Zacchei}, A. and {Zonca}, A.},
 doi = {10.1051/0004-6361/201833910},
 eid = {A6},
 eprint = {1807.06209},
 journal = {\aap},
 month = {September},
 pages = {A6},
 primaryclass = {astro-ph.CO},
 title = {{Planck 2018 results. VI. Cosmological parameters}},
 volume = {641},
 year = {2020}
}

@article{namaster,
 adsurl = {https://ui.adsabs.harvard.edu/abs/2019MNRAS.484.4127A},
 archiveprefix = {arXiv},
 author = {{Alonso}, David and {Sanchez}, Javier and {Slosar}, An{\v{z}}e and {LSST Dark Energy Science Collaboration}},
 doi = {10.1093/mnras/stz093},
 eprint = {1809.09603},
 journal = {\mnras},
 month = {April},
 number = {3},
 pages = {4127-4151},
 primaryclass = {astro-ph.CO},
 title = {{A unified pseudo-C$_{{\ensuremath{\ell}}}$ framework}},
 volume = {484},
 year = {2019},
 url = {https://namaster.readthedocs.io/}
}

@article{ccl,
 adsurl = {https://ui.adsabs.harvard.edu/abs/2019ApJS..242....2C},
 archiveprefix = {arXiv},
 author = {{Chisari}, Nora Elisa and {Alonso}, David and {Krause}, Elisabeth and {Leonard}, C. Danielle and {Bull}, Philip and {Neveu}, J{\'e}r{\'e}my and {Villarreal}, Antonia Sierra and {Singh}, Sukhdeep and {McClintock}, Thomas and {Ellison}, John and {Du}, Zilong and {Zuntz}, Joe and {Mead}, Alexander and {Joudaki}, Shahab and {Lorenz}, Christiane S. and {Tr{\"o}ster}, Tilman and {Sanchez}, Javier and {Lanusse}, Francois and {Ishak}, Mustapha and {Hlozek}, Ren{\'e}e and {Blazek}, Jonathan and {Campagne}, Jean-Eric and {Almoubayyed}, Husni and {Eifler}, Tim and {Kirby}, Matthew and {Kirkby}, David and {Plaszczynski}, St{\'e}phane and {Slosar}, An{\v{z}}e and {Vrastil}, Michal and {Wagoner}, Erika L. and {LSST Dark Energy Science Collaboration}},
 doi = {10.3847/1538-4365/ab1658},
 eid = {2},
 eprint = {1812.05995},
 journal = {\apjs},
 month = {May},
 number = {1},
 pages = {2},
 primaryclass = {astro-ph.CO},
 title = {{Core Cosmology Library: Precision Cosmological Predictions for LSST}},
 volume = {242},
 year = {2019},
 url = {https://ccl.readthedocs.io/}
}

@article{1904.08859,
 adsurl = {https://ui.adsabs.harvard.edu/abs/2019JCAP...09..010C},
 archiveprefix = {arXiv},
 author = {{Castorina}, Emanuele and {Hand}, Nick and {Seljak}, Uro{\v{s}} and {Beutler}, Florian and {Chuang}, Chia-Hsun and {Zhao}, Cheng and {Gil-Mar{\'\i}n}, H{\'e}ctor and {Percival}, Will J. and {Ross}, Ashley J. and {Choi}, Peter Doohyun and {Dawson}, Kyle and {de la Macorra}, Axel and {Rossi}, Graziano and {Ruggeri}, Rossana and {Schneider}, Donald and {Zhao}, Gong-Bo},
 doi = {10.1088/1475-7516/2019/09/010},
 eid = {010},
 eprint = {1904.08859},
 journal = {\jcap},
 month = {September},
 number = {9},
 pages = {010},
 primaryclass = {astro-ph.CO},
 title = {{Redshift-weighted constraints on primordial non-Gaussianity from the clustering of the eBOSS DR14 quasars in Fourier space}},
 volume = {2019},
 year = {2019}
}

@article{1905.02078,
 adsurl = {https://ui.adsabs.harvard.edu/abs/2019PhRvD.100b3543C},
 archiveprefix = {arXiv},
 author = {{Chisari}, Nora Elisa and {Pontzen}, Andrew},
 doi = {10.1103/PhysRevD.100.023543},
 eid = {023543},
 eprint = {1905.02078},
 journal = {\prd},
 month = {July},
 number = {2},
 pages = {023543},
 primaryclass = {astro-ph.CO},
 title = {{Unequal time correlators and the Zel'dovich approximation}},
 volume = {100},
 year = {2019}
}

@article{1905.05697,
 adsurl = {https://ui.adsabs.harvard.edu/abs/2020A&A...641A...9P},
 archiveprefix = {arXiv},
 author = {{Planck Collaboration} and {Akrami}, Y. and {Arroja}, F. and {Ashdown}, M. and {Aumont}, J. and {Baccigalupi}, C. and {Ballardini}, M. and {Banday}, A.~J. and {Barreiro}, R.~B. and {Bartolo}, N. and {Basak}, S. and {Benabed}, K. and {Bernard}, J. -P. and {Bersanelli}, M. and {Bielewicz}, P. and {Bond}, J.~R. and {Borrill}, J. and {Bouchet}, F.~R. and {Bucher}, M. and {Burigana}, C. and {Butler}, R.~C. and {Calabrese}, E. and {Cardoso}, J. -F. and {Casaponsa}, B. and {Challinor}, A. and {Chiang}, H.~C. and {Colombo}, L.~P.~L. and {Combet}, C. and {Crill}, B.~P. and {Cuttaia}, F. and {de Bernardis}, P. and {de Rosa}, A. and {de Zotti}, G. and {Delabrouille}, J. and {Delouis}, J. -M. and {Di Valentino}, E. and {Diego}, J.~M. and {Dor{\'e}}, O. and {Douspis}, M. and {Ducout}, A. and {Dupac}, X. and {Dusini}, S. and {Efstathiou}, G. and {Elsner}, F. and {En{\ss}lin}, T.~A. and {Eriksen}, H.~K. and {Fantaye}, Y. and {Fergusson}, J. and {Fernandez-Cobos}, R. and {Finelli}, F. and {Frailis}, M. and {Fraisse}, A.~A. and {Franceschi}, E. and {Frolov}, A. and {Galeotta}, S. and {Galli}, S. and {Ganga}, K. and {G{\'e}nova-Santos}, R.~T. and {Gerbino}, M. and {Gonz{\'a}lez-Nuevo}, J. and {G{\'o}rski}, K.~M. and {Gratton}, S. and {Gruppuso}, A. and {Gudmundsson}, J.~E. and {Hamann}, J. and {Handley}, W. and {Hansen}, F.~K. and {Herranz}, D. and {Hivon}, E. and {Huang}, Z. and {Jaffe}, A.~H. and {Jones}, W.~C. and {Jung}, G. and {Keih{\"a}nen}, E. and {Keskitalo}, R. and {Kiiveri}, K. and {Kim}, J. and {Krachmalnicoff}, N. and {Kunz}, M. and {Kurki-Suonio}, H. and {Lamarre}, J. -M. and {Lasenby}, A. and {Lattanzi}, M. and {Lawrence}, C.~R. and {Le Jeune}, M. and {Levrier}, F. and {Lewis}, A. and {Liguori}, M. and {Lilje}, P.~B. and {Lindholm}, V. and {L{\'o}pez-Caniego}, M. and {Ma}, Y. -Z. and {Mac{\'\i}as-P{\'e}rez}, J.~F. and {Maggio}, G. and {Maino}, D. and {Mandolesi}, N. and {Marcos-Caballero}, A. and {Maris}, M. and {Martin}, P.~G. and {Mart{\'\i}nez-Gonz{\'a}lez}, E. and {Matarrese}, S. and {Mauri}, N. and {McEwen}, J.~D. and {Meerburg}, P.~D. and {Meinhold}, P.~R. and {Melchiorri}, A. and {Mennella}, A. and {Migliaccio}, M. and {Miville-Desch{\^e}nes}, M. -A. and {Molinari}, D. and {Moneti}, A. and {Montier}, L. and {Morgante}, G. and {Moss}, A. and {M{\"u}nchmeyer}, M. and {Natoli}, P. and {Oppizzi}, F. and {Pagano}, L. and {Paoletti}, D. and {Partridge}, B. and {Patanchon}, G. and {Perrotta}, F. and {Pettorino}, V. and {Piacentini}, F. and {Polenta}, G. and {Puget}, J. -L. and {Rachen}, J.~P. and {Racine}, B. and {Reinecke}, M. and {Remazeilles}, M. and {Renzi}, A. and {Rocha}, G. and {Rubi{\~n}o-Mart{\'\i}n}, J.~A. and {Ruiz-Granados}, B. and {Salvati}, L. and {Savelainen}, M. and {Scott}, D. and {Shellard}, E.~P.~S. and {Shiraishi}, M. and {Sirignano}, C. and {Sirri}, G. and {Smith}, K. and {Spencer}, L.~D. and {Stanco}, L. and {Sunyaev}, R. and {Suur-Uski}, A. -S. and {Tauber}, J.~A. and {Tavagnacco}, D. and {Tenti}, M. and {Toffolatti}, L. and {Tomasi}, M. and {Trombetti}, T. and {Valiviita}, J. and {Van Tent}, B. and {Vielva}, P. and {Villa}, F. and {Vittorio}, N. and {Wandelt}, B.~D. and {Wehus}, I.~K. and {Zacchei}, A. and {Zonca}, A.},
 doi = {10.1051/0004-6361/201935891},
 eid = {A9},
 eprint = {1905.05697},
 journal = {\aap},
 month = {September},
 pages = {A9},
 primaryclass = {astro-ph.CO},
 title = {{Planck 2018 results. IX. Constraints on primordial non-Gaussianity}},
 volume = {641},
 year = {2020}
}

@article{1906.11765,
 adsurl = {https://ui.adsabs.harvard.edu/abs/2019JCAP...11..043G},
 archiveprefix = {arXiv},
 author = {{Garc{\'\i}a-Garc{\'\i}a}, Carlos and {Alonso}, David and {Bellini}, Emilio},
 doi = {10.1088/1475-7516/2019/11/043},
 eid = {043},
 eprint = {1906.11765},
 journal = {\jcap},
 month = {November},
 number = {11},
 pages = {043},
 primaryclass = {astro-ph.CO},
 title = {{Disconnected pseudo-C$_{l}$ covariances for projected large-scale structure data}},
 volume = {2019},
 year = {2019}
}

@article{1909.05444,
 adsurl = {https://ui.adsabs.harvard.edu/abs/2019PASP..131l4504M},
 archiveprefix = {arXiv},
 author = {{Meisner}, A.~M. and {Lang}, D. and {Schlafly}, E.~F. and {Schlegel}, D.~J.},
 doi = {10.1088/1538-3873/ab3df4},
 eprint = {1909.05444},
 journal = {\pasp},
 month = {December},
 number = {1006},
 pages = {124504},
 primaryclass = {astro-ph.IM},
 title = {{unWISE Coadds: The Five-year Data Set}},
 volume = {131},
 year = {2019}
}

@article{1911.11947,
 adsurl = {https://ui.adsabs.harvard.edu/abs/2020JCAP...05..010F},
 archiveprefix = {arXiv},
 author = {{Fang}, Xiao and {Krause}, Elisabeth and {Eifler}, Tim and {MacCrann}, Niall},
 doi = {10.1088/1475-7516/2020/05/010},
 eid = {010},
 eprint = {1911.11947},
 journal = {\jcap},
 month = {May},
 number = {5},
 pages = {010},
 primaryclass = {astro-ph.CO},
 title = {{Beyond Limber: efficient computation of angular power spectra for galaxy clustering and weak lensing}},
 volume = {2020},
 year = {2020}
}

@article{1953ApJ...117..134L,
 adsurl = {https://ui.adsabs.harvard.edu/abs/1953ApJ...117..134L},
 author = {{Limber}, D. Nelson},
 doi = {10.1086/145672},
 journal = {\apj},
 month = {January},
 pages = {134},
 title = {{The Analysis of Counts of the Extragalactic Nebulae in Terms of a Fluctuating Density Field.}},
 volume = {117},
 year = {1953}
}

@article{1974ApJ...187..425P,
 adsurl = {https://ui.adsabs.harvard.edu/abs/1974ApJ...187..425P},
 author = {{Press}, William H. and {Schechter}, Paul},
 doi = {10.1086/152650},
 journal = {\apj},
 month = {February},
 pages = {425-438},
 title = {{Formation of Galaxies and Clusters of Galaxies by Self-Similar Gravitational Condensation}},
 volume = {187},
 year = {1974}
}

@article{1992ApJ...398..169R,
 adsurl = {https://ui.adsabs.harvard.edu/abs/1992ApJ...398..169R},
 author = {{Rybicki}, George B. and {Press}, William H.},
 doi = {10.1086/171845},
 journal = {\apj},
 month = {October},
 pages = {169},
 title = {{Interpolation, Realization, and Reconstruction of Noisy, Irregularly Sampled Data}},
 volume = {398},
 year = {1992}
}

@article{2006.09368,
 adsurl = {https://ui.adsabs.harvard.edu/abs/2020JCAP...12..013B},
 archiveprefix = {arXiv},
 author = {{Barreira}, Alexandre and {Cabass}, Giovanni and {Schmidt}, Fabian and {Pillepich}, Annalisa and {Nelson}, Dylan},
 doi = {10.1088/1475-7516/2020/12/013},
 eid = {013},
 eprint = {2006.09368},
 journal = {\jcap},
 month = {December},
 number = {12},
 pages = {013},
 primaryclass = {astro-ph.CO},
 title = {{Galaxy bias and primordial non-Gaussianity: insights from galaxy formation simulations with IllustrisTNG}},
 volume = {2020},
 year = {2020}
}

@article{2007.09000,
 adsurl = {https://ui.adsabs.harvard.edu/abs/2020MNRAS.498.2354R},
 archiveprefix = {arXiv},
 author = {{Ross}, Ashley J. and {Bautista}, Julian and {Tojeiro}, Rita and {Alam}, Shadab and {Bailey}, Stephen and {Burtin}, Etienne and {Comparat}, Johan and {Dawson}, Kyle S. and {de Mattia}, Arnaud and {du Mas des Bourboux}, H{\'e}lion and {Gil-Mar{\'\i}n}, H{\'e}ctor and {Hou}, Jiamin and {Kong}, Hui and {Lyke}, Brad W. and {Mohammad}, Faizan G. and {Moustakas}, John and {Mueller}, Eva-Maria and {Myers}, Adam D. and {Percival}, Will J. and {Raichoor}, Anand and {Rezaie}, Mehdi and {Seo}, Hee-Jong and {Smith}, Alex and {Tinker}, Jeremy L. and {Zarrouk}, Pauline and {Zhao}, Cheng and {Zhao}, Gong-Bo and {Bizyaev}, Dmitry and {Brinkmann}, Jonathan and {Brownstein}, Joel R. and {Rosell}, Aurelio Carnero and {Chabanier}, Sol{\`e}ne and {Choi}, Peter D. and {Chuang}, Chia-Hsun and {Cruz-Gonzalez}, Irene and {de la Macorra}, Axel and {de la Torre}, Sylvain and {Escoffier}, Stephanie and {Fromenteau}, Sebastien and {Higley}, Alexandra and {Jullo}, Eric and {Kneib}, Jean-Paul and {McLane}, Jacob N. and {Mu{\~n}oz-Guti{\'e}rrez}, Andrea and {Neveux}, Richard and {Newman}, Jeffrey A. and {Nitschelm}, Christian and {Palanque-Delabrouille}, Nathalie and {Paviot}, Romain and {Pullen}, Anthony R. and {Rossi}, Graziano and {Ruhlmann-Kleider}, Vanina and {Schneider}, Donald P. and {Maga{\~n}a}, Mariana Vargas and {Vivek}, M. and {Zhang}, Yucheng},
 doi = {10.1093/mnras/staa2416},
 eprint = {2007.09000},
 journal = {\mnras},
 month = {October},
 number = {2},
 pages = {2354-2371},
 primaryclass = {astro-ph.CO},
 title = {{The Completed SDSS-IV extended Baryon Oscillation Spectroscopic Survey: Large-scale structure catalogues for cosmological analysis}},
 volume = {498},
 year = {2020}
}

@article{2010.14523,
 adsurl = {https://ui.adsabs.harvard.edu/abs/2021JCAP...05..015M},
 archiveprefix = {arXiv},
 author = {{Moradinezhad Dizgah}, Azadeh and {Biagetti}, Matteo and {Sefusatti}, Emiliano and {Desjacques}, Vincent and {Nore{\~n}a}, Jorge},
 doi = {10.1088/1475-7516/2021/05/015},
 eid = {015},
 eprint = {2010.14523},
 journal = {\jcap},
 month = {May},
 number = {5},
 pages = {015},
 primaryclass = {astro-ph.CO},
 title = {{Primordial non-Gaussianity from biased tracers: likelihood analysis of real-space power spectrum and bispectrum}},
 volume = {2021},
 year = {2021}
}

@article{2106.13725,
 adsurl = {https://ui.adsabs.harvard.edu/abs/2021arXiv210613725M},
 archiveprefix = {arXiv},
 author = {{Mueller}, Eva-Maria and {Rezaie}, Mehdi and {Percival}, Will J. and {Ross}, Ashley J. and {Ruggeri}, Rossana and {Seo}, Hee-Jong and {Gil-Mar{\i}n}, Hector and {Bautista}, Julian and {Brownstein}, Joel R. and {Dawson}, Kyle and {de la Macorra}, Axel and {Palanque-Delabrouille}, Nathalie and {Rossi}, Graziano and {Schneider}, Donald P. and {Yeche}, Christophe},
 doi = {10.48550/arXiv.2106.13725},
 eid = {arXiv:2106.13725},
 eprint = {2106.13725},
 journal = {arXiv e-prints},
 month = {June},
 pages = {arXiv:2106.13725},
 primaryclass = {astro-ph.CO},
 title = {{The clustering of galaxies in the completed SDSS-IV extended Baryon Oscillation Spectroscopic Survey: Primordial non-Gaussianity in Fourier Space}},
 year = {2021}
}

@article{2107.06887,
 adsurl = {https://ui.adsabs.harvard.edu/abs/2022JCAP...01..033B},
 archiveprefix = {arXiv},
 author = {{Barreira}, Alexandre},
 doi = {10.1088/1475-7516/2022/01/033},
 eid = {033},
 eprint = {2107.06887},
 journal = {\jcap},
 month = {January},
 number = {1},
 pages = {033},
 primaryclass = {astro-ph.CO},
 title = {{Predictions for local PNG bias in the galaxy power spectrum and bispectrum and the consequences for f $_{NL}$ constraints}},
 volume = {2022},
 year = {2022}
}

@article{2201.11518,
 adsurl = {https://ui.adsabs.harvard.edu/abs/2025PhRvD.111f3514D},
 archiveprefix = {arXiv},
 author = {{D'Amico}, Guido and {Lewandowski}, Matthew and {Senatore}, Leonardo and {Zhang}, Pierre},
 doi = {10.1103/PhysRevD.111.063514},
 eid = {063514},
 eprint = {2201.11518},
 journal = {\prd},
 month = {March},
 number = {6},
 pages = {063514},
 primaryclass = {astro-ph.CO},
 title = {{Limits on primordial non-Gaussianities from BOSS galaxy-clustering data}},
 volume = {111},
 year = {2025}
}

@article{2204.01781,
 adsurl = {https://ui.adsabs.harvard.edu/abs/2022PhRvD.106d3506C},
 archiveprefix = {arXiv},
 author = {{Cabass}, Giovanni and {Ivanov}, Mikhail M. and {Philcox}, Oliver H.~E. and {Simonovi{\'c}}, Marko and {Zaldarriaga}, Matias},
 doi = {10.1103/PhysRevD.106.043506},
 eid = {043506},
 eprint = {2204.01781},
 journal = {\prd},
 month = {August},
 number = {4},
 pages = {043506},
 primaryclass = {astro-ph.CO},
 title = {{Constraints on multifield inflation from the BOSS galaxy survey}},
 volume = {106},
 year = {2022}
}

@article{2206.07773,
 adsurl = {https://ui.adsabs.harvard.edu/abs/2022JCAP...09..039C},
 archiveprefix = {arXiv},
 author = {{Carron}, Julien and {Mirmelstein}, Mark and {Lewis}, Antony},
 doi = {10.1088/1475-7516/2022/09/039},
 eid = {039},
 eprint = {2206.07773},
 journal = {\jcap},
 month = {September},
 number = {9},
 pages = {039},
 primaryclass = {astro-ph.CO},
 title = {{CMB lensing from Planck PR4 maps}},
 volume = {2022},
 year = {2022}
}

@article{2208.00211,
 adsurl = {https://ui.adsabs.harvard.edu/abs/2023A&A...674A...1G},
 archiveprefix = {arXiv},
 author = {{Gaia Collaboration} and {Vallenari}, A. and {Brown}, A.~G.~A. and {Prusti}, T. and {de Bruijne}, J.~H.~J. and {Arenou}, F. and {Babusiaux}, C. and {Biermann}, M. and {Creevey}, O.~L. and {Ducourant}, C. and {Evans}, D.~W. and {Eyer}, L. and {Guerra}, R. and {Hutton}, A. and {Jordi}, C. and {Klioner}, S.~A. and {Lammers}, U.~L. and {Lindegren}, L. and {Luri}, X. and {Mignard}, F. and {Panem}, C. and {Pourbaix}, D. and {Randich}, S. and {Sartoretti}, P. and {Soubiran}, C. and {Tanga}, P. and {Walton}, N.~A. and {Bailer-Jones}, C.~A.~L. and {Bastian}, U. and {Drimmel}, R. and {Jansen}, F. and {Katz}, D. and {Lattanzi}, M.~G. and {van Leeuwen}, F. and {Bakker}, J. and {Cacciari}, C. and {Casta{\~n}eda}, J. and {De Angeli}, F. and {Fabricius}, C. and {Fouesneau}, M. and {Fr{\'e}mat}, Y. and {Galluccio}, L. and {Guerrier}, A. and {Heiter}, U. and {Masana}, E. and {Messineo}, R. and {Mowlavi}, N. and {Nicolas}, C. and {Nienartowicz}, K. and {Pailler}, F. and {Panuzzo}, P. and {Riclet}, F. and {Roux}, W. and {Seabroke}, G.~M. and {Sordo}, R. and {Th{\'e}venin}, F. and {Gracia-Abril}, G. and {Portell}, J. and {Teyssier}, D. and {Altmann}, M. and {Andrae}, R. and {Audard}, M. and {Bellas-Velidis}, I. and {Benson}, K. and {Berthier}, J. and {Blomme}, R. and {Burgess}, P.~W. and {Busonero}, D. and {Busso}, G. and {C{\'a}novas}, H. and {Carry}, B. and {Cellino}, A. and {Cheek}, N. and {Clementini}, G. and {Damerdji}, Y. and {Davidson}, M. and {de Teodoro}, P. and {Nu{\~n}ez Campos}, M. and {Delchambre}, L. and {Dell'Oro}, A. and {Esquej}, P. and {Fern{\'a}ndez-Hern{\'a}ndez}, J. and {Fraile}, E. and {Garabato}, D. and {Garc{\'\i}a-Lario}, P. and {Gosset}, E. and {Haigron}, R. and {Halbwachs}, J. -L. and {Hambly}, N.~C. and {Harrison}, D.~L. and {Hern{\'a}ndez}, J. and {Hestroffer}, D. and {Hodgkin}, S.~T. and {Holl}, B. and {Jan{\ss}en}, K. and {Jevardat de Fombelle}, G. and {Jordan}, S. and {Krone-Martins}, A. and {Lanzafame}, A.~C. and {L{\"o}ffler}, W. and {Marchal}, O. and {Marrese}, P.~M. and {Moitinho}, A. and {Muinonen}, K. and {Osborne}, P. and {Pancino}, E. and {Pauwels}, T. and {Recio-Blanco}, A. and {Reyl{\'e}}, C. and {Riello}, M. and {Rimoldini}, L. and {Roegiers}, T. and {Rybizki}, J. and {Sarro}, L.~M. and {Siopis}, C. and {Smith}, M. and {Sozzetti}, A. and {Utrilla}, E. and {van Leeuwen}, M. and {Abbas}, U. and {{\'A}brah{\'a}m}, P. and {Abreu Aramburu}, A. and {Aerts}, C. and {Aguado}, J.~J. and {Ajaj}, M. and {Aldea-Montero}, F. and {Altavilla}, G. and {{\'A}lvarez}, M.~A. and {Alves}, J. and {Anders}, F. and {Anderson}, R.~I. and {Anglada Varela}, E. and {Antoja}, T. and {Baines}, D. and {Baker}, S.~G. and {Balaguer-N{\'u}{\~n}ez}, L. and {Balbinot}, E. and {Balog}, Z. and {Barache}, C. and {Barbato}, D. and {Barros}, M. and {Barstow}, M.~A. and {Bartolom{\'e}}, S. and {Bassilana}, J. -L. and {Bauchet}, N. and {Becciani}, U. and {Bellazzini}, M. and {Berihuete}, A. and {Bernet}, M. and {Bertone}, S. and {Bianchi}, L. and {Binnenfeld}, A. and {Blanco-Cuaresma}, S. and {Blazere}, A. and {Boch}, T. and {Bombrun}, A. and {Bossini}, D. and {Bouquillon}, S. and {Bragaglia}, A. and {Bramante}, L. and {Breedt}, E. and {Bressan}, A. and {Brouillet}, N. and {Brugaletta}, E. and {Bucciarelli}, B. and {Burlacu}, A. and {Butkevich}, A.~G. and {Buzzi}, R. and {Caffau}, E. and {Cancelliere}, R. and {Cantat-Gaudin}, T. and {Carballo}, R. and {Carlucci}, T. and {Carnerero}, M.~I. and {Carrasco}, J.~M. and {Casamiquela}, L. and {Castellani}, M. and {Castro-Ginard}, A. and {Chaoul}, L. and {Charlot}, P. and {Chemin}, L. and {Chiaramida}, V. and {Chiavassa}, A. and {Chornay}, N. and {Comoretto}, G. and {Contursi}, G. and {Cooper}, W.~J. and {Cornez}, T. and {Cowell}, S. and {Crifo}, F. and {Cropper}, M. and {Crosta}, M. and {Crowley}, C. and {Dafonte}, C. and {Dapergolas}, A. and {David}, M. and {David}, P. and {de Laverny}, P. and {De Luise}, F. and {De March}, R.},
 doi = {10.1051/0004-6361/202243940},
 eid = {A1},
 eprint = {2208.00211},
 journal = {\aap},
 month = {June},
 pages = {A1},
 primaryclass = {astro-ph.GA},
 title = {{Gaia Data Release 3. Summary of the content and survey properties}},
 volume = {674},
 year = {2023}
}

@article{2210.01049,
 adsurl = {https://ui.adsabs.harvard.edu/abs/2023PhRvD.108h3522M},
 archiveprefix = {arXiv},
 author = {{McCarthy}, Fiona and {Madhavacheril}, Mathew S. and {Maniyar}, Abhishek S.},
 doi = {10.1103/PhysRevD.108.083522},
 eid = {083522},
 eprint = {2210.01049},
 journal = {\prd},
 month = {October},
 number = {8},
 pages = {083522},
 primaryclass = {astro-ph.CO},
 title = {{Constraints on primordial non-Gaussianity from halo bias measured through CMB lensing cross-correlations}},
 volume = {108},
 year = {2023}
}

@article{2212.04291,
 adsurl = {https://ui.adsabs.harvard.edu/abs/2023OJAp....6E...8L},
 archiveprefix = {arXiv},
 author = {{Leonard}, C. Danielle and {Ferreira}, Tassia and {Fang}, Xiao and {Reischke}, Robert and {Schoeneberg}, Nils and {Tr{\"o}ster}, Tilman and {Alonso}, David and {Campagne}, Jean-Eric and {Lanusse}, Fran{\c{c}}ois and {Slosar}, An{\v{z}}e and {Ishak}, Mustapha},
 doi = {10.21105/astro.2212.04291},
 eid = {8},
 eprint = {2212.04291},
 journal = {The Open Journal of Astrophysics},
 month = {February},
 pages = {8},
 primaryclass = {astro-ph.CO},
 title = {{The N5K Challenge: Non-Limber Integration for LSST Cosmology}},
 volume = {6},
 year = {2023}
}

@article{2305.07650,
 adsurl = {https://ui.adsabs.harvard.edu/abs/2024JCAP...03..021K},
 archiveprefix = {arXiv},
 author = {{Krolewski}, Alex and {Percival}, Will J. and {Ferraro}, Simone and {Chaussidon}, Edmond and {Rezaie}, Mehdi and {Aguilar}, Jessica Nicole and {Ahlen}, Steven and {Brooks}, David and {Dawson}, Kyle and {de la Macorra}, Axel and {Doel}, Peter and {Fanning}, Kevin and {Font-Ribera}, Andreu and {a Gontcho}, Satya Gontcho and {Guy}, Julien and {Honscheid}, Klaus and {Kehoe}, Robert and {Kisner}, Theodore and {Kremin}, Anthony and {Landriau}, Martin and {Levi}, Michael E. and {Martini}, Paul and {Meisner}, Aaron M. and {Miquel}, Ramon and {Nie}, Jundan and {Poppett}, Claire and {Ross}, Ashley J. and {Rossi}, Graziano and {Schubnell}, Michael and {Seo}, Hee-Jong and {Tarl{\'e}}, Gregory and {Vargas-Maga{\~n}a}, Mariana and {Weaver}, Benjamin Alan and {Y{\`e}che}, Christophe and {Zhou}, Rongpu and {Zhou}, Zhimin},
 doi = {10.1088/1475-7516/2024/03/021},
 eid = {021},
 eprint = {2305.07650},
 journal = {\jcap},
 month = {March},
 number = {3},
 pages = {021},
 primaryclass = {astro-ph.CO},
 title = {{Constraining primordial non-Gaussianity from DESI quasar targets and Planck CMB lensing}},
 volume = {2024},
 year = {2024}
}

@article{2306.03926,
 adsurl = {https://ui.adsabs.harvard.edu/abs/2023ApJ...958..118C},
 archiveprefix = {arXiv},
 author = {{Chiang}, Yi-Kuan},
 doi = {10.3847/1538-4357/acf4a1},
 eid = {118},
 eprint = {2306.03926},
 journal = {\apj},
 month = {December},
 number = {2},
 pages = {118},
 primaryclass = {astro-ph.GA},
 title = {{Corrected SFD: A More Accurate Galactic Dust Map with Minimal Extragalactic Contamination}},
 volume = {958},
 year = {2023}
}

@article{2306.17748,
 adsurl = {https://ui.adsabs.harvard.edu/abs/2023JCAP...11..043A},
 archiveprefix = {arXiv},
 author = {{Alonso}, David and {Fabbian}, Giulio and {Storey-Fisher}, Kate and {Eilers}, Anna-Christina and {Garc{\'\i}a-Garc{\'\i}a}, Carlos and {Hogg}, David W. and {Rix}, Hans-Walter},
 doi = {10.1088/1475-7516/2023/11/043},
 eid = {043},
 eprint = {2306.17748},
 journal = {\jcap},
 month = {November},
 number = {11},
 pages = {043},
 primaryclass = {astro-ph.CO},
 title = {{Constraining cosmology with the Gaia-unWISE Quasar Catalog and CMB lensing: structure growth}},
 volume = {2023},
 year = {2023}
}

@article{2306.17749,
 adsurl = {https://ui.adsabs.harvard.edu/abs/2024ApJ...964...69S},
 archiveprefix = {arXiv},
 author = {{Storey-Fisher}, Kate and {Hogg}, David W. and {Rix}, Hans-Walter and {Eilers}, Anna-Christina and {Fabbian}, Giulio and {Blanton}, Michael R. and {Alonso}, David},
 doi = {10.3847/1538-4357/ad1328},
 eid = {69},
 eprint = {2306.17749},
 journal = {\apj},
 month = {March},
 number = {1},
 pages = {69},
 primaryclass = {astro-ph.GA},
 title = {{Quaia, the Gaia-unWISE Quasar Catalog: An All-sky Spectroscopic Quasar Sample}},
 volume = {964},
 year = {2024}
}

@article{2309.15814,
 adsurl = {https://ui.adsabs.harvard.edu/abs/2024JCAP...08..036C},
 archiveprefix = {arXiv},
 author = {{Cagliari}, Marina S. and {Castorina}, Emanuele and {Bonici}, Marco and {Bianchi}, Davide},
 doi = {10.1088/1475-7516/2024/08/036},
 eid = {036},
 eprint = {2309.15814},
 journal = {\jcap},
 month = {August},
 number = {8},
 pages = {036},
 primaryclass = {astro-ph.CO},
 title = {{Optimal constraints on Primordial non-Gaussianity with the eBOSS DR16 quasars in Fourier space}},
 volume = {2024},
 year = {2024}
}

@article{2402.05761,
 adsurl = {https://ui.adsabs.harvard.edu/abs/2024JCAP...06..012P},
 archiveprefix = {arXiv},
 author = {{Piccirilli}, Giulia and {Fabbian}, Giulio and {Alonso}, David and {Storey-Fisher}, Kate and {Carron}, Julien and {Lewis}, Antony and {Garc{\'\i}a-Garc{\'\i}a}, Carlos},
 doi = {10.1088/1475-7516/2024/06/012},
 eid = {012},
 eprint = {2402.05761},
 journal = {\jcap},
 month = {June},
 number = {6},
 pages = {012},
 primaryclass = {astro-ph.CO},
 title = {{Growth history and quasar bias evolution at z < 3 from Quaia}},
 volume = {2024},
 year = {2024}
}

@article{2409.07980,
 adsurl = {https://ui.adsabs.harvard.edu/abs/2025OJAp....8E...6H},
 archiveprefix = {arXiv},
 author = {{Harscouet}, Lea and {Cowell}, Jessica A. and {Ereza}, Julia and {Alonso}, David and {Camacho}, Hugo and {Nicola}, Andrina and {Slosar}, An{\v{z}}e},
 doi = {10.33232/001c.128309},
 eid = {6},
 eprint = {2409.07980},
 journal = {The Open Journal of Astrophysics},
 month = {January},
 pages = {6},
 primaryclass = {astro-ph.CO},
 title = {{Fast Projected Bispectra: the filter-square approach}},
 volume = {8},
 year = {2025}
}

@article{2410.24134,
       author = {{Alonso}, David and {Hetmantsev}, Oleksandr and {Fabbian}, Giulio and {Slosar}, Anze and {Storey-Fisher}, Kate},
        title = "{Measurement of the power spectrum turnover scale from the cross-correlation between CMB lensing and Quaia}",
      journal = {The Open Journal of Astrophysics},
         year = 2025,
        month = apr,
       volume = {8},
          eid = {42},
        pages = {42},
          doi = {10.33232/001c.136891},
archivePrefix = {arXiv},
       eprint = {2410.24134},
 primaryClass = {astro-ph.CO},
       adsurl = {https://ui.adsabs.harvard.edu/abs/2025OJAp....8E..42A}
}

@article{2411.17623,
       author = {{Chaussidon}, E. and {Y{\`e}che}, C. and {de Mattia}, A. and {Payerne}, C. and {McDonald}, P. and {Ross}, A.~J. and {Ahlen}, S. and {Bianchi}, D. and {Brooks}, D. and {Burtin}, E. and {Claybaugh}, T. and {de la Macorra}, A. and {Doel}, P. and {Ferraro}, S. and {Font-Ribera}, A. and {Forero-Romero}, J.~E. and {Gazta{\~n}aga}, E. and {Gil-Mar{\'\i}n}, H. and {Gontcho}, S. Gontcho A. and {Gutierrez}, G. and {Guy}, J. and {Honscheid}, K. and {Howlett}, C. and {Huterer}, D. and {Kehoe}, R. and {Kirkby}, D. and {Kisner}, T. and {Kremin}, A. and {Le Guillou}, L. and {Levi}, M.~E. and {Manera}, M. and {Meisner}, A. and {Miquel}, R. and {Moustakas}, J. and {Newman}, J.~A. and {Niz}, G. and {Palanque-Delabrouille}, N. and {Percival}, W.~J. and {Prada}, F. and {P{\'e}rez-R{\`a}fols}, I. and {Ravoux}, C. and {Rossi}, G. and {Sanchez}, E. and {Schlegel}, D. and {Schubnell}, M. and {Seo}, H. and {Sprayberry}, D. and {Tarl{\'e}}, G. and {Vargas-Maga{\~n}a}, M. and {Weaver}, B.~A. and {Zhao}, C. and {Zou}, H.},
        title = "{Constraining primordial non-Gaussianity with DESI 2024 LRG and QSO samples}",
      journal = {\jcap},
         year = 2025,
        month = jun,
       volume = {2025},
       number = {6},
          eid = {029},
        pages = {029},
          doi = {10.1088/1475-7516/2025/06/029},
archivePrefix = {arXiv},
       eprint = {2411.17623},
 primaryClass = {astro-ph.CO},
       adsurl = {https://ui.adsabs.harvard.edu/abs/2025JCAP...06..029C}
}

@article{2412.06553,
       author = {{Guedezounme}, S{\^e}cloka L. and {Jolicoeur}, Sheean and {Maartens}, Roy},
        title = "{Primordial non-Gaussianity {\textemdash} the effects of relativistic and wide-angle corrections to the power spectrum}",
      journal = {\jcap},
         year = 2025,
        month = jul,
       volume = {2025},
       number = {7},
          eid = {063},
        pages = {063},
          doi = {10.1088/1475-7516/2025/07/063},
archivePrefix = {arXiv},
       eprint = {2412.06553},
 primaryClass = {astro-ph.CO},
       adsurl = {https://ui.adsabs.harvard.edu/abs/2025JCAP...07..063G}
}

@article{2412.10279,
       author = {{Bermejo-Climent}, J.~R. and {Demina}, R. and {Krolewski}, A. and {Chaussidon}, E. and {Rezaie}, M. and {Ahlen}, S. and {Bailey}, S. and {Bianchi}, D. and {Brooks}, D. and {Burtin}, E. and {Claybaugh}, T. and {de la Macorra}, A. and {Dey}, A. and {Doel}, P. and {Farren}, G. and {Ferraro}, S. and {Forero-Romero}, J.~E. and {Gazta{\~n}aga}, E. and {Gontcho A Gontcho}, S. and {Gutierrez}, G. and {Hahn}, C. and {Honscheid}, K. and {Howlett}, C. and {Kehoe}, R. and {Kirkby}, D. and {Kisner}, T. and {Landriau}, M. and {Le Guillou}, L. and {Levi}, M.~E. and {Manera}, M. and {Meisner}, A. and {Miquel}, R. and {Moustakas}, J. and {Newman}, J.~A. and {Niz}, G. and {Palanque-Delabrouille}, N. and {Percival}, W.~J. and {Prada}, F. and {P{\'e}rez-R{\`a}fols}, I. and {Rabinowitz}, D. and {Ross}, A.~J. and {Rossi}, G. and {Sanchez}, E. and {Schlegel}, D. and {Sprayberry}, D. and {Tarl{\'e}}, G. and {Weaver}, B.~A. and {White}, M. and {Y{\`e}che}, C. and {Zarrouk}, P.},
        title = "{Constraints on primordial non-Gaussianity from the cross-correlation of DESI luminous red galaxies and Planck CMB lensing}",
      journal = {\aap},
         year = 2025,
        month = jun,
       volume = {698},
          eid = {A177},
        pages = {A177},
          doi = {10.1051/0004-6361/202453446},
archivePrefix = {arXiv},
       eprint = {2412.10279},
 primaryClass = {astro-ph.CO},
       adsurl = {https://ui.adsabs.harvard.edu/abs/2025A&A...698A.177B}
}

@article{2502.14758,
      author = {{Cagliari}, Marina S. and {Barberi-Squarotti}, Matilde and {Pardede}, Kevin and {Castorina}, Emanuele and {D'Amico}, Guido},
        title = "{Bispectrum constraints on Primordial Non-Gaussianities with the eBOSS DR16 quasars}",
      journal = {\jcap},
         year = 2025,
        month = jul,
       volume = {2025},
       number = {7},
          eid = {043},
        pages = {043},
          doi = {10.1088/1475-7516/2025/07/043},
archivePrefix = {arXiv},
       eprint = {2502.14758},
 primaryClass = {astro-ph.CO},
       adsurl = {https://ui.adsabs.harvard.edu/abs/2025JCAP...07..043C}
}

@article{2504.00884,
 adsurl = {https://ui.adsabs.harvard.edu/abs/2025arXiv250400884J},
 archiveprefix = {arXiv},
 author = {{Jung}, Gabriel and {Citran}, Michele and {van Tent}, Bartjan and {Dumilly}, L{\'e}a and {Aghanim}, Nabila},
 doi = {10.48550/arXiv.2504.00884},
 eid = {arXiv:2504.00884},
 eprint = {2504.00884},
 journal = {arXiv e-prints},
 month = {April},
 pages = {arXiv:2504.00884},
 primaryclass = {astro-ph.CO},
 title = {{Constraints on primordial non-Gaussianity from Planck PR4 data}},
 year = {2025}
}

@article{astro-ph/0105302,
 adsurl = {https://ui.adsabs.harvard.edu/abs/2002ApJ...567....2H},
 archiveprefix = {arXiv},
 author = {{Hivon}, Eric and {G{\'o}rski}, Krzysztof M. and {Netterfield}, C. Barth and {Crill}, Brendan P. and {Prunet}, Simon and {Hansen}, Frode},
 doi = {10.1086/338126},
 eprint = {astro-ph/0105302},
 journal = {\apj},
 month = {March},
 number = {1},
 pages = {2-17},
 primaryclass = {astro-ph},
 title = {{MASTER of the Cosmic Microwave Background Anisotropy Power Spectrum: A Fast Method for Statistical Analysis of Large and Complex Cosmic Microwave Background Data Sets}},
 volume = {567},
 year = {2002}
}

@article{astro-ph/0406398,
 adsurl = {https://ui.adsabs.harvard.edu/abs/2004PhR...402..103B},
 archiveprefix = {arXiv},
 author = {{Bartolo}, N. and {Komatsu}, E. and {Matarrese}, S. and {Riotto}, A.},
 doi = {10.1016/j.physrep.2004.08.022},
 eprint = {astro-ph/0406398},
 journal = {\physrep},
 month = {November},
 number = {3-4},
 pages = {103-266},
 primaryclass = {astro-ph},
 title = {{Non-Gaussianity from inflation: theory and observations}},
 volume = {402},
 year = {2004}
}

@article{astro-ph/0407059,
 adsurl = {https://ui.adsabs.harvard.edu/abs/2004JCAP...10..006C},
 archiveprefix = {arXiv},
 author = {{Creminelli}, Paolo and {Zaldarriaga}, Matias},
 doi = {10.1088/1475-7516/2004/10/006},
 eid = {006},
 eprint = {astro-ph/0407059},
 journal = {\jcap},
 month = {October},
 number = {10},
 pages = {006},
 primaryclass = {astro-ph},
 title = {{A single-field consistency relation for the three-point function}},
 volume = {2004},
 year = {2004}
}

@article{astro-ph/0608064,
 adsurl = {https://ui.adsabs.harvard.edu/abs/2007A&A...464..399H},
 archiveprefix = {arXiv},
 author = {{Hartlap}, J. and {Simon}, P. and {Schneider}, P.},
 doi = {10.1051/0004-6361:20066170},
 eprint = {astro-ph/0608064},
 journal = {\aap},
 month = {March},
 number = {1},
 pages = {399-404},
 primaryclass = {astro-ph},
 title = {{Why your model parameter confidences might be too optimistic. Unbiased estimation of the inverse covariance matrix}},
 volume = {464},
 year = {2007}
}

@article{astro-ph/9611174,
 adsurl = {https://ui.adsabs.harvard.edu/abs/1997PhRvD..55.5895T},
 archiveprefix = {arXiv},
 author = {{Tegmark}, Max},
 doi = {10.1103/PhysRevD.55.5895},
 eprint = {astro-ph/9611174},
 journal = {\prd},
 month = {May},
 number = {10},
 pages = {5895-5907},
 primaryclass = {astro-ph},
 title = {{How to measure CMB power spectra without losing information}},
 volume = {55},
 year = {1997}
}

@article{astro-ph/9706198,
 adsurl = {https://ui.adsabs.harvard.edu/abs/1997PhRvL..79.3806T},
 archiveprefix = {arXiv},
 author = {{Tegmark}, Max},
 doi = {10.1103/PhysRevLett.79.3806},
 eprint = {astro-ph/9706198},
 journal = {\prl},
 month = {November},
 number = {20},
 pages = {3806-3809},
 primaryclass = {astro-ph},
 title = {{Measuring Cosmological Parameters with Galaxy Surveys}},
 volume = {79},
 year = {1997}
}

@ARTICLE{1910.02608,
       author = {{Polarbear Collaboration} and {Adachi}, S. and {Aguilar Fa{\'u}ndez}, M.~A.~O. and {Arnold}, K. and {Baccigalupi}, C. and {Barron}, D. and {Beck}, D. and {Beckman}, S. and {Bianchini}, F. and {Boettger}, D. and {Borrill}, J. and {Carron}, J. and {Chapman}, S. and {Cheung}, K. and {Chinone}, Y. and {Crowley}, K. and {Cukierman}, A. and {Dobbs}, M. and {El Bouhargani}, H. and {Elleflot}, T. and {Errard}, J. and {Fabbian}, G. and {Feng}, C. and {Fujino}, T. and {Galitzki}, N. and {Goeckner-Wald}, N. and {Groh}, J. and {Hall}, G. and {Halverson}, N. and {Hamada}, T. and {Hasegawa}, M. and {Hazumi}, M. and {Hill}, C.~A. and {Howe}, L. and {Inoue}, Y. and {Jaehnig}, G. and {Jeong}, O. and {Kaneko}, D. and {Katayama}, N. and {Keating}, B. and {Keskitalo}, R. and {Kikuchi}, S. and {Kisner}, T. and {Krachmalnicoff}, N. and {Kusaka}, A. and {Lee}, A.~T. and {Leon}, D. and {Linder}, E. and {Lowry}, L.~N. and {Mangu}, A. and {Matsuda}, F. and {Minami}, Y. and {Navaroli}, M. and {Nishino}, H. and {Pham}, A.~T.~P. and {Poletti}, D. and {Puglisi}, G. and {Reichardt}, C.~L. and {Segawa}, Y. and {Silva-Feaver}, M. and {Siritanasak}, P. and {Stebor}, N. and {Stompor}, R. and {Suzuki}, A. and {Tajima}, O. and {Takakura}, S. and {Takatori}, S. and {Tanabe}, D. and {Teply}, G.~P. and {Tsai}, C. and {Verges}, C. and {Westbrook}, B. and {Zhou}, Y.},
        title = "{A Measurement of the Degree-scale CMB B-mode Angular Power Spectrum with POLARBEAR}",
      journal = {\apj},
         year = 2020,
        month = jul,
       volume = {897},
       number = {1},
          eid = {55},
        pages = {55},
          doi = {10.3847/1538-4357/ab8f24},
archivePrefix = {arXiv},
       eprint = {1910.02608},
 primaryClass = {astro-ph.CO},
       adsurl = {https://ui.adsabs.harvard.edu/abs/2020ApJ...897...55P}
}

@ARTICLE{2502.00946,
       author = {{Herv{\'\i}as-Caimapo}, Carlos and {Wolz}, Kevin and {La Posta}, Adrien and {Azzoni}, Susanna and {Alonso}, David and {Arnold}, Kam and {Baccigalupi}, Carlo and {Biquard}, Simon and {Brown}, Michael L. and {Calabrese}, Erminia and {Chinone}, Yuji and {Day-Weiss}, Samuel and {Dunkley}, Jo and {D{\"u}nner}, Rolando and {Errard}, Josquin and {Fabbian}, Giulio and {Ganga}, Ken and {Giardiello}, Serena and {Hertig}, Emilie and {Huffenberger}, Kevin M. and {Johnson}, Bradley R. and {Jost}, Baptiste and {Keskitalo}, Reijo and {Kisner}, Theodore S. and {Louis}, Thibaut and {Morshed}, Magdy and {Page}, Lyman A. and {Reichardt}, Christian L. and {Rosenberg}, Erik and {Silva-Feaver}, Max and {Sohn}, Wuhyun and {Sueno}, Yoshinori and {Thomas}, Dan B. and {Tsang King Sang}, Ema and {Villarrubia-Aguilar}, Amalia and {Yamada}, Kyohei},
        title = "{The Simons Observatory: validation of reconstructed power spectra from simulated filtered maps for the small aperture telescope survey}",
      journal = {\jcap},
         year = 2025,
        month = jun,
       volume = {2025},
       number = {6},
          eid = {055},
        pages = {055},
          doi = {10.1088/1475-7516/2025/06/055},
archivePrefix = {arXiv},
       eprint = {2502.00946},
 primaryClass = {astro-ph.CO},
       adsurl = {https://ui.adsabs.harvard.edu/abs/2025JCAP...06..055H}
}

@ARTICLE{2503.14452,
       author = {{Louis}, Thibaut and {La Posta}, Adrien and {Atkins}, Zachary and {Jense}, Hidde T. and {Abril-Cabezas}, Irene and {Addison}, Graeme E. and {Ade}, Peter A.~R. and {Aiola}, Simone and {Alford}, Tommy and {Alonso}, David and {Amiri}, Mandana and {An}, Rui and {Austermann}, Jason E. and {Barbavara}, Eleonora and {Battaglia}, Nicholas and {Battistelli}, Elia Stefano and {Beall}, James A. and {Bean}, Rachel and {Beheshti}, Ali and {Beringue}, Benjamin and {Bhandarkar}, Tanay and {Biermann}, Emily and {Bolliet}, Boris and {Bond}, J Richard and {Calabrese}, Erminia and {Capalbo}, Valentina and {Carrero}, Felipe and {Chen}, Stephen and {Chesmore}, Grace and {Cho}, Hsiao-mei and {Choi}, Steve K. and {Clark}, Susan E. and {Cothard}, Nicholas F. and {Coughlin}, Kevin and {Coulton}, William and {Crichton}, Devin and {Crowley}, Kevin T. and {Darwish}, Omar and {Devlin}, Mark J. and {Dicker}, Simon and {Duell}, Cody J. and {Duff}, Shannon M. and {Duivenvoorden}, Adriaan J. and {Dunkley}, Jo and {Dunner}, Rolando and {Embil Villagra}, Carmen and {Fankhanel}, Max and {Farren}, Gerrit S. and {Ferraro}, Simone and {Foster}, Allen and {Freundt}, Rodrigo and {Fuzia}, Brittany and {Gallardo}, Patricio A. and {Garrido}, Xavier and {Gerbino}, Martina and {Giardiello}, Serena and {Gill}, Ajay and {Givans}, Jahmour and {Gluscevic}, Vera and {Goldstein}, Samuel and {Golec}, Joseph E. and {Gong}, Yulin and {Guan}, Yilun and {Halpern}, Mark and {Harrison}, Ian and {Hasselfield}, Matthew and {Healy}, Erin and {Henderson}, Shawn and {Hensley}, Brandon and {Herv{\'\i}as-Caimapo}, Carlos and {Hill}, J. Colin and {Hilton}, Gene C. and {Hilton}, Matt and {Hincks}, Adam D. and {Hlo{\v{z}}ek}, Ren{\'e}e and {Ho}, Shuay-Pwu Patty and {Hood}, John and {Hornecker}, Erika and {Huber}, Zachary B. and {Hubmayr}, Johannes and {Huffenberger}, Kevin M. and {Hughes}, John P. and {Ikape}, Margaret and {Irwin}, Kent and {Isopi}, Giovanni and {Joshi}, Neha and {Keller}, Ben and {Kim}, Joshua and {Knowles}, Kenda and {Koopman}, Brian J. and {Kosowsky}, Arthur and {Kramer}, Darby and {Kusiak}, Aleksandra and {Lague}, Alex and {Lakey}, Victoria and {Lee}, Eunseong and {Li}, Yaqiong and {Li}, Zack and {Limon}, Michele and {Lokken}, Martine and {Lungu}, Marius and {MacCrann}, Niall and {MacInnis}, Amanda and {Madhavacheril}, Mathew S. and {Maldonado}, Diego and {Maldonado}, Felipe and {Mallaby-Kay}, Maya and {Marques}, Gabriela A. and {van Marrewijk}, Joshiwa and {McCarthy}, Fiona and {McMahon}, Jeff and {Mehta}, Yogesh and {Menanteau}, Felipe and {Moodley}, Kavilan and {Morris}, Thomas W. and {Mroczkowski}, Tony and {Naess}, Sigurd and {Namikawa}, Toshiya and {Nati}, Federico and {Nerval}, Simran K. and {Newburgh}, Laura and {Nicola}, Andrina and {Niemack}, Michael D. and {Nolta}, Michael R. and {Orlowski-Scherer}, John and {Pagano}, Luca and {Page}, Lyman A. and {Pandey}, Shivam and {Partridge}, Bruce and {Perez Sarmiento}, Karen and {Prince}, Heather and {Puddu}, Roberto and {Qu}, Frank J. and {Ragavan}, Damien C. and {Ried Guachalla}, Bernardita and {Rogers}, Keir K. and {Rojas}, Felipe and {Sakuma}, Tai and {Schaan}, Emmanuel and {Schmitt}, Benjamin L. and {Sehgal}, Neelima and {Shaikh}, Shabbir and {Sherwin}, Blake D. and {Sierra}, Carlos and {Sievers}, Jon and {Sif{\'o}n}, Crist{\'o}bal and {Simon}, Sara and {Sonka}, Rita and {Spergel}, David N. and {Staggs}, Suzanne T. and {Storer}, Emilie and {Surrao}, Kristen and {Switzer}, Eric R. and {Tampier}, Niklas and {Thornton}, Robert and {Trac}, Hy and {Tucker}, Carole and {Ullom}, Joel and {Vale}, Leila R. and {Van Engelen}, Alexander and {Van Lanen}, Jeff and {Vargas}, Cristian and {Vavagiakis}, Eve M. and {Wagoner}, Kasey and {Wang}, Yuhan and {Wenzl}, Lukas and {Wollack}, Edward J. and {Zheng}, Kaiwen},
        title = "{The Atacama Cosmology Telescope: DR6 Power Spectra, Likelihoods and $\Lambda$CDM Parameters}",
      journal = {arXiv e-prints},
         year = 2025,
        month = mar,
          eid = {},
        pages = {},
          doi = {10.48550/arXiv.2503.14452},
archivePrefix = {arXiv},
       eprint = {2503.14452},
 primaryClass = {astro-ph.CO},
       adsurl = {https://ui.adsabs.harvard.edu/abs/2025arXiv250314452L}
}

@ARTICLE{2311.14938,
       author = {{Mittal}, Vasudev and {Oayda}, Oliver T. and {Lewis}, Geraint F.},
        title = "{The cosmic dipole in the Quaia sample of quasars: a Bayesian analysis}",
      journal = {\mnras},
         year = 2024,
        month = jan,
       volume = {527},
       number = {3},
        pages = {8497-8510},
          doi = {10.1093/mnras/stad3706},
archivePrefix = {arXiv},
       eprint = {2311.14938},
 primaryClass = {astro-ph.CO},
       adsurl = {https://ui.adsabs.harvard.edu/abs/2024MNRAS.527.8497M}
}

@ARTICLE{1911.07754,
       author = {{Gratton}, Steven and {Challinor}, Anthony},
        title = "{Understanding parameter differences between analyses employing nested data subsets}",
      journal = {\mnras},
         year = 2020,
        month = dec,
       volume = {499},
       number = {3},
        pages = {3410-3416},
          doi = {10.1093/mnras/staa2996},
archivePrefix = {arXiv},
       eprint = {1911.07754},
 primaryClass = {astro-ph.IM},
       adsurl = {https://ui.adsabs.harvard.edu/abs/2020MNRAS.499.3410G}
}

@ARTICLE{1807.06210,
       author = {{Planck Collaboration} and {Aghanim}, N. and {Akrami}, Y. and {Ashdown}, M. and {Aumont}, J. and {Baccigalupi}, C. and {Ballardini}, M. and {Banday}, A.~J. and {Barreiro}, R.~B. and {Bartolo}, N. and {Basak}, S. and {Benabed}, K. and {Bernard}, J. -P. and {Bersanelli}, M. and {Bielewicz}, P. and {Bock}, J.~J. and {Bond}, J.~R. and {Borrill}, J. and {Bouchet}, F.~R. and {Boulanger}, F. and {Bucher}, M. and {Burigana}, C. and {Calabrese}, E. and {Cardoso}, J. -F. and {Carron}, J. and {Challinor}, A. and {Chiang}, H.~C. and {Colombo}, L.~P.~L. and {Combet}, C. and {Crill}, B.~P. and {Cuttaia}, F. and {de Bernardis}, P. and {de Zotti}, G. and {Delabrouille}, J. and {Di Valentino}, E. and {Diego}, J.~M. and {Dor{\'e}}, O. and {Douspis}, M. and {Ducout}, A. and {Dupac}, X. and {Efstathiou}, G. and {Elsner}, F. and {En{\ss}lin}, T.~A. and {Eriksen}, H.~K. and {Fantaye}, Y. and {Fernandez-Cobos}, R. and {Finelli}, F. and {Forastieri}, F. and {Frailis}, M. and {Fraisse}, A.~A. and {Franceschi}, E. and {Frolov}, A. and {Galeotta}, S. and {Galli}, S. and {Ganga}, K. and {G{\'e}nova-Santos}, R.~T. and {Gerbino}, M. and {Ghosh}, T. and {Gonz{\'a}lez-Nuevo}, J. and {G{\'o}rski}, K.~M. and {Gratton}, S. and {Gruppuso}, A. and {Gudmundsson}, J.~E. and {Hamann}, J. and {Handley}, W. and {Hansen}, F.~K. and {Herranz}, D. and {Hivon}, E. and {Huang}, Z. and {Jaffe}, A.~H. and {Jones}, W.~C. and {Karakci}, A. and {Keih{\"a}nen}, E. and {Keskitalo}, R. and {Kiiveri}, K. and {Kim}, J. and {Knox}, L. and {Krachmalnicoff}, N. and {Kunz}, M. and {Kurki-Suonio}, H. and {Lagache}, G. and {Lamarre}, J. -M. and {Lasenby}, A. and {Lattanzi}, M. and {Lawrence}, C.~R. and {Le Jeune}, M. and {Levrier}, F. and {Lewis}, A. and {Liguori}, M. and {Lilje}, P.~B. and {Lindholm}, V. and {L{\'o}pez-Caniego}, M. and {Lubin}, P.~M. and {Ma}, Y. -Z. and {Mac{\'\i}as-P{\'e}rez}, J.~F. and {Maggio}, G. and {Maino}, D. and {Mandolesi}, N. and {Mangilli}, A. and {Marcos-Caballero}, A. and {Maris}, M. and {Martin}, P.~G. and {Mart{\'\i}nez-Gonz{\'a}lez}, E. and {Matarrese}, S. and {Mauri}, N. and {McEwen}, J.~D. and {Melchiorri}, A. and {Mennella}, A. and {Migliaccio}, M. and {Miville-Desch{\^e}nes}, M. -A. and {Molinari}, D. and {Moneti}, A. and {Montier}, L. and {Morgante}, G. and {Moss}, A. and {Natoli}, P. and {Pagano}, L. and {Paoletti}, D. and {Partridge}, B. and {Patanchon}, G. and {Perrotta}, F. and {Pettorino}, V. and {Piacentini}, F. and {Polastri}, L. and {Polenta}, G. and {Puget}, J. -L. and {Rachen}, J.~P. and {Reinecke}, M. and {Remazeilles}, M. and {Renzi}, A. and {Rocha}, G. and {Rosset}, C. and {Roudier}, G. and {Rubi{\~n}o-Mart{\'\i}n}, J.~A. and {Ruiz-Granados}, B. and {Salvati}, L. and {Sandri}, M. and {Savelainen}, M. and {Scott}, D. and {Sirignano}, C. and {Sunyaev}, R. and {Suur-Uski}, A. -S. and {Tauber}, J.~A. and {Tavagnacco}, D. and {Tenti}, M. and {Toffolatti}, L. and {Tomasi}, M. and {Trombetti}, T. and {Valiviita}, J. and {Van Tent}, B. and {Vielva}, P. and {Villa}, F. and {Vittorio}, N. and {Wandelt}, B.~D. and {Wehus}, I.~K. and {White}, M. and {White}, S.~D.~M. and {Zacchei}, A. and {Zonca}, A.},
        title = "{Planck 2018 results. VIII. Gravitational lensing}",
      journal = {\aap},
         year = 2020,
        month = sep,
       volume = {641},
          eid = {A8},
        pages = {A8},
          doi = {10.1051/0004-6361/201833886},
archivePrefix = {arXiv},
       eprint = {1807.06210},
 primaryClass = {astro-ph.CO},
       adsurl = {https://ui.adsabs.harvard.edu/abs/2020A&A...641A...8P}
}

@ARTICLE{2307.01753,
       author = {{Rezaie}, Mehdi and {Ross}, Ashley J. and {Seo}, Hee-Jong and {Kong}, Hui and {Porredon}, Anna and {Samushia}, Lado and {Chaussidon}, Edmond and {Krolewski}, Alex and {de Mattia}, Arnaud and {Beutler}, Florian and {Aguilar}, Jessica Nicole and {Ahlen}, Steven and {Alam}, Shadab and {Avila}, Santiago and {Bahr-Kalus}, Benedict and {Bermejo-Climent}, Jose and {Brooks}, David and {Claybaugh}, Todd and {Cole}, Shaun and {Dawson}, Kyle and {de la Macorra}, Axel and {Doel}, Peter and {Font-Ribera}, Andreu and {Forero-Romero}, Jaime E. and {Gontcho}, Satya Gontcho A. and {Guy}, Julien and {Honscheid}, Klaus and {Huterer}, Dragan and {Kisner}, Theodore and {Landriau}, Martin and {Levi}, Michael and {Manera}, Marc and {Meisner}, Aaron and {Miquel}, Ramon and {Mueller}, Eva-Maria and {Myers}, Adam and {Newman}, Jeffrey A. and {Nie}, Jundan and {Palanque-Delabrouille}, Nathalie and {Percival}, Will and {Poppett}, Claire and {Rossi}, Graziano and {Sanchez}, Eusebio and {Schubnell}, Michael and {Tarl{\'e}}, Gregory and {Weaver}, Benjamin Alan and {Y{\`e}che}, Christophe and {Zhou}, Zhimin and {Zou}, Hu},
        title = "{Local primordial non-Gaussianity from the large-scale clustering of photometric DESI luminous red galaxies}",
      journal = {\mnras},
         year = 2024,
        month = aug,
       volume = {532},
       number = {2},
        pages = {1902-1928},
          doi = {10.1093/mnras/stae886},
archivePrefix = {arXiv},
       eprint = {2307.01753},
 primaryClass = {astro-ph.CO},
       adsurl = {https://ui.adsabs.harvard.edu/abs/2024MNRAS.532.1902R}
}

@article{scipy,
    author = {Virtanen, Pauli and Gommers, Ralf and Oliphant, Travis E. and others},
    title = {SciPy 1.0: Fundamental Algorithms for Scientific Computing in Python},
    journal = {Nature Methods},
    volume = {17},
    pages = {261--272},
    year = {2020},
    doi = {10.1038/s41592-019-0686-2},
    url = {https://scipy.org/}
}

@article{numpy,
    author = {Harris, Charles R. and Millman, K. Jarrod and van der Walt, Stéfan J. and others},
    title = {Array programming with {NumPy}},
    journal = {Nature},
    volume = {585},
    pages = {357--362},
    year = {2020},
    doi = {10.1038/s41586-020-2649-2},
    url = {https://numpy.org/}
}

@article{matplotlib,
    author = {Hunter, J. D.},
    title = {Matplotlib: A 2D Graphics Environment},
    journal = {Computing in Science \& Engineering},
    volume = {9},
    number = {3},
    pages = {90--95},
    year = {2007},
    doi = {10.1109/MCSE.2007.55},
    url = {https://matplotlib.org/}
}

@article{healpy,
    author = {Zonca, Andrea and Singer, Leo and Lenz, Daniel and Reinecke, Martin and Rosset, Cyrille and Hivon, Eric and Gorski, Krzysztof},
    title = {healpy: equal area pixelization and spherical harmonics transforms for data on the sphere in Python},
    journal = {Journal of Open Source Software},
    volume = {4},
    number = {35},
    pages = {1298},
    year = {2019},
    doi = {10.21105/joss.01298},
    url = {https://healpy.readthedocs.io/}
}

@misc{xqml,
    author = {Vanneste, Simon and Hamilton, Jean-Christophe and Grain, Julien},
    title = {xQML: A Pure Python Implementation of the QML Method},
    year = {2021},
    url = {https://gitlab.in2p3.fr/xQML/xQML}
}

@article{healpix,
       author = {{G{\'o}rski}, K.~M. and {Hivon}, E. and {Banday}, A.~J. and {Wandelt}, B.~D. and {Hansen}, F.~K. and {Reinecke}, M. and {Bartelmann}, M.},
        title = "{HEALPix: A Framework for High-Resolution Discretization and Fast Analysis of Data Distributed on the Sphere}",
      journal = {\apj},
         year = 2005,
        month = apr,
       volume = {622},
       number = {2},
        pages = {759-771},
          doi = {10.1086/427976},
archivePrefix = {arXiv},
       eprint = {astro-ph/0409513},
 primaryClass = {astro-ph},
       adsurl = {https://ui.adsabs.harvard.edu/abs/2005ApJ...622..759G}
}

@ARTICLE{byrnes2010,
       author = {{Byrnes}, Christian T. and {Choi}, Ki-Young},
        title = "{Review of Local Non-Gaussianity from Multifield Inflation}",
      journal = {Advances in Astronomy},
         year = 2010,
        month = jan,
       volume = {2010},
          eid = {724525},
        pages = {724525},
          doi = {10.1155/2010/724525},
archivePrefix = {arXiv},
       eprint = {1002.3110},
 primaryClass = {astro-ph.CO},
       adsurl = {https://ui.adsabs.harvard.edu/abs/2010AdAst2010E..76B}
}

@ARTICLE{deputter2017,
       author = {{de Putter}, Roland and {Gleyzes}, J{\'e}r{\^o}me and {Dor{\'e}}, Olivier},
        title = "{Next non-Gaussianity frontier: What can a measurement with {\ensuremath{\sigma}} (f$_{NL}$){\ensuremath{\lesssim}}1 tell us about multifield inflation?}",
      journal = {\prd},
         year = 2017,
        month = jun,
       volume = {95},
       number = {12},
          eid = {123507},
        pages = {123507},
          doi = {10.1103/PhysRevD.95.123507},
archivePrefix = {arXiv},
       eprint = {1612.05248},
 primaryClass = {astro-ph.CO},
       adsurl = {https://ui.adsabs.harvard.edu/abs/2017PhRvD..95l3507D}
}

@ARTICLE{png-lss-review,
       author = {{Desjacques}, V. and {Seljak}, U.},
        title = "{Primordial non-Gaussianity from the large-scale structure}",
      journal = {Classical and Quantum Gravity},
         year = 2010,
        month = jun,
       volume = {27},
       number = {12},
          eid = {124011},
        pages = {124011},
          doi = {10.1088/0264-9381/27/12/124011},
archivePrefix = {arXiv},
       eprint = {1003.5020},
 primaryClass = {astro-ph.CO},
       adsurl = {https://ui.adsabs.harvard.edu/abs/2010CQGra..27l4011D}
}

@ARTICLE{weaverdyck2021,
       author = {{Weaverdyck}, Noah and {Huterer}, Dragan},
        title = "{Mitigating contamination in LSS surveys: a comparison of methods}",
      journal = {\mnras},
         year = 2021,
        month = may,
       volume = {503},
       number = {4},
        pages = {5061-5084},
          doi = {10.1093/mnras/stab709},
archivePrefix = {arXiv},
       eprint = {2007.14499},
 primaryClass = {astro-ph.CO},
       adsurl = {https://ui.adsabs.harvard.edu/abs/2021MNRAS.503.5061W}
}

@ARTICLE{loverde2008,
       author = {{LoVerde}, Marilena and {Afshordi}, Niayesh},
        title = "{Extended Limber approximation}",
      journal = {\prd},
         year = 2008,
        month = dec,
       volume = {78},
       number = {12},
          eid = {123506},
        pages = {123506},
          doi = {10.1103/PhysRevD.78.123506},
archivePrefix = {arXiv},
       eprint = {0809.5112},
 primaryClass = {astro-ph},
       adsurl = {https://ui.adsabs.harvard.edu/abs/2008PhRvD..78l3506L}
}

@ARTICLE{simon2007,
       author = {{Simon}, P.},
        title = "{How accurate is Limber's equation?}",
      journal = {\aap},
         year = 2007,
        month = oct,
       volume = {473},
       number = {3},
        pages = {711-714},
          doi = {10.1051/0004-6361:20066352},
archivePrefix = {arXiv},
       eprint = {astro-ph/0609165},
 primaryClass = {astro-ph},
       adsurl = {https://ui.adsabs.harvard.edu/abs/2007A&A...473..711S}
}

@ARTICLE{so,
       author = {{Ade}, Peter and {Aguirre}, James and {Ahmed}, Zeeshan and {Aiola}, Simone and {Ali}, Aamir and {Alonso}, David and {Alvarez}, Marcelo A. and {Arnold}, Kam and {Ashton}, Peter and {Austermann}, Jason and {Awan}, Humna and {Baccigalupi}, Carlo and {Baildon}, Taylor and {Barron}, Darcy and {Battaglia}, Nick and {Battye}, Richard and {Baxter}, Eric and {Bazarko}, Andrew and {Beall}, James A. and {Bean}, Rachel and {Beck}, Dominic and {Beckman}, Shawn and {Beringue}, Benjamin and {Bianchini}, Federico and {Boada}, Steven and {Boettger}, David and {Bond}, J. Richard and {Borrill}, Julian and {Brown}, Michael L. and {Bruno}, Sarah Marie and {Bryan}, Sean and {Calabrese}, Erminia and {Calafut}, Victoria and {Calisse}, Paolo and {Carron}, Julien and {Challinor}, Anthony and {Chesmore}, Grace and {Chinone}, Yuji and {Chluba}, Jens and {Cho}, Hsiao-Mei Sherry and {Choi}, Steve and {Coppi}, Gabriele and {Cothard}, Nicholas F. and {Coughlin}, Kevin and {Crichton}, Devin and {Crowley}, Kevin D. and {Crowley}, Kevin T. and {Cukierman}, Ari and {D'Ewart}, John M. and {D{\"u}nner}, Rolando and {de Haan}, Tijmen and {Devlin}, Mark and {Dicker}, Simon and {Didier}, Joy and {Dobbs}, Matt and {Dober}, Bradley and {Duell}, Cody J. and {Duff}, Shannon and {Duivenvoorden}, Adri and {Dunkley}, Jo and {Dusatko}, John and {Errard}, Josquin and {Fabbian}, Giulio and {Feeney}, Stephen and {Ferraro}, Simone and {Flux{\`a}}, Pedro and {Freese}, Katherine and {Frisch}, Josef C. and {Frolov}, Andrei and {Fuller}, George and {Fuzia}, Brittany and {Galitzki}, Nicholas and {Gallardo}, Patricio A. and {Tomas Galvez Ghersi}, Jose and {Gao}, Jiansong and {Gawiser}, Eric and {Gerbino}, Martina and {Gluscevic}, Vera and {Goeckner-Wald}, Neil and {Golec}, Joseph and {Gordon}, Sam and {Gralla}, Megan and {Green}, Daniel and {Grigorian}, Arpi and {Groh}, John and {Groppi}, Chris and {Guan}, Yilun and {Gudmundsson}, Jon E. and {Han}, Dongwon and {Hargrave}, Peter and {Hasegawa}, Masaya and {Hasselfield}, Matthew and {Hattori}, Makoto and {Haynes}, Victor and {Hazumi}, Masashi and {He}, Yizhou and {Healy}, Erin and {Henderson}, Shawn W. and {Hervias-Caimapo}, Carlos and {Hill}, Charles A. and {Hill}, J. Colin and {Hilton}, Gene and {Hilton}, Matt and {Hincks}, Adam D. and {Hinshaw}, Gary and {Hlo{\v{z}}ek}, Ren{\'e}e and {Ho}, Shirley and {Ho}, Shuay-Pwu Patty and {Howe}, Logan and {Huang}, Zhiqi and {Hubmayr}, Johannes and {Huffenberger}, Kevin and {Hughes}, John P. and {Ijjas}, Anna and {Ikape}, Margaret and {Irwin}, Kent and {Jaffe}, Andrew H. and {Jain}, Bhuvnesh and {Jeong}, Oliver and {Kaneko}, Daisuke and {Karpel}, Ethan D. and {Katayama}, Nobuhiko and {Keating}, Brian and {Kernasovskiy}, Sarah S. and {Keskitalo}, Reijo and {Kisner}, Theodore and {Kiuchi}, Kenji and {Klein}, Jeff and {Knowles}, Kenda and {Koopman}, Brian and {Kosowsky}, Arthur and {Krachmalnicoff}, Nicoletta and {Kuenstner}, Stephen E. and {Kuo}, Chao-Lin and {Kusaka}, Akito and {Lashner}, Jacob and {Lee}, Adrian and {Lee}, Eunseong and {Leon}, David and {Leung}, Jason S. -Y. and {Lewis}, Antony and {Li}, Yaqiong and {Li}, Zack and {Limon}, Michele and {Linder}, Eric and {Lopez-Caraballo}, Carlos and {Louis}, Thibaut and {Lowry}, Lindsay and {Lungu}, Marius and {Madhavacheril}, Mathew and {Mak}, Daisy and {Maldonado}, Felipe and {Mani}, Hamdi and {Mates}, Ben and {Matsuda}, Frederick and {Maurin}, Lo{\"\i}c and {Mauskopf}, Phil and {May}, Andrew and {McCallum}, Nialh and {McKenney}, Chris and {McMahon}, Jeff and {Meerburg}, P. Daniel and {Meyers}, Joel and {Miller}, Amber and {Mirmelstein}, Mark and {Moodley}, Kavilan and {Munchmeyer}, Moritz and {Munson}, Charles and {Naess}, Sigurd and {Nati}, Federico and {Navaroli}, Martin and {Newburgh}, Laura and {Nguyen}, Ho Nam and {Niemack}, Michael and {Nishino}, Haruki and {Orlowski-Scherer}, John and {Page}, Lyman and {Partridge}, Bruce and {Peloton}, Julien and {Perrotta}, Francesca and {Piccirillo}, Lucio and {Pisano}, Giampaolo and {Poletti}, Davide and {Puddu}, Roberto and {Puglisi}, Giuseppe and {Raum}, Chris and {Reichardt}, Christian L. and {Remazeilles}, Mathieu and {Rephaeli}, Yoel and {Riechers}, Dominik and {Rojas}, Felipe and {Roy}, Anirban and {Sadeh}, Sharon and {Sakurai}, Yuki and {Salatino}, Maria and {Sathyanarayana Rao}, Mayuri and {Schaan}, Emmanuel and {Schmittfull}, Marcel and {Sehgal}, Neelima and {Seibert}, Joseph},
        title = "{The Simons Observatory: science goals and forecasts}",
      journal = {\jcap},
         year = 2019,
        month = feb,
       volume = {2019},
       number = {2},
          eid = {056},
        pages = {056},
          doi = {10.1088/1475-7516/2019/02/056},
archivePrefix = {arXiv},
       eprint = {1808.07445},
 primaryClass = {astro-ph.CO},
       adsurl = {https://ui.adsabs.harvard.edu/abs/2019JCAP...02..056A}
}

@ARTICLE{aso,
       author = {{Abitbol}, M. and {Abril-Cabezas}, I. and {Adachi}, S. and {Ade}, P. and {Adler}, A.~E. and {Agrawal}, P. and {Aguirre}, J. and {Ahmed}, Z. and {Aiola}, S. and {Alford}, T. and {Ali}, A. and {Alonso}, D. and {Alvarez}, M.~A. and {An}, R. and {Arnold}, K. and {Ashton}, P. and {Atkins}, Z. and {Austermann}, J. and {Azzoni}, S. and {Baccigalupi}, C. and {Baleato Lizancos}, A. and {Barron}, D. and {Barry}, P. and {Bartlett}, J. and {Battaglia}, N. and {Battye}, R. and {Baxter}, E. and {Bazarko}, A. and {Beall}, J.~A. and {Bean}, R. and {Beck}, D. and {Beckman}, S. and {Begin}, J. and {Beheshti}, A. and {Beringue}, B. and {Bhandarkar}, T. and {Bhimani}, S. and {Bianchini}, F. and {Biermann}, E. and {Biquard}, S. and {Bixler}, B. and {Boada}, S. and {Boettger}, D. and {Bolliet}, B. and {Bond}, J.~R. and {Borrill}, J. and {Borrow}, J. and {Braithwaite}, C. and {Brien}, T.~L.~R. and {Brown}, M.~L. and {Bruno}, S.~M. and {Bryan}, S. and {Bustos}, R. and {Cai}, H. and {Calabrese}, E. and {Calafut}, V. and {Carl}, F.~M. and {Carones}, A. and {Carron}, J. and {Challinor}, A. and {Chanial}, P. and {Chen}, N. and {Cheung}, K. and {Chiang}, B. and {Chinone}, Y. and {Chluba}, J. and {Cho}, H.~S. and {Choi}, S.~K. and {Chu}, M. and {Clancy}, J. and {Clark}, S.~E. and {Clarke}, P. and {Cleary}, J. and {Clements}, D.~L. and {Connors}, J. and {Contaldi}, C. and {Coppi}, G. and {Corbett}, L. and {Cothard}, N.~F. and {Coulton}, W. and {Crowley}, K.~D. and {Crowley}, K.~T. and {Cukierman}, A. and {D'Ewart}, J.~M. and {Dachlythra}, K. and {Datta}, R. and {Day-Weiss}, S. and {de Haan}, T. and {Devlin}, M. and {Di Mascolo}, L. and {Dicker}, S. and {Dober}, B. and {Doux}, C. and {Dow}, P. and {Doyle}, S. and {Duell}, C.~J. and {Duff}, S.~M. and {Duivenvoorden}, A.~J. and {Dunkley}, J. and {Dutcher}, D. and {D{\"u}nner}, R. and {Edenton}, M. and {El Bouhargani}, H. and {Errard}, J. and {Fabbian}, G. and {Fanfani}, V. and {Farren}, G.~S. and {Fergusson}, J. and {Ferraro}, S. and {Flauger}, R. and {Foster}, A. and {Freese}, K. and {Frisch}, J.~C. and {Frolov}, A. and {Fuller}, G. and {Galitzki}, N. and {Gallardo}, P.~A. and {Galvez Ghersi}, J.~T. and {Ganga}, K. and {Gao}, J. and {Garrido}, X. and {Gawiser}, E. and {Gerbino}, M. and {Gerras}, R. and {Giardiello}, S. and {Gill}, A. and {Gilles}, V. and {Giri}, U. and {Gleave}, E. and {Gluscevic}, V. and {Goeckner-Wald}, N. and {Golec}, J.~E. and {Gordon}, S. and {Gralla}, M. and {Gratton}, S. and {Green}, D. and {Groh}, J.~C. and {Groppi}, C. and {Guan}, Y. and {Gupta}, N. and {Gudmundsson}, J.~E. and {Hagstotz}, S. and {Hargrave}, P. and {Haridas}, S. and {Harrington}, K. and {Harrison}, I. and {Hasegawa}, M. and {Hasselfield}, M. and {Haynes}, V. and {Hazumi}, M. and {He}, A. and {Healy}, E. and {Henderson}, S.~W. and {Hensley}, B.~S. and {Hertig}, E. and {Herv{\'\i}as-Caimapo}, C. and {Higuchi}, M. and {Hill}, C.~A. and {Hill}, J.~C. and {Hilton}, G. and {Hilton}, M. and {Hincks}, A.~D. and {Hinshaw}, G. and {Hlo{\v{z}}ek}, R. and {Ho}, A.~Y.~Q. and {Ho}, S. and {Ho}, S.~P. and {Hoang}, T.~D. and {Hoh}, J. and {Hornecker}, E. and {Hornsby}, A.~L. and {Hotinli}, S.~C. and {Huang}, Z. and {Huber}, Z.~B. and {Hubmayr}, J. and {Huffenberger}, K. and {Hughes}, J.~P. and {Idicherian Lonappan}, A. and {Ikape}, M. and {Irwin}, K. and {Iuliano}, J. and {Jaffe}, A.~H. and {Jain}, B. and {Jense}, H.~T. and {Jeong}, O. and {Johnson}, A. and {Johnson}, B.~R. and {Johnson}, M. and {Jones}, M. and {Jost}, B. and {Kaneko}, D. and {Karpel}, E.~D. and {Kasai}, Y. and {Katayama}, N. and {Keating}, B. and {Keller}, B. and {Keskitalo}, R. and {Kim}, J. and {Kisner}, T. and {Kiuchi}, K.},
        title = "{The Simons Observatory: science goals and forecasts for the enhanced Large Aperture Telescope}",
      journal = {\jcap},
         year = 2025,
        month = aug,
       volume = {2025},
       number = {8},
          eid = {034},
        pages = {034},
          doi = {10.1088/1475-7516/2025/08/034},
archivePrefix = {arXiv},
       eprint = {2503.00636},
 primaryClass = {astro-ph.IM},
       adsurl = {https://ui.adsabs.harvard.edu/abs/2025JCAP...08..034A}
}

@ARTICLE{embil-villagra2025,
       author = {{Embil Villagra}, Carmen and {Farren}, Gerrit and {Fabbian}, Giulio and {Bolliet}, Boris and {Abril-Cabezas}, Irene and {Alonso}, David and {Challinor}, Anthony and {Dunkley}, Jo and {Kim}, Joshua and {MacCrann}, Niall and {McCarthy}, Fiona and {Moodley}, Kavilan and {Qu}, Frank J. and {Sherwin}, Blake and {Sifon}, Cristobal and {van Engelen}, Alexander and {Wollack}, Edward J.},
        title = "{The Atacama Cosmology Telescope: High-redshift measurement of structure growth from the cross-correlation of Quaia quasars and CMB lensing from ACT DR6 and $\textit{Planck}$ PR4}",
      journal = {arXiv e-prints},
         year = 2025,
        month = jul,
          eid = {arXiv:2507.08798},
        pages = {arXiv:2507.08798},
          doi = {10.48550/arXiv.2507.08798},
archivePrefix = {arXiv},
       eprint = {2507.08798},
 primaryClass = {astro-ph.CO},
       adsurl = {https://ui.adsabs.harvard.edu/abs/2025arXiv250708798E}
}

@ARTICLE{unwise,
       author = {{Schlafly}, Edward F. and {Meisner}, Aaron M. and {Green}, Gregory M.},
        title = "{The unWISE Catalog: Two Billion Infrared Sources from Five Years of WISE Imaging}",
      journal = {\apjs},
         year = 2019,
        month = feb,
       volume = {240},
       number = {2},
          eid = {30},
        pages = {30},
          doi = {10.3847/1538-4365/aafbea},
archivePrefix = {arXiv},
       eprint = {1901.03337},
 primaryClass = {astro-ph.IM},
       adsurl = {https://ui.adsabs.harvard.edu/abs/2019ApJS..240...30S}
}

@ARTICLE{wise,
       author = {{Wright}, Edward L. and {Eisenhardt}, Peter R.~M. and {Mainzer}, Amy K. and {Ressler}, Michael E. and {Cutri}, Roc M. and {Jarrett}, Thomas and {Kirkpatrick}, J. Davy and {Padgett}, Deborah and {McMillan}, Robert S. and {Skrutskie}, Michael and {Stanford}, S.~A. and {Cohen}, Martin and {Walker}, Russell G. and {Mather}, John C. and {Leisawitz}, David and {Gautier}, III, Thomas N. and {McLean}, Ian and {Benford}, Dominic and {Lonsdale}, Carol J. and {Blain}, Andrew and {Mendez}, Bryan and {Irace}, William R. and {Duval}, Valerie and {Liu}, Fengchuan and {Royer}, Don and {Heinrichsen}, Ingolf and {Howard}, Joan and {Shannon}, Mark and {Kendall}, Martha and {Walsh}, Amy L. and {Larsen}, Mark and {Cardon}, Joel G. and {Schick}, Scott and {Schwalm}, Mark and {Abid}, Mohamed and {Fabinsky}, Beth and {Naes}, Larry and {Tsai}, Chao-Wei},
        title = "{The Wide-field Infrared Survey Explorer (WISE): Mission Description and Initial On-orbit Performance}",
      journal = {\aj},
         year = 2010,
        month = dec,
       volume = {140},
       number = {6},
        pages = {1868-1881},
          doi = {10.1088/0004-6256/140/6/1868},
archivePrefix = {arXiv},
       eprint = {1008.0031},
 primaryClass = {astro-ph.IM},
       adsurl = {https://ui.adsabs.harvard.edu/abs/2010AJ....140.1868W}
}

\end{document}